\documentclass[superscriptaddress,twocolumn,secnumarabic,
nobibnotes,aps,prd,nofootinbib]{revtex4-2}

\usepackage{graphicx}
\usepackage{dcolumn}
\usepackage{epsf}
\usepackage[colorlinks=true, linkcolor=blue , citecolor=YellowOrange, urlcolor=magenta]{hyperref}
 \usepackage{bm}
 \usepackage{epstopdf}
 \usepackage{natbib}
 \usepackage{float}
\usepackage{amsmath}
 \usepackage{amssymb}
 \usepackage{amsfonts}
 \usepackage{resizegather}
 \usepackage{caption}
 \usepackage{color}
 \usepackage{subfigure}
\usepackage{soul}
 \usepackage{blindtext}
 \usepackage{url}
\usepackage[dvipsnames,table]{xcolor}

\usepackage[capitalize]{cleveref}
\usepackage[newcommands]{ragged2e}
\usepackage{physics}

\usepackage{lipsum}
\usepackage{graphicx}
   
\usepackage{array,mathtools,amssymb,booktabs}
\newcolumntype{C}{>{$}c<{$}}

\begin{document}

\title{Do we need wavelets in the late Universe?}

\author{Luis A. Escamilla}
\email{luis.escamilla@icf.unam.mx}
\affiliation{School of Mathematics and Statistics, University of Sheffield, Hounsfield Road, Sheffield S3 7RH, United Kingdom}
\affiliation{Instituto de Ciencias F\'isicas, Universidad Nacional Aut\'onoma de M\'exico, Cuernavaca, Morelos, 62210, M\'exico}

\author{Emre \"{O}z\"{u}lker}
\email{ozulker17@itu.edu.tr}
\affiliation{Department of Physics, Istanbul Technical University, Maslak 34469 Istanbul, T\"{u}rkiye}
\affiliation{School of Mathematics and Statistics, University of Sheffield, Hounsfield Road, Sheffield S3 7RH, United Kingdom}

\author{\"{O}zg\"{u}r Akarsu}
\email{akarsuo@itu.edu.tr}
\affiliation{Department of Physics, Istanbul Technical University, Maslak 34469 Istanbul, T\"{u}rkiye}

\author{Eleonora Di Valentino}
\email{e.divalentino@sheffield.ac.uk}
\affiliation{School of Mathematics and Statistics, University of Sheffield, Hounsfield Road, Sheffield S3 7RH, United Kingdom}

\author{J. A. V\'azquez}
\email{javazquez@icf.unam.mx}
\affiliation{Instituto de Ciencias F\'isicas, Universidad Nacional Aut\'onoma de M\'exico, Cuernavaca, Morelos, 62210, M\'exico}


\begin{abstract}

 
We parameterize the Hubble function by adding Hermitian wavelets to the Hubble radius of $\Lambda$CDM. This allows us to build Hubble functions that oscillate around $\Lambda$CDM at late times without modifying its angular diameter distance to last scattering. We perform parameter inference and model selection procedures on these new Hubble functions at the background level. 
In our analyses consisting of a wide variety of cosmological observations, we find that baryon acoustic oscillations (BAO) data play a crucial role in determining the constraints on the wavelet parameters. In particular, we focus on the differences between SDSS- and DESI-BAO datasets and find that wavelets provide a better fit to the data when either of the BAO datasets is present. 
However, DESI-BAO has a preference for the center of the wavelets to be around $z \sim 0.7$, while SDSS-BAO prefers higher redshifts of $z > 1$. This difference appears to be driven by the discrepancies between these two datasets in their $D_H / r_{\rm d}$ measurements at $z = 0.51$ and $z \sim 2.3$. Finally, we also derive the consequences of the wavelets on a dark energy component. We find that the dark energy density oscillates by construction and also attains negative values at large redshifts ($z\gtrsim2$) as a consequence of the SDSS-BAO data. We conclude that while the early universe and the constraints on the matter density and the Hubble constant remain unchanged, wavelets are favored in the late universe by the BAO data. Specifically, there is a significant improvement at more than $3\sigma$ in the fit when new DESI-BAO data are included in the analysis.
\\


\end{abstract}

\maketitle

\section{INTRODUCTION}\label{section:intro}

The standard model of cosmology, the $\Lambda$ cold dark matter (CDM) model dubbed $\Lambda$CDM, provides excellent accuracy in explaining the majority of high-precision cosmological observations despite the simplicity of the model~\cite{SupernovaSearchTeam:1998fmf, SupernovaCosmologyProject:1998vns, Amanullah:2010vv, SupernovaCosmologyProject:1997zqe, Planck:2015mrs, WMAP:2012fli, WMAP:2010qai, Planck:2018vyg, 2013PhR...530...87W, Li:2012dt, Reid:2009xm, SDSS:2009ocz, SDSS:2008tqn,Mossa:2020gjc,ACT:2020gnv,eBOSS:2020yzd,SPT-3G:2022hvq}. Yet, it faces numerous challenges, both theoretical and statistical. The most prominent of these are the Cosmological Constant Problem~\cite{Weinberg:1988cp, Sahni:2002kh, Wang:1999fa, Sahni:2002kh} and the Coincidence Problem~\cite{Sahni:2004ai, Alam:2004jy, Bull:2015stt} on the theoretical side, and the $H_0$ tension~\cite{Verde:2019ivm, DiValentino:2020zio, DiValentino:2021izs, Riess:2021jrx,Knox:2019rjx,Kamionkowski:2022pkx,Verde:2023lmm,DiValentino:2024yew,Breuval:2024lsv,Li:2024yoe,Murakami:2023xuy,Perivolaropoulos:2024yxv} and the $S_8$ tension~\cite{DiValentino:2020vvd, Douspis:2018xlj,DES:2021wwk,DiValentino:2018gcu,Kilo-DegreeSurvey:2023gfr,Troster:2019ean,Heymans:2020gsg,Dalal:2023olq,Chen:2024vvk,ACT:2024okh,DES:2024oud,Harnois-Deraps:2024ucb,Dvornik:2022xap,Adil:2023jtu} on the observational side; see also Refs.~\cite{Abdalla:2022yfr, Perivolaropoulos:2021jda,DiValentino:2022fjm} for reviews of the tensions and anomalies in cosmology.

A plethora of alternative models to $\Lambda$CDM have been proposed to provide a better description of the observed universe~\cite{Murgia:2016ccp,Pourtsidou:2016ico,Nunes:2016dlj,DiValentino:2016hlg,Kumar:2016zpg,Kumar:2017dnp,DiValentino:2017iww,DiValentino:2017gzb,DiValentino:2017rcr,DiValentino:2017oaw,Dutta:2018vmq,Yang:2018uae,Poulin:2018cxd,DiValentino:2019exe,Visinelli:2019qqu,Pan:2019hac,DiValentino:2019ffd,Smith:2019ihp,Niedermann:2019olb,vonMarttens:2019ixw,Akarsu:2019hmw,Ye:2020btb,Yang:2020zuk,Yang:2020uga,Krishnan:2020obg,Lucca:2020zjb,DiValentino:2020naf,Calderon:2020hoc,DiValentino:2020leo,DiValentino:2020vnx,Yang:2021flj,Kumar:2021eev,Yang:2021eud,Nunes:2021zzi,Schoneberg:2021qvd,Ye:2021iwa,Akarsu:2021fol,Poulin:2021bjr,DiValentino:2021rjj,Alestas:2021luu,Gariazzo:2021qtg,Niedermann:2021vgd,Sen:2021wld,Heisenberg:2022lob,Anchordoqui:2022gmw,DiGennaro:2022ykp,Akarsu:2022typ,Ong:2022wrs,Lee:2022gzh,Bernui:2023byc,Mishra:2023ueo,Tiwari:2023jle,deSouza:2023sqp,Poulin:2023lkg,vanderWesthuizen:2023hcl,Zhai:2023yny,Cruz:2023lmn,Giare:2023xoc,Ruchika:2023ugh,Greene:2023cro,Adil:2023exv,Niedermann:2023ssr,Frion:2023xwq,Liu:2023kce,Akarsu:2023mfb,Hoerning:2023hks,Vagnozzi:2023nrq,Hogas:2023pjz,Gomez-Valent:2023uof,Lapi:2023plb,Pan:2023mie,Efstathiou:2023fbn,Wen:2023wes,Cervantes-Cota:2023wet,Castello:2023zjr,Forconi:2023hsj,Anchordoqui:2023woo,Pitrou:2023swx,Pan:2023frx,Yao:2023jau,Akarsu:2024qiq,Akarsu:2024qsi,Garcia-Arroyo:2024tqq,Benisty:2024lmj,Halder:2024uao,Greene:2024qis,Krolewski:2024jwj,Silva:2024ift,Giare:2024ytc,Garny:2024ums,Giare:2024akf,Montani:2024pou,Allali:2024anb,Anchordoqui:2024gfa,Bousis:2024rnb,Baryakhtar:2024rky,Seto:2024cgo,Co:2024oek,Gonzalez:2024qjs,Bagherian:2024obh,Giare:2024syw,Akarsu:2024eoo,Lynch:2024hzh,Yadav:2024duq,Aboubrahim:2024spa,Toda:2024ncp,Pooya:2024wsq,Dwivedi:2024okk,Tang:2024gtq,Li:2024qso,Schoneberg:2024ynd,Poulin:2024ken,Jiang:2024xnu,Ruchika:2024ymt,Pedrotti:2024kpn,Kochappan:2024jyf,Manoharan:2024thb} 
and many studies work on reconstructing the expansion history or relevant cosmological functions from the available cosmological data~\cite{BOSS:2014hhw, Wang:2018fng, Poulin:2018zxs, Tamayo:2019gqj, Bonilla:2020wbn, Escamilla:2021uoj, Colgain:2021pmf, Raveri:2021dbu, Pogosian:2021mcs, Bernardo:2021cxi, Escamilla:2023shf,DESI:2024aqx, Sabogal:2024qxs}. However, it is still far from clear what it takes for a model to overcome the challenges faced by $\Lambda$CDM.
Due to the success of the $\Lambda$CDM model in accurately explaining most observations, a considerable number of proposed alternatives act as extensions or slight deviations from the standard model. In this way, many of the characteristics contributing to the success of the standard model are preserved.
It has recently been shown in Ref.~\cite{Akarsu:2022lhx} that, due to the very high precision of the measurement of the angular scale of the sound horizon at last scattering from the cosmic microwave background (CMB) observations by the \textit{Planck} collaboration~\cite{Planck:2018vyg}, if an alternative model is to retain the consistency of $\Lambda$CDM with this measurement without modifying the comoving size of the sound horizon and the $H_0$ parameter, the Hubble radius $H^{-1}(z)$ of the alternative model should deviate from that of the $\Lambda$CDM model in the form of \textit{admissible wavelets}, which are localized oscillatory functions with a vanishing integral.

Wavelets can also be conveniently built by differentiating probability density functions, and they provide an arbitrary number of oscillations through differentiation without increasing the number of free parameters. These deviations from $\Lambda$CDM at the background level can then be mapped to underlying physics, such as an evolving dark energy (DE) density or time-varying gravitational coupling~\cite{Akarsu:2022lhx}. In this regard, to gain insight into potential solutions to the inherent problems of the standard model, a wavelet extension of $\Lambda$CDM can be characterized using parameter inference techniques such as Markov Chain Monte Carlo (MCMC) or Nested Sampling.

In this paper, we study how a wavelet-extension scenario of the standard model performs compared to $\Lambda$CDM when applying parameter-inference and model-selection procedures. The paper is organized as follows: in~\cref{section:wavelet_model}, we describe the wavelet deviations in the Hubble radius studied in this paper; in~\cref{section:data}, we present the datasets and methodology used; in~\cref{section:results}, we discuss the results; and in~\cref{section:conclusions}, we conclude with our findings.

\section{Wavelet deviations in the Hubble radius}\label{section:wavelet_model}

Wavelets are localized oscillatory functions, where ``localized" means that they either have compact support or they approximate compact support by decaying rapidly.\footnote{This behavior can be observed in~\cref{fig:wavelet_varying_params}, where the deviations from $\Lambda$CDM (dashed line) characterized by underlying wavelets are oscillatory and approximately vanish outside of a finite range of redshift.} In this paper, we will use them to write down Hubble radii, $H^{-1}(z)$, whose deviations from the Hubble radius of the $\Lambda$CDM model are wavelets. That is, for a given wavelet $\psi(z)$, 
we write
\begin{equation}
    \frac{1}{H(z)} = \frac{1}{\mathcal{H}(z)} + \psi(z)
    \label{eq:wavelet}
\end{equation}
where $\mathcal{H}(z)$ is the usual Hubble function of the $\Lambda$CDM model, which reads 
\begin{equation}
3\mathcal{H}^2(z) = \rho_{\rm m,0}(1+z)^3 + \rho_{\rm r,0}(1+z)^4 + \rho_{\Lambda},
\end{equation}
with $z$ being the redshift and $\rho_i$ denoting the energy density of the matter (m), radiation (r), and cosmological constant ($\Lambda$) components using the unit convention $8\pi G = c = 1$. Hereafter, the index ``0" denotes the present-day value of any quantity. This framework implicitly assumes the Robertson-Walker (RW) metric from which the Hubble functions are defined, and as usual, the metric associated with the $\Lambda$CDM model is spatially flat.

These sorts of deviations from the Hubble radius of $\Lambda$CDM, defined by~\cref{eq:wavelet}, were first considered in Ref.~\cite{Akarsu:2022lhx}. There, it was shown that for a given $\mathcal{H}(z)$ in agreement with CMB measurements of the angular scale of the sound horizon at last scattering, any deviation in the Hubble radius that retains this agreement must be an admissible wavelet\footnote{Wavelets that satisfy the admissibility condition have a vanishing integral over their whole domain; hence, the angular diameter distance to the last scattering surface is preserved in the deviated scenario.} unless the value of $H_0$ and/or the comoving length of the sound horizon is also modified. 
It is crucial to clarify that, in the present paper, despite making use of the same wavelet framework described by~\cref{eq:wavelet}, we do not require $\mathcal{H}(z)$ itself to be in agreement with CMB measurements; rather, we simply use wavelets on top of $\mathcal{H}(z)^{-1}$ to construct Hubble functions that oscillate around the functional form of $\Lambda$CDM due to the ability of wavelets to provide an arbitrary number of oscillations with a limited number of free parameters.

It is useful to relate the parameters of $\mathcal{H}(z)$ and $H(z)$. A conveniently simple relation exists between the deceleration parameters of the Hubble functions, that is,
\begin{equation}
q(z) = \mathcal{Q}(z) + \dv{\psi(z)}{t},
\end{equation}
where $q(z)$ and $\mathcal{Q}(z)$ are the usual deceleration parameters for the metric solutions associated with $H(z)$ and $\mathcal{H}(z)$, respectively. A special case of particular interest is a scenario where all of the deviations of $H(z)$ from $\mathcal{H}(z)$ arise due to a dynamical dark energy density, $\rho_{\rm DE}(z)$, that replaces the $\rho_\Lambda$ of the $\Lambda$CDM model, namely, the scenario with 
\begin{equation}
3H^2(z) = \rho_{\rm m,0}(1+z)^3 + \rho_{\rm r,0}(1+z)^4 + \rho_{\rm DE}(z).
\end{equation}
Defining the normalized deviation between the Hubble functions,
\begin{equation}
    \delta(z) \equiv \frac{H(z) - \mathcal{H}(z)}{\mathcal{H}(z)}, 
\end{equation}
one has
\begin{equation}
    \rho_{\rm DE}(z) = \rho_{\rm DE,0} + 3 \mathcal{H}^2(z)\delta(z)[2 + \delta(z)].
    \label{eq:dynamicDE}%
\end{equation}
Moreover, we can obtain a DE equation of state (EoS) parameter as 
\begin{equation}
    w_{\rm DE}(z) = -1 + \frac{2(1+z)\mathcal{H}^2}{\rho_{\rm DE}} \qty( \frac{\mathcal{H}'}{\mathcal{H}} \delta(\delta+2) + \delta'(\delta+1)),
\end{equation}
and from here one can define the DE inertial mass density, $\varrho_{\rm DE} \equiv \rho_{\rm DE}(1+w_{\rm DE})$, which reads
\begin{equation}
    \varrho_{\rm DE}(z) = 2(1+z)\mathcal{H}^2 \qty( \frac{\mathcal{H}'}{\mathcal{H}} \delta(\delta+2) + \delta'(\delta+1)),
\end{equation}
where the prime denotes differentiation with respect to the redshift (we have dropped the redshift dependence of the functions for better readability).

Notice that, so far, we have not made any assumptions about $\psi(z)$ other than it being a wavelet. A reliable way to obtain wavelets is by differentiating probability density functions. In parallel to Ref.~\cite{Akarsu:2022lhx}, for the analyses in this paper, we will consider the family of Hermitian wavelets obtained from the derivatives of the Gaussian distribution—in fact, wavelets obtained from probability density functions generically satisfy the admissibility condition, which was a crucial point in Ref.~\cite{Akarsu:2022lhx}, but not here. Let us define a Gaussian distribution
\begin{equation}
    G(z) = -\frac{\alpha_h}{2\beta_h}e^{-\beta_h(z-z^\dagger)^2},
\end{equation}
where $\beta_h > 0$ and $\alpha_h$ can take any value, with the minus sign in front of the function being our convention (since this function can be negative depending on the sign of $\alpha_h$, strictly speaking, it is not a probability density function).
Then, one can define a family of wavelets where the $n^{\rm th}$ member of the family corresponds to the $n^{\rm th}$ derivative of $G(z)$ for $n > 0$, i.e.,
\begin{equation}
    \psi_n(z) \equiv \dv[n]{G(z)}{z},
\end{equation}
yielding
\begin{subequations}
    \begin{align}
    &\psi_1(z) = -2\beta_h\qty[z - z^\dagger]G(z),\\
    &\psi_2(z) = 4\beta_h\qty[\beta_h(z - z^\dagger)^2 - \frac{1}{2}]G(z),\\
    &\psi_3(z) = -8\beta_h^2\qty[\beta_h(z - z^\dagger)^3 - \frac{3}{2}(z - z^\dagger)]G(z),\\
    &\psi_4(z) = 16\beta_h^2\qty[\beta_h^2(z - z^\dagger)^4 - 3\beta_h(z - z^\dagger)^2 + \frac{3}{4}]G(z),
    \end{align}
    \label{eq:all_psi}%
\end{subequations}
and so on.
This method of generating wavelets has the convenience of allowing one to control the number of nodes (crossings of zero) of the wavelet, hence the number of oscillations. While $G(z)$ can be considered the zeroth-order wavelet, it has no nodes and does not oscillate. In comparison, every differentiation introduces one node to the wavelet, e.g., $\psi_1$ has a single node corresponding to one bump and one dip in the functional form; see~\cref{fig:wavelet_varying_params}. 
In the following sections, we make use of the first four Hermitian wavelets given in~\cref{eq:all_psi} to conduct our observational analyses.

\section{DATA SETS AND METHODOLOGY}\label{section:data}

To compare the standard model against our wavelet-extension of it, we will utilize the Bayesian framework in the form of a parameter inference procedure, which requires the use of data and a sampler code.

Our datasets consist of various combinations of observations related to the cosmic microwave background (CMB), baryon acoustic oscillations (BAO), type Ia supernovae (SN Ia) light curves, cosmic chronometers, and SN Ia absolute magnitudes. Specifically, the following data related to the above observations are used. We use the collection of 31 cosmic chronometers~\cite{Jimenez:2003iv, Simon:2004tf, Stern:2009ep, Moresco:2012by, Zhang:2012mp, Moresco:2015cya, Moresco:2016mzx}, found within the repository~\cite{hz}, in combination with the catalogue from the Pantheon+ SN Ia sample, covering a redshift range of $0.001 < z < 2.26$ with 1701 light curves of 1550 distinguishable SN Ia~\cite{Pan-STARRS1:2017jku}. The associated full covariance matrix is comprised of a statistical and a systematic part, which, along with the data, is provided in the repository~\cite{pantheon_data}. These two datasets will always be used together and will be conjointly referred to as \textbf{SN}. We employ two different BAO datasets. 
The first dataset contains SDSS, BOSS, and eBOSS surveys~\cite{eBOSS:2020yzd}, in which the Lyman-$\alpha$ forest BAO measurements that cover a redshift up to $z \approx 2.34$ are also included. This dataset will be referred to as \textbf{SB}, which stands for ``SDSS-BAO". The other dataset consists of BAO distance measurements from the first year of the Dark Energy Spectroscopic Instrument (DESI)~\cite{DESI:2024mwx}, hereafter referred to as \textbf{DB} (for DESI-BAO); this dataset also includes Lyman-$\alpha$ measurements. To avoid double-counting, these measurements will not be combined with their \textbf{SB} counterparts; rather, we investigate in the next section the consequences of using two different BAO datasets. 
In some instances, we will also use a Gaussian prior of $H_0 = 73.04 \pm 1.04 \, \mathrm{km} \, \mathrm{s}^{-1}\, \mathrm{Mpc}^{-1}$ from the local distance ladder measurements of the SH0ES team, relying on the calibration of the SN Ia absolute magnitude with Cepheid variables~\cite{Riess:2021jrx}. When this prior is in use, we denote it with the addition of \textbf{H0} in the name of the dataset combination.
Finally, we will also make use of some background information from the Cosmic Microwave Background (CMB) from the Planck satellite in the form of a ``BAO data point'' at a redshift of $z \sim 1100$. The background-level information of the CMB can be encapsulated by the three parameters~\cite{BOSS:2014hhw}: $\omega_{\rm b}$ (physical baryon density parameter), $\omega_{\rm m}$ (physical matter density parameter), and $D_{\rm M}(\sim 1100)/r_{\rm d}$, where $r_{\rm d}$ is the comoving size of the sound horizon at the drag epoch and $D_{\rm M}(z)$ is the comoving angular diameter distance to $z$. This dataset will be referred to as \textbf{Pl}.

To calculate the $\chi^2$ for each data sample, we have
\begin{equation}
    \chi^2_{\rm data} = (d_{i,m} - d_{i,{\rm obs}})^T C_{ij,{\rm data}}^{-1} (d_{j,m} - d_{j,{\rm obs}}),
\end{equation}
where $d_{m}$ and $d_{\rm obs}$ are our model predictions and the observables, respectively, and $C_{\rm data}$ is the covariance matrix associated with each of the datasets. Since observations of each dataset are independent of each other, the joint $\chi^2$ can be computed as
\begin{equation}
    \chi^2_{\rm total} = \chi^2_{\rm Pl} + \chi^2_{\rm SN} + \chi^2_{\rm BAO} + \chi^2_{\rm H0},
\end{equation}
where some terms may not be present depending on the dataset combination used, and the index of BAO will refer to either \textbf{SB} or \textbf{DB}.

For the determination of the optimal parameter values of our model, we employ an adapted version of a Monte Carlo Markov Chain (MCMC) code known as SimpleMC (see Ref.~\cite{simplemc}). SimpleMC is designed to calculate expansion rates and distances based on a given Friedmann equation, allowing the analysis of the data at the background level. More information about the application of this code can be found in the Bayesian inference review presented in Ref.~\cite{Padilla:2019mgi}. Additionally, our implementation uses \texttt{dynesty}, a nested sampling library described in Ref.~\cite{Speagle:2019ivv}, to efficiently compute the Bayesian evidence $E(D|M)$, where $D$ is the data used and $M$ is the sampled model. This evidence can be used to compare two distinct models by using the following relationship:
\begin{equation}
    B_{12} \equiv \frac{E(D|M_1)}{E(D|M_2)},
\end{equation}
where $B_{12}$ is known as the \textit{Bayes Factor}, with $M_1$ and $M_2$ being the two models to be compared. By taking the natural logarithm of this quotient, together with the Jeffreys' scale shown in~\cref{jeffreys}~\cite{Trotta:2008qt, efron2001scales}, a useful empirical tool for performing model selection arises. That is, we can use it to gain insight into how well model $M_1$ compares to model $M_2$.

\begin{table}
\captionsetup{justification=Justified,singlelinecheck=false,font=footnotesize}
\footnotesize
\scalebox{1.2}{%
\begin{tabular}{cccc} 
\cline{1-4}\noalign{\smallskip}
\vspace{0.15cm}
$\ln{B_{12}}$ & Odds  & Probability & Strength of Evidence \\
\hline
\hline
\vspace{0.15cm}
$<$ 1.0 & $<$ 3:1 & $<$ 0.75 & Inconclusive \\
\vspace{0.15cm}
1.0 & $\sim$ 3:1 & 0.750 & Weak Evidence \\
\vspace{0.15cm}
2.5 & $\sim$ 12:1 & 0.923 & Moderate Evidence \\
\vspace{0.15cm}
5.0 & $\sim$ 150:1 & 0.993 & Strong Evidence \\
\hline
\hline
\end{tabular}}
\caption{Jeffreys' scale for model selection based on the logarithm of the Bayes' factor. A positive value of $\ln{B_{12}}$ indicates evidence in favor of $M_1$, while a negative value indicates evidence in favor of $M_2$, following the convention from~\cite{Trotta:2008qt}.}
\label{jeffreys}
\end{table}




This combination of tools enables us to perform parameter estimation and model comparison. Regarding the flat priors used for the parameters of the model: $\Omega_{\rm m} = [0.1, 0.5]$ for the matter density parameter, $\Omega_{\rm b} h^2 = [0.02, 0.025]$ for the physical baryon density, and $h = [0.4, 0.9]$ for the dimensionless reduced Hubble constant (where $h = H_0 /100 \, \text{km} \, \text{s}^{-1} \, \text{Mpc}^{-1} $). For the Hermitian wavelets, we need to be more careful with how we select our priors, and our selection requires some explanation.

\begin{figure}
\captionsetup{justification=Justified,font=footnotesize}
    \centering
    \includegraphics[trim = {15mm 20mm 0mm 15mm}, clip, 
    width=9.75cm,
    height=10.cm]{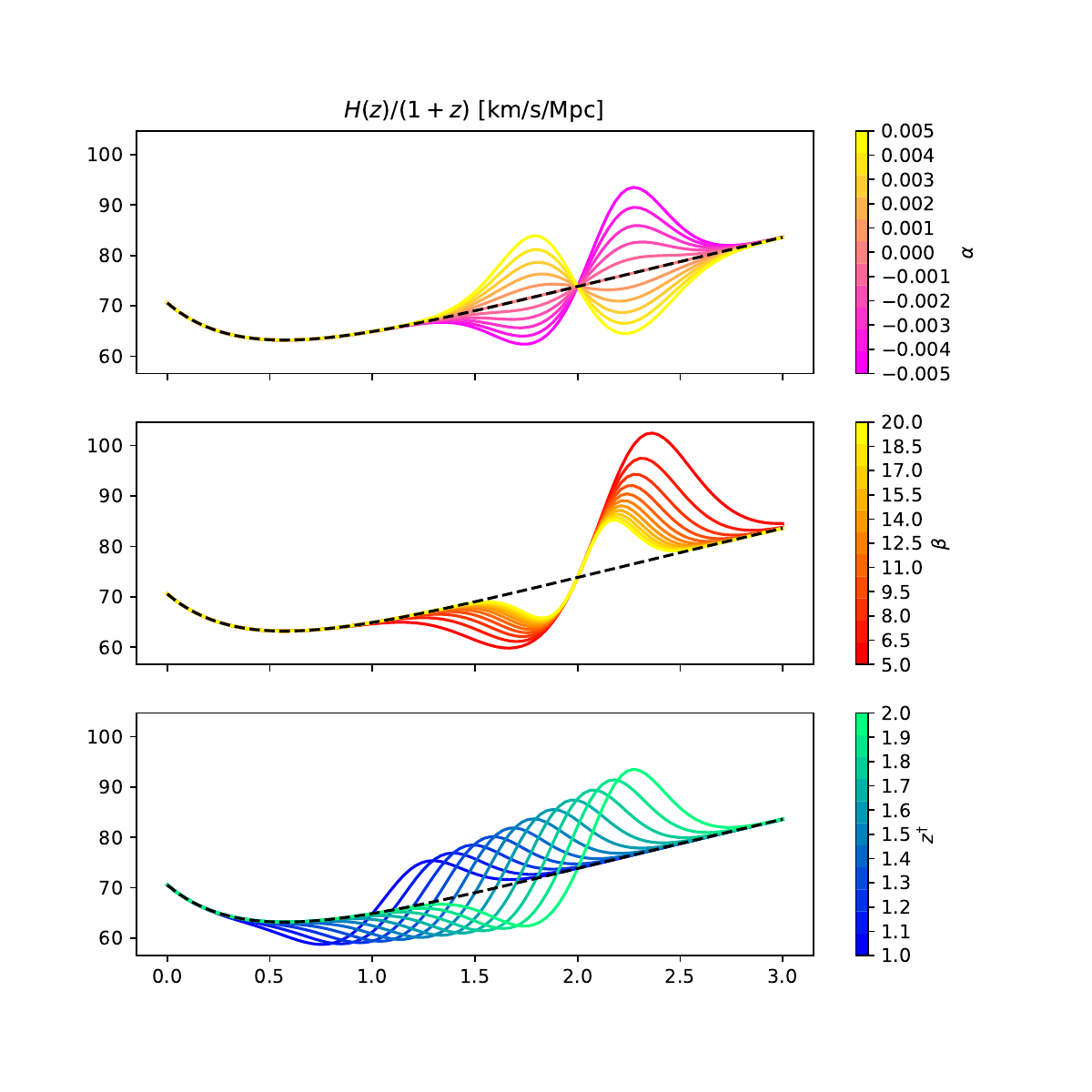}
    \caption{Visual representation of how the underlying Hermitian wavelet (specifically $\psi_1$) affects the co-moving Hubble parameter $H(z)/(1+z)$, where $H(z) = \frac{H_{\Lambda \text{CDM}}(z)}{1+\psi(z) H_{\Lambda \text{CDM}}(z)}$, when varying each parameter individually. The upper plot corresponds to the parameter $\alpha_h$ in the interval $[-0.005, 0.005]$, the middle one to $\beta_h=[5,20]$, and the lower one to $z^\dagger=[1.0, 2.0]$. The black dashed line represents $H_{\Lambda \text{CDM}}(z)/(1+z)$.}
    \label{fig:wavelet_varying_params}
\end{figure}

As seen in~\cref{eq:all_psi}, each Hermitian wavelet has three free parameters: $\alpha_h$, $\beta_h$, and $z^\dagger$. Each parameter governs a distinct behavior of the wavelet: the parameter $\alpha_h$ controls the amplitude; the parameter $\beta_h$ controls both the ``width'' and the amplitude; and $z^\dagger$ determines the center of the wavelet. For a visual reference, see~\cref{fig:wavelet_varying_params}, where we show the influence of each parameter on $H(z)/(1+z)$.
One important thing to note is that the sensitivity of the amplitude of wavelets with respect to the parameter $\alpha_h$ increases with higher-order wavelets. Thus, its prior range diminishes by an order of magnitude when going from $\psi_i$ to $\psi_{i+1}$. Therefore, our priors for each parameter are: $\beta_h : [0.01, 25.0]$ and $z^\dagger : [0.0, 5.0]$ for every wavelet; meanwhile, we will set $\alpha_h : [-0.01, 0.01]$ for $\psi_1$, $\alpha_h : [-0.001, 0.001]$ for $\psi_2$, $\alpha_h : [-0.0001, 0.0001]$ for $\psi_3$, and $\alpha_h : [-0.00001, 0.00001]$ for $\psi_4$ (for reference, they can be found in~\cref{tabla_priors}).

\begin{table}[t!]
\captionsetup{justification=justified, singlelinecheck=false, font=footnotesize}
\footnotesize
\scalebox{1.1}{%
\begin{tabular}{ccc} 
\cline{1-2}\noalign{\smallskip}
 \vspace{0.15cm}
   Models &   Priors   \\
\hline
$\psi_1$, $\psi_2$, $\psi_3$, $\psi_4$ & $\beta_h : [0.01, 25.0]$     \\
$\psi_1$, $\psi_2$, $\psi_3$, $\psi_4$ & $z^\dagger : [0.0, 5.0]$     \\
$\psi_1$ &  $\alpha_h : [-0.01, 0.01]$  \\   
$\psi_2$ &  $\alpha_h : [-0.001, 0.001]$  \\   
$\psi_3$ &  $\alpha_h : [-0.0001, 0.0001]$  \\   
$\psi_4$ &  $\alpha_h : [-0.00001, 0.00001]$  \\   
\hline
\hline
\end{tabular}
}
\caption{Flat prior ranges for the parameters of the Hermitian wavelets used in the analysis.}
\label{tabla_priors}
\end{table}

\section{RESULTS}\label{section:results}

\begin{table}[t!]
\captionsetup{justification=Justified, singlelinecheck=false, font=footnotesize}
\footnotesize
\scalebox{0.82}{%
\begin{tabular}{cccccc} 
\cline{1-6}\noalign{\smallskip}
 \vspace{0.15cm}
Model & Data Sets &  $h$ &  $\Omega_{\rm m,0}$  &   $\ln B_{\Lambda \text{CDM},i}$  &  $-2\Delta\ln \mathcal{L}_{\rm max}$ \\
\hline
 \vspace{0.15cm}
$\Lambda$CDM & \textbf{SB+Pl} &  0.679 (0.006) &  0.309 (0.007) &  0 &  0 \\
\hline
\hline
\vspace{0.15cm}
$\psi_1$ &  & 0.688 (0.006) & 0.301 (0.007) &  -0.69 (0.34) & -5.08 \\
\vspace{0.15cm}
$\psi_2$ &  & 0.688 (0.006) & 0.302 (0.008) &  -1.02 (0.35) & -4.94 \\
\vspace{0.15cm}
$\psi_3$ &  & 0.689 (0.006) & 0.301 (0.007) &  -0.89 (0.35)  & -4.54 \\
\vspace{0.15cm}
$\psi_4$ &  & 0.689 (0.006) & 0.301 (0.008) &  -0.75 (0.34)  & -4.81 \\
\hline
\hline
 \vspace{0.15cm}
$\Lambda$CDM & \textbf{SB+SN} &   0.686 (0.013) &  0.306 (0.013) &  0   &  0 \\
\hline
\hline
\vspace{0.15cm}
$\psi_1$ &  & 0.692 (0.014) & 0.315 (0.015) &  -0.61 (0.17)  & -7.42 \\
\vspace{0.15cm}
$\psi_2$ &  & 0.692 (0.014) & 0.318 (0.016) &  -1.22 (0.17)  & -8.59 \\
\vspace{0.15cm}
$\psi_3$ &  & 0.693 (0.014) & 0.315 (0.015) &  -0.83 (0.18)  & -8.07 \\
\vspace{0.15cm}
$\psi_4$ &  & 0.692 (0.014) & 0.314 (0.014) &  -0.71 (0.18)  & -8.61 \\
\hline
\hline
 \vspace{0.15cm}
$\Lambda$CDM & \textbf{SB+SN+Pl} & 0.676 (0.006) & 0.312 (0.007) & 0 & 0 \\
\hline
\hline
\vspace{0.15cm}
$\psi_1$ &  & 0.684 (0.005) & 0.306 (0.007) &  -0.22 (0.35) & -5.22 \\
\vspace{0.15cm}
$\psi_2$ &  & 0.685 (0.005) & 0.306 (0.006) &  -0.51 (0.34)  & -4.72 \\
\vspace{0.15cm}
$\psi_3$ &  & 0.684 (0.005) & 0.307 (0.007) &  -0.54 (0.34) & -5.29 \\
\vspace{0.15cm}
$\psi_4$ &  & 0.685 (0.006) & 0.306 (0.007) &  -0.66 (0.34)  & -6.58 \\
\hline
\hline
 \vspace{0.15cm}
$\Lambda$CDM & \textbf{SB+SN+H0} & 0.709 (0.014) &  0.311 (0.013) & 0  & 0 \\
\hline
\hline
\vspace{0.15cm}
$\psi_1$ &  & 0.705 (0.012) & 0.315 (0.015) &  0.14 (0.24)  & -2.91 \\
\vspace{0.15cm}
$\psi_2$ &  & 0.706 (0.012) & 0.318 (0.015) &  0.34 (0.24)  & -3.17 \\
\vspace{0.15cm}
$\psi_3$ &  & 0.705 (0.012) & 0.314 (0.015) &  0.55 (0.23)  & -3.89 \\
\vspace{0.15cm}
$\psi_4$ &  & 0.705 (0.011) & 0.314 (0.015) &  -0.11 (0.23)  & -3.23 \\
\hline
\hline
 \vspace{0.15cm}
$\Lambda$CDM & \textbf{SB+SN+Pl+H0} &  0.679 (0.005) & 0.308 (0.007) & 0  & 0 \\
\hline
\hline
\vspace{0.15cm}
$\psi_1$ &  & 0.687 (0.005) & 0.303 (0.007) &  -1.38 (0.34) & -6.18 \\
\vspace{0.15cm}
$\psi_2$ &  & 0.688 (0.005) & 0.302 (0.007) &  -1.52 (0.34) & -6.34 \\
\vspace{0.15cm}
$\psi_3$ &  & 0.688 (0.006) & 0.302 (0.007) &  -1.48 (0.35) & -7.16 \\
\vspace{0.15cm}
$\psi_4$ &  & 0.688 (0.005) & 0.302 (0.006) &  -1.21 (0.34)  & -7.78 \\
\hline
\hline
 \vspace{0.15cm}
$\Lambda$CDM & \textbf{SN+Pl} & 0.671 (0.006) &  0.319 (0.008) &  0  &  0 \\
\hline
\hline
\vspace{0.15cm}
$\psi_1$ &  & 0.681 (0.006) & 0.311 (0.008) &  -0.02 (0.34) & -2.82 \\
\vspace{0.15cm}
$\psi_2$ &  & 0.681 (0.006) & 0.309 (0.008) &  0.22 (0.35) & -3.53 \\
\vspace{0.15cm}
$\psi_3$ &  & 0.681 (0.006) & 0.310 (0.008) &  0.52 (0.34) & -3.54 \\
\vspace{0.15cm}
$\psi_4$ &  & 0.681 (0.006) & 0.310 (0.008) &  0.28 (0.35)  & -4.42 \\
\hline
\hline
 \vspace{0.15cm}
$\Lambda$CDM & \textbf{SN+Pl+H0} & 0.675 (0.006) &  0.314 (0.008) &  0  &  0 \\
\hline
\hline
\vspace{0.15cm}
$\psi_1$ &  & 0.685 (0.006) & 0.306 (0.007) &  -0.61 (0.34) & -5.56 \\
\vspace{0.15cm}
$\psi_2$ &  & 0.685 (0.006) & 0.306 (0.007) &  -0.69 (0.35) & -6.51 \\
\vspace{0.15cm}
$\psi_3$ &  & 0.685 (0.006) & 0.306 (0.008) &  -0.82 (0.35)  & -6.27 \\
\vspace{0.15cm}
$\psi_4$ &  & 0.684 (0.006) & 0.306 (0.007) &  -0.85 (0.35)  & -6.37 \\
\hline
\hline
\end{tabular}}
\caption{First half of the mean values and standard deviations for two of the parameters used in the reconstruction, $\Omega_{\rm m,0}$ and $h$. For each model, the last two columns present the natural logarithm of the Bayes Factor, $\ln B_{\Lambda \text{CDM},i}$, and $-2\Delta\ln \mathcal{L}_{\rm max}$, defined as $-2\ln(\mathcal{L}_{\rm max,i} / \mathcal{L}_{\rm max,\Lambda \text{CDM}})$, for a comparison of the fit of the data. When BAO data is included, it is denoted as \textbf{SB} in this table.}
\label{tabla_evidencias2}
\end{table}

\begin{table}[t!]
\captionsetup{justification=Justified, singlelinecheck=false, font=footnotesize}
\footnotesize
\scalebox{0.82}{%
\begin{tabular}{cccccc} 
\cline{1-6}\noalign{\smallskip}
 \vspace{0.15cm}

Model & datasets &  $h$ &  $\Omega_{\rm m,0}$  &   $\ln B_{\Lambda \text{CDM},i}$  &  $-2\Delta\ln \mathcal{L_{\rm max}}$ \\
\hline
 \vspace{0.15cm}
$\Lambda$CDM & \textbf{DB+Pl} &   0.677 (0.006) &  0.311 (0.008) &  0   &  0     \\
\hline
\hline
\vspace{0.15cm}
$\psi_1$ &  & 0.686 (0.008) & 0.304 (0.009) &  -0.82 (0.34)  & -8.18  \\
\vspace{0.15cm}
$\psi_2$ &  &  0.686 (0.006) & 0.303 (0.008) &  -0.78 (0.33)  & -10.12 \\
\vspace{0.15cm}
$\psi_3$ &  & 0.687 (0.006) & 0.303 (0.007) &  -0.84 (0.32)  & -10.24  \\
\vspace{0.15cm}
$\psi_4$ &  & 0.688 (0.006) & 0.302 (0.007) &  -0.81 (0.34)  & -10.48  \\

\hline
\hline
 \vspace{0.15cm}
$\Lambda$CDM & \textbf{DB+SN} &   0.683 (0.017) &  0.313 (0.012) &  0  &  0   \\
\hline
\hline
\vspace{0.15cm}
$\psi_1$ &  & 0.679 (0.014) & 0.309 (0.012) &  -0.12 (0.23)  & -11.42  \\
\vspace{0.15cm}
$\psi_2$ &  & 0.681 (0.013) & 0.308 (0.012) &  -0.21 (0.23)  & -11.08  \\
\vspace{0.15cm}
$\psi_3$ &  & 0.679 (0.014) & 0.308 (0.012) &  0.05 (0.24) & -10.92  \\
\vspace{0.15cm}
$\psi_4$ &  & 0.679 (0.013) & 0.309 (0.012) &  0.02 (0.22)  & -11.22  \\

\hline
\hline
 \vspace{0.15cm}
$\Lambda$CDM &\textbf{ DB+SN+Pl} &  0.675 (0.005) &  0.314 (0.007) &  0   &  0   \\
\hline
\hline
\vspace{0.15cm}
$\psi_1$ &  & 0.683 (0.005) & 0.308 (0.007) & 0.01 (0.32)  &  -11.61 \\
\vspace{0.15cm}
$\psi_2$ &  & 0.683 (0.005) & 0.307 (0.007) &  0.05 (0.33)  & -10.11  \\
\vspace{0.15cm}
$\psi_3$ &  & 0.683 (0.006) & 0.307 (0.007) &  -0.21 (0.32)  & -10.65  \\
\vspace{0.15cm}
$\psi_4$ &  & 0.683 (0.005) & 0.308 (0.007) &  0.15 (0.31)  & -10.25  \\

\hline
\hline
 \vspace{0.15cm}
$\Lambda$CDM & \textbf{DB+SN+H0} &  0.702 (0.014) &  0.311 (0.012) &  0  &  0  \\
\hline
\hline
\vspace{0.15cm}
$\psi_1$ &  & 0.703 (0.013) & 0.311 (0.011) &  -0.14 (0.27)  & -11.33  \\
\vspace{0.15cm}
$\psi_2$ &  & 0.704 (0.013) & 0.312 (0.011) &  -0.31 (0.23)  & -10.86 \\
\vspace{0.15cm}
$\psi_3$ &  & 0.703 (0.013) & 0.311 (0.011) &  -0.67 (0.26)  & -10.14  \\
\vspace{0.15cm}
$\psi_4$ &  & 0.704 (0.013) & 0.311 (0.012) &  -0.56 (0.25)  & -10.81  \\

\hline
\hline
 \vspace{0.15cm}
$\Lambda$CDM & \textbf{DB+SN+Pl+H0} &  0.678 (0.005) &  0.311 (0.007) &  0  &  0  \\
\hline
\hline
\vspace{0.15cm}
$\psi_1$ &  & 0.686 (0.005) & 0.304 (0.007) &  -1.61 (0.35) & -13.25  \\
\vspace{0.15cm}
$\psi_2$ &  & 0.686 (0.005) & 0.304 (0.007) &  -1.41 (0.34)  & -11.87 \\
\vspace{0.15cm}
$\psi_3$ &  & 0.686 (0.005) & 0.304 (0.006) &  -1.99 (0.34)  & -12.48  \\
\vspace{0.15cm}
$\psi_4$ &  & 0.686 (0.005) & 0.304 (0.007) &  -1.31 (0.34)  & -12.01 \\

\hline
\hline
 \vspace{0.15cm}
$\Lambda$CDM & \textbf{SN} &  0.674 (0.027) &  0.331 (0.017) &  0  &  0  \\
\hline
\hline
\vspace{0.15cm}
$\psi_1$ &  & 0.676 (0.026) & 0.331 (0.017) &  0.58 (0.23) & -3.71  \\
\vspace{0.15cm}
$\psi_2$ &  & 0.675 (0.026) & 0.332 (0.018) &  0.12 (0.24) & -3.85 \\
\vspace{0.15cm}
$\psi_3$ &  & 0.676 (0.028) & 0.332 (0.018) &  0.13 (0.22) & -5.21 \\
\vspace{0.15cm}
$\psi_4$ &  & 0.676 (0.026) & 0.330 (0.017) &  0.42 (0.23) & -5.41 \\

\hline
\hline
 \vspace{0.15cm}
$\Lambda$CDM & \textbf{SN+H0} & 0.711 (0.019) &  0.322 (0.017) & 0 &  0 \\
\hline
\hline
\vspace{0.15cm}
$\psi_1$ &  & 0.711 (0.017) & 0.323 (0.016) &  -0.28 (0.22) & -3.88  \\
\vspace{0.15cm}
$\psi_2$ &  & 0.712 (0.017) & 0.323 (0.017) &  -0.12 (0.22) & -4.18 \\
\vspace{0.15cm}
$\psi_3$ &  & 0.712 (0.017) & 0.324 (0.016) &  -0.69 (0.22) & -5.04 \\
\vspace{0.15cm}
$\psi_4$ &  & 0.712 (0.018) & 0.322 (0.017) &  -0.09 (0.24) & -5.66 \\

\hline
\hline
\end{tabular}}
\caption{Second half of the mean values and standard deviations for two of the parameters used in the reconstruction, $\Omega_{\rm m,0}$ and $h$. In this table, when BAO data is included, it is denoted as \textbf{DB}. For each model, the last two columns present the natural logarithm of the Bayes Factor, $\ln B_{\Lambda \text{CDM},i}$, and $-2\Delta\ln \mathcal{L}_{\rm max}$, defined as $-2\ln(\mathcal{L}_{\rm max,i} / \mathcal{L}_{\rm max,\Lambda \text{CDM}})$, for a comparison of the fit of the data.}
\label{tabla_evidencias}
\end{table}

In this section, we present the results of our sampling process, with a particular focus on the differences between the two BAO datasets. Due to the high number of dataset combinations used, we visualize only a selection of the plots and comment on the others as necessary. For completeness, the best-fit values of $h$ and $\Omega_{\rm m,0}$, as well as the Bayes' factors and $-2\Delta\ln \mathcal{L_{\rm max}}$, for all analyses are provided in~\cref{tabla_evidencias,tabla_evidencias2}.

\begin{figure*}[t!]
\captionsetup{justification=Justified,singlelinecheck=false,font=footnotesize}
    \centering
    \makebox[11cm][c]{
    \includegraphics[trim = 0mm  0mm 0mm 0mm, clip, width=6.0cm, height=6.0cm]{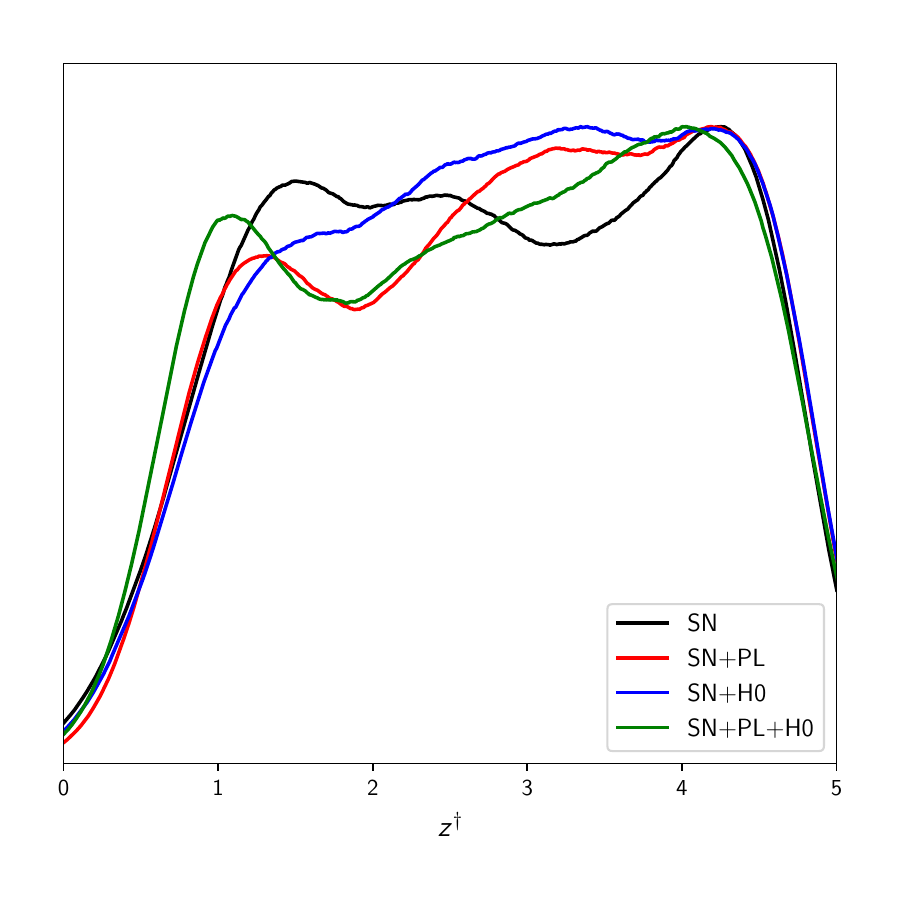}
        \includegraphics[trim = 0mm  0mm 0mm 0mm, clip, width=6.0cm, height=6.0cm]{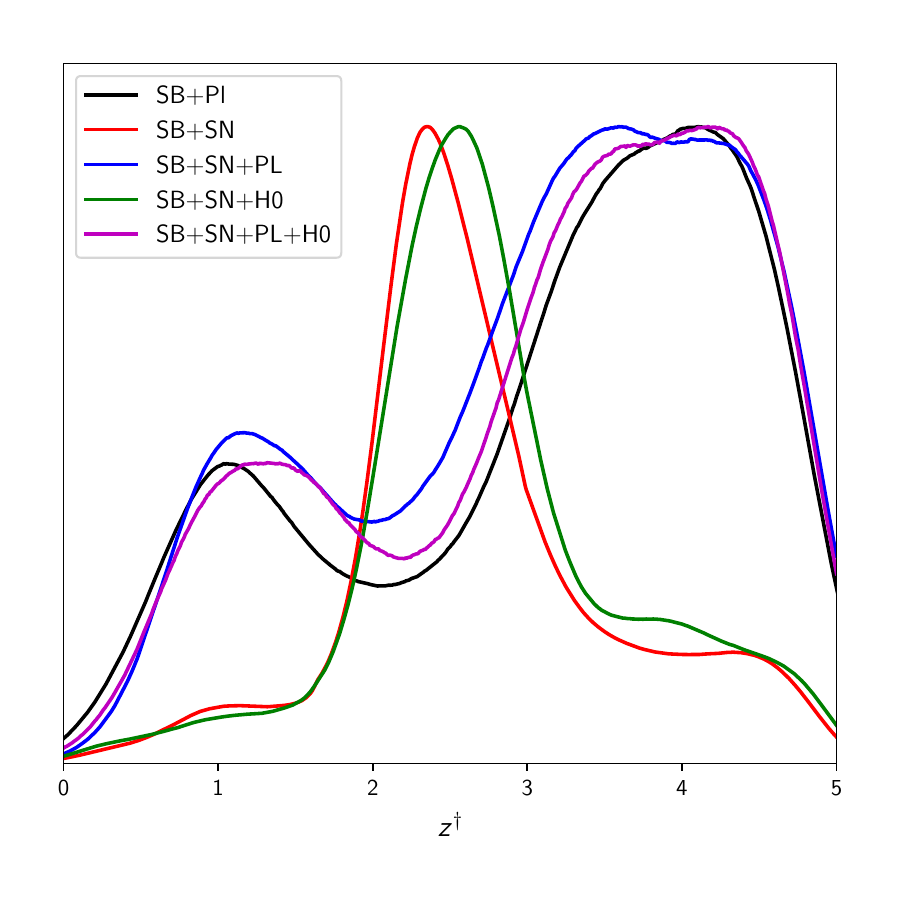}
     \includegraphics[trim = 0mm  0mm 0mm 0mm, clip, width=6.0cm, height=6.0cm]{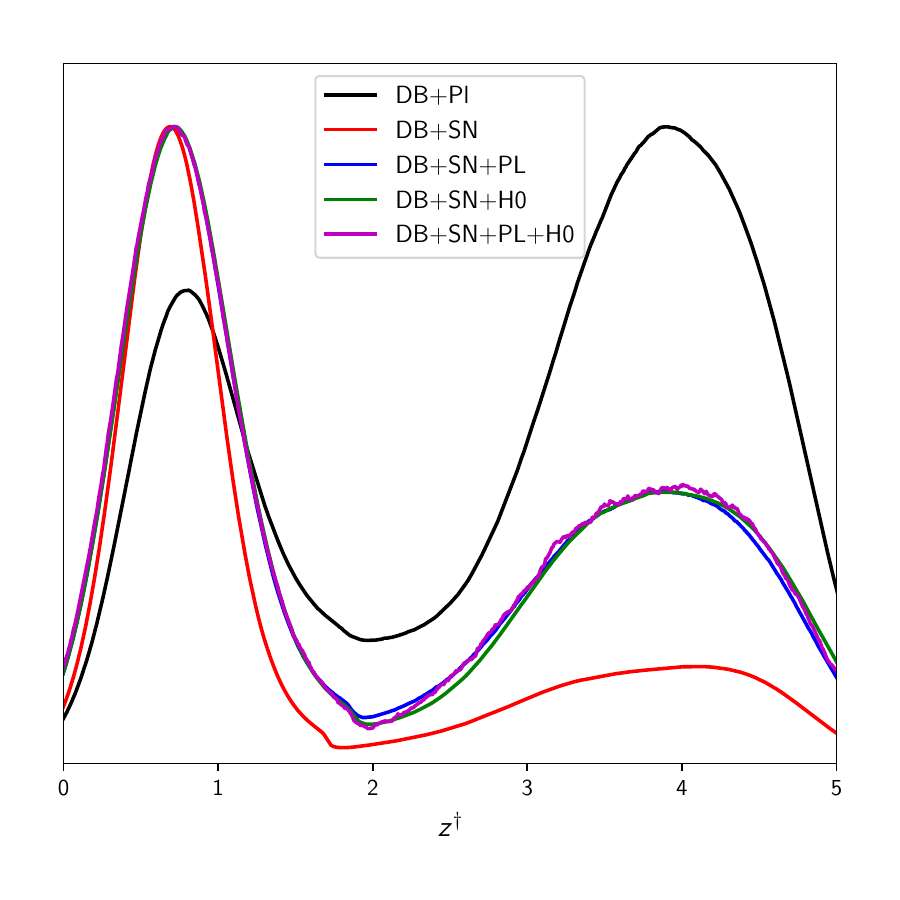}}
    \caption{1D marginalized posterior distributions for the reconstructed parameter $z^\dagger$ for $\psi_1(z)$. We present the results in three different panels to show the effect of BAO datasets on the posteriors. These plots were made using the Python library \texttt{getdist}~\cite{Lewis:2019xzd}. 
}\label{fig:1d_comparison}
\end{figure*}
\begin{figure*}[t!]
\captionsetup{justification=Justified,singlelinecheck=false,font=footnotesize}
    \centering
    \makebox[11cm][c]{
    \includegraphics[trim = 0mm  0mm 0mm 0mm, clip, width=5.cm, height=4.cm]{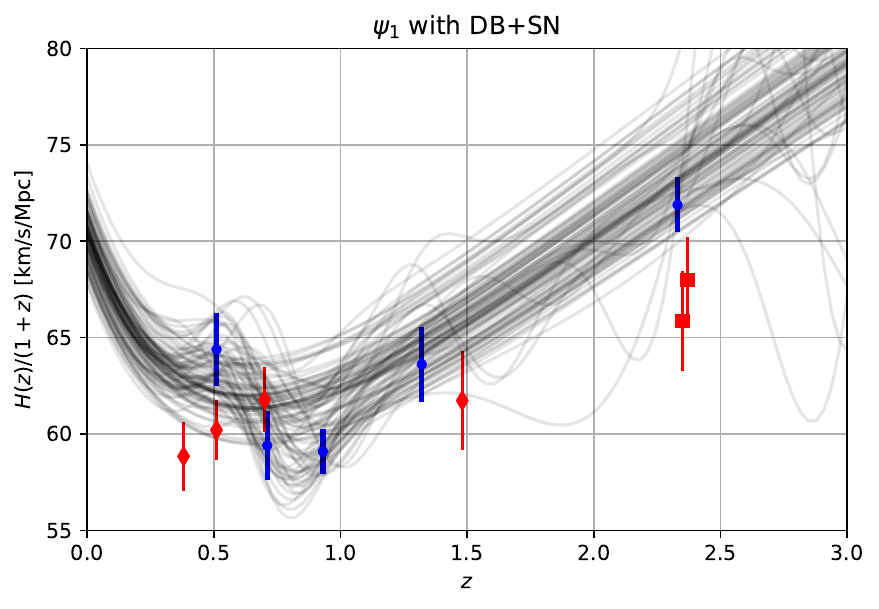}
    \includegraphics[trim = 12mm  0mm 0mm 0mm, clip, width=4.88cm, height=4.cm]{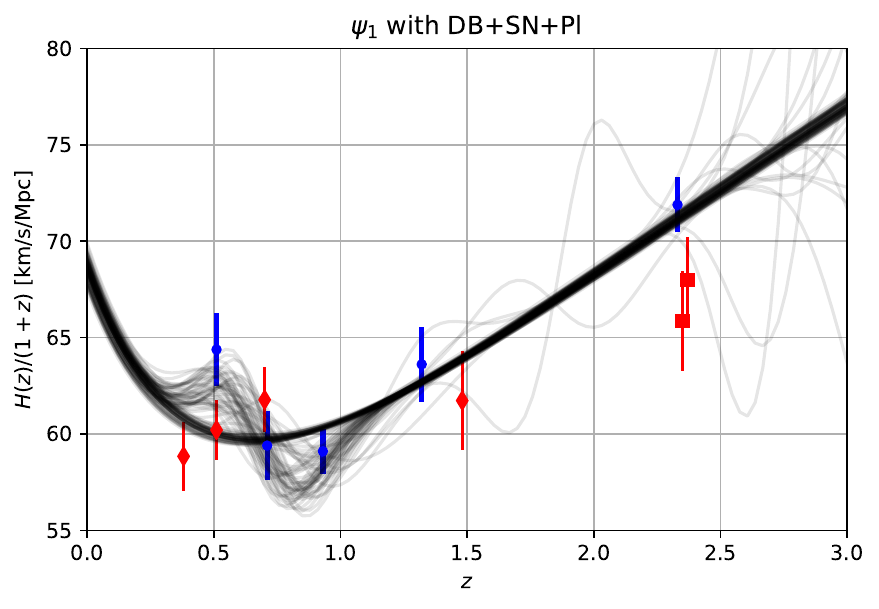}
    \includegraphics[trim = 12mm  0mm 0mm 0mm, clip, width=4.88cm, height=4.cm]{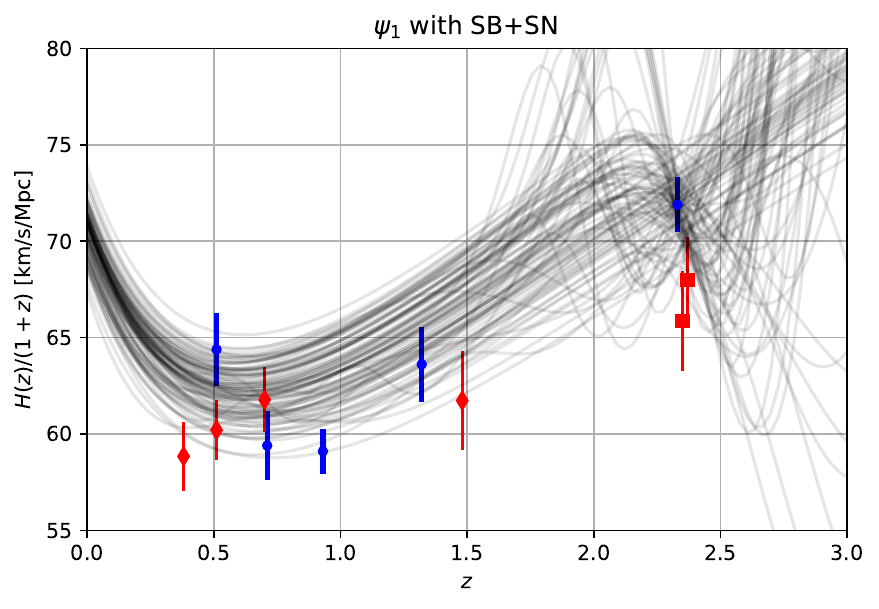}
    \includegraphics[trim = 12mm  0mm 0mm 0mm, clip, width=4.88cm, height=4.cm]{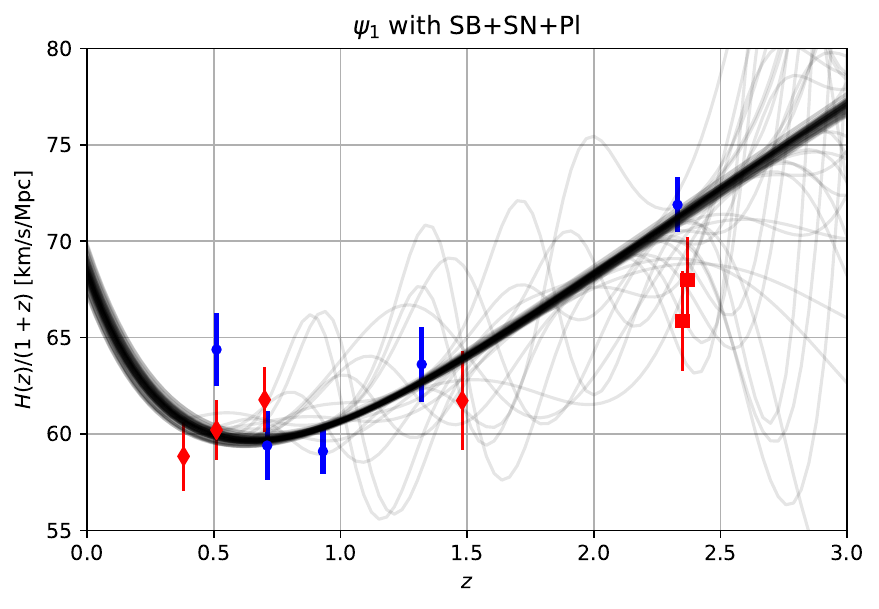}
    }
    \caption{Functional posterior for $H(z)/(1+z)$ for $\psi_1(z)$ using four distinct data combinations: \textbf{DB+SN}, \textbf{DB+SN+Pl}, \textbf{SB+SN}, and \textbf{SB+SN+Pl}. The red (blue) error bars correspond to the SDSS (DESI) BAO distance $D_H(z)/r_{\rm d}$ measurements. The size of the sound horizon was fixed to the robust value of $147$ Mpc to show the BAO data in the figure. The plots were made using the Python library \texttt{fgivenx}~\cite{Handley_2018}.
}\label{fig:hz_different_datasets}
\end{figure*}

In~\cref{fig:1d_comparison}, we plot the 1D marginalized posterior distributions of $z^\dagger$ for the first-order wavelet $\psi_1$. While we show only one wavelet, similar features are observed across the others (see~\cref{appendix_a} and~\cref{fig:1d_comparison_other_psi}). A notable feature is the absence of a clear peak in the posterior distribution without BAO data, as seen in the leftmost panel. This situation changes in the middle panel with the inclusion of \textbf{SB}, where a distinct peak around $z^\dagger \sim 2.5$ appears when \textbf{Pl} is not included (red and green plots). When \textbf{Pl} is also included, a bimodal distribution is observed, with a smaller peak around $z^\dagger \sim 1$ and a larger accumulation of probability for $z^\dagger \gtrsim 3$. It is important to note that \textbf{Pl} is the only dataset used beyond the redshift of $2.34$, and the impact of a wavelet at high redshifts does not lead to significant differences when compared with any combination of BAO, \textbf{SN}, or \textbf{H0} data. 
Moreover, even \textbf{Pl} is not affected by a high redshift wavelet as we only do a background analysis for which admissible wavelets induce no changes over $\Lambda$CDM when CMB data is considered (see Ref.~\cite{Akarsu:2022lhx}). In fact, a high redshift wavelet well-approximates $\Lambda$CDM where data are present; this can be seen by comparing, say, the rightmost panel of~\cref{fig:hz_different_datasets} with~\cref{fig:wavelet_varying_params}, where the shape of the $\Lambda$CDM curve can be seen in dashed lines. Since the \textbf{Pl} data is also not sensitive to high-redshift wavelets, we interpret the change in the posterior of $z^\dagger$ to be due to the restrictions imposed by \textbf{Pl} on the baseline $\Lambda$CDM parameters ($\Omega_{\rm m}$, $h$, and $\Omega_{\rm b}h^2$), which in turn, cannot be combined with a wavelet at low redshifts in the presence of low-redshift data. This restriction on the baseline parameters becomes evident by comparing the third and fourth panels of~\cref{fig:hz_different_datasets}, where the $\Lambda$CDM-like curves become denser resulting in the removal of some freedom in the expansion history. Finally, we note that the inclusion of \textbf{H0} in the dataset has almost no impact on the constraints on $z^\dagger$ with or without \textbf{SB}.

Moving on to the last panel of~\cref{fig:1d_comparison}, we see that when \textbf{DB} is included instead of \textbf{SB}, the results are more stable with respect to the addition of \textbf{Pl} and the bimodal distribution is present for all dataset combinations. More interestingly, the peak at low redshifts moves to around $z^\dagger\sim0.7$ and the inclusion of the \textbf{SN} data strengthens this peak, significantly decreasing the probability around a $\Lambda$CDM-like cosmology within the redshift range where data are present. This result is in line with the findings of the DESI collaboration in Ref.~\cite{DESI:2024mwx} that note a $\sim\!2.6\sigma$ preference for dynamical DE (when the DE is allowed to evolve with the Chevallier-Polarski-Linder parametrization) when DESI-BAO is combined with CMB data, which goes up to $\sim3.9\sigma$ when SN Ia data are also included; however, note that they still find only $\sim\!2.5\sigma$ evidence when their chosen SN Ia dataset is Pantheon+ (in comparison to Union3~\cite{Rubin:2023ovl} or DESY5~\cite{DES:2024tys}), whereas, here we find that the preference for a dynamical DE [if the wavelets are assumed to originate from a dynamical DE component as described by~\cref{eq:dynamicDE}] is strengthened with the inclusion of \textbf{SN} that consists of Pantheon+ and cosmic chronometers. And as before, the constraints on $z^\dagger$ are not very sensitive to the inclusion of \textbf{H0}.

Regarding the constraints on the parameters $\alpha_h$ and $\beta_h$, we consider these results to be less relevant (especially when compared to $z^\dagger$) for inclusion in the main text. However, their 1D posteriors and a brief discussion can be found in~\cref{appendix_a}.

\begin{figure}
\captionsetup{justification=Justified,font=footnotesize}
    \centering
    \includegraphics[trim = {0mm 0mm 0mm 0mm}, clip, width=8.5cm, height=6.cm]{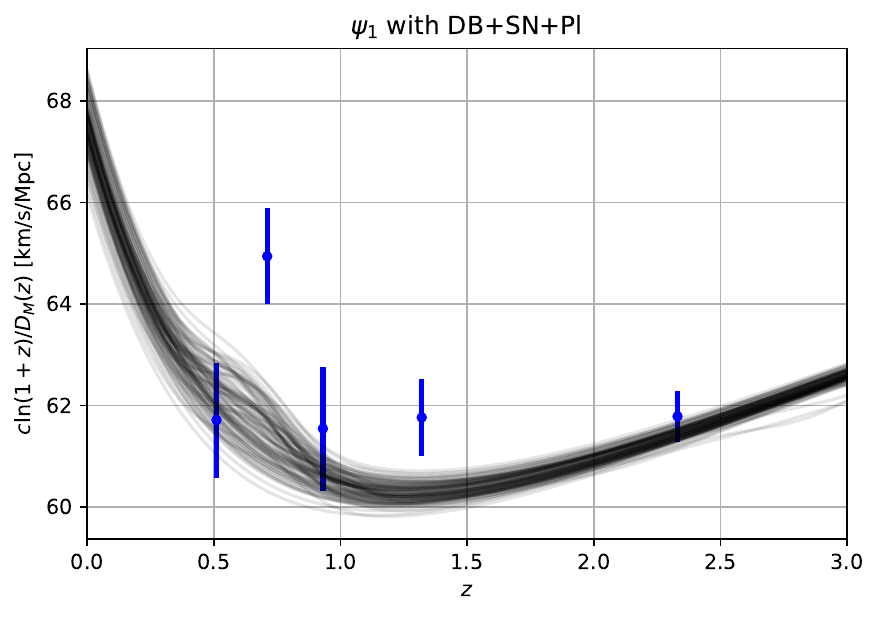}
    \caption{
    Functional posterior of the inverted and scaled comoving angular diameter distance, $c\ln(1+z)/D_{\rm M}(z)$, where $D_{\rm M}(z)=c\int_{0}^{z}dz'/H(z')$; note that this scaled function approaches $H_0$ as $z\to0$. The blue error bars correspond to the re-scaled $D_{\rm M}$ measurements from DESI, and the functional posterior was made with the reconstructed wavelet of first order, $\psi_1$, using the dataset combination \textbf{DB+SN+Pl}. The size of the sound horizon was fixed to the robust value of $147$ Mpc to show the BAO data in the figure. The plot was made using the Python library \texttt{fgivenx}~\cite{Handley_2018}.
    }
    \label{fig:dm_psi1}
\end{figure}

In~\cref{fig:hz_different_datasets} (see also~\cref{fig:hz_different_datasets_v2}), we show the Hubble functions corresponding to $\psi_1$ for BAO+\textbf{SN} dataset combinations with and without \textbf{Pl}, and using both \textbf{SB} and \textbf{DB}. In a similar manner to the 1D posterior plots for $z^\dagger$, we limit this discussion to $\psi_1$ since the behaviors are similar across all wavelets, as shown in~\cref{fig:hz_different_datasets_appendix}.. Again, we see the lower-redshift dynamics preferred by \textbf{DB} with $z^\dagger\sim0.7$ in comparison to the preference for higher $z^\dagger$ values by \textbf{SB}. The inclusion of \textbf{Pl} causes the $\Lambda$CDM parameters to be strictly constrained, and when \textbf{SB} is present, in contrast to the \textbf{DB} case, it also pushes the wavelets to redshifts mostly irrelevant to our low-redshift datasets. We see that the presence of the wavelets allows a better fit to the BAO data as argued in Ref.~\cite{Akarsu:2022lhx}. However, the $D_H(z)/r_{\rm d}$ values from the BAO data are accompanied also by $D_{\rm M}(z)/r_{\rm d}$ values (also $D_V(z)/r_{\rm d}$, but we do not discuss them).\footnote{For reference, we note that $D_H(z)=c/H(z)$, $D_{\rm M}(z)=c\int_{0}^{z}dz'/H(z')$ for a spatially flat RW metric, and $D_V(z)=[zD_{\rm M}(z)^2D_H(z)]^{1/3}$.} In~\cref{fig:dm_psi1}, we also plot the inverted and scaled $D_{\rm M}(z)$ function for $\psi_1$ for the dataset combination selected in the second panel of~\cref{fig:hz_different_datasets}. It is clear from~\cref{fig:hz_different_datasets,fig:dm_psi1} that even for this case, where the oscillatory features of wavelets are readily apparent in the $H(z)/(1+z)$ plots, the $D_{\rm M}(z)$ function is barely affected by the wavelets. This is due to the $D_{\rm M}(z)$ function being an integral of the Hubble radius which smooths out the wavelet deviations. Thus, we can safely argue that it is the $D_H(z)/r_{\rm d}$ measurements from the BAO data that drive the preference for wavelets and not the $D_{\rm M}(z)/r_{\rm d}$ measurements. In particular,~\cref{fig:hz_different_datasets} indicates that it is the Lyman-$\alpha$ $D_H(z)/r_{\rm d}$ measurement for \textbf{SB}, and the $D_H(z)/r_{\rm d}$ measurement at $z=0.51$ for the \textbf{DB} with the ``anomalous" $D_{\rm M}(z)/r_{\rm d}$ measurement of \textbf{DB} having little to no effect in driving the wavelets.


As shown in~\cref{tabla_evidencias,tabla_evidencias2}, while the early universe constraints on the matter density and the Hubble constant remain unchanged, every case with wavelets provides a better fit to the data, as indicated by the negative $-2\Delta\ln \mathcal{L_{\rm max}}$ values. This improvement exceeds the expected $-3$ difference due to the additional three free parameters associated with the wavelets compared to $\Lambda$CDM. Notably, when using \textbf{DB}, wavelets are favored at over $3\sigma$ ($\sqrt{\Delta\chi^2}$) in all cases except one ($\psi_1$ with \textbf{DB+Pl}, as seen in~\cref{tabla_evidencias}), even with the Planck likelihood.\footnote{This is again in line with the findings of the DESI collaboration as explained in the previous paragraph of the main text.} This advantage diminishes when \textbf{SB} is used as the BAO dataset.
This is impressive as the Planck likelihood imposes heavy restrictions on both $\Omega_{\rm m,0}$ and $h$ (as seen in their inferred values and $1\sigma$ when the \textbf{Pl} dataset is present and in~\cref{fig:hz_different_datasets} when \textbf{Pl} is used) and this in turn could act against our wavelet extension of the standard model. A clear way to verify this is by analyzing~\cref{fig:1d_comparison}. When the Planck likelihood is present, we observe a rise of an apparent bimodality centered in $z^\dagger \sim 4$. Given the fact that our data spans a redshift up to $2.34$, if the wavelet is centered way beyond this point (which in this case means centered in $z > 3$), we can assume that the datasets used for the parameter inference stand against the inclusion of the wavelet in the redshift interval covered by them. This is especially true when SDSS-BAO is included, although not with DESI-BAO, which can be verified by referring back to~\cref{fig:hz_different_datasets} and the $\Delta\ln\mathcal{L}_{\rm max}$ values in~\cref{tabla_evidencias,tabla_evidencias2}. By looking at these functional posteriors, it is clear that, when using \textbf{SB} (\textbf{DB}), the center of the wavelet is localized in $z \sim 2.3$ ($z \sim 0.7$), and by adding \textbf{Pl}, we observe that the wavelet behavior is fully displaced with $\Lambda$CDM-like behavior and unconstrained when \textbf{SB} is present.

Focusing now on the Bayesian evidence through the Bayes' factor $\ln B_{\Lambda \text{CDM},i}$ in~\cref{tabla_evidencias,tabla_evidencias2}, we can see some interesting results. Almost every case is as good ($\abs{\ln B}<1$) or better (with weak evidence in favor of the wavelets according to Jeffreys' scale) than the standard model at explaining the data. The best cases obtained are \textbf{DB+SN+Pl+H0} and \textbf{SB+SN+Pl+H0}, the reason being that a wavelet-like behavior seems to be a good alternative to $\Lambda$CDM at reconciling the localized dynamical behavior preferred by both of the different BAO datasets while preserving the good fit of $\Lambda$CDM to the measurements of the first peak of the CMB power spectrum.

\begin{figure*}
\captionsetup{justification=Justified,singlelinecheck=false,font=footnotesize}
    \centering
    \makebox[11cm][c]{
    \includegraphics[trim = 5mm  0mm 25mm 0mm, clip, width=5.cm, height=4.5cm]{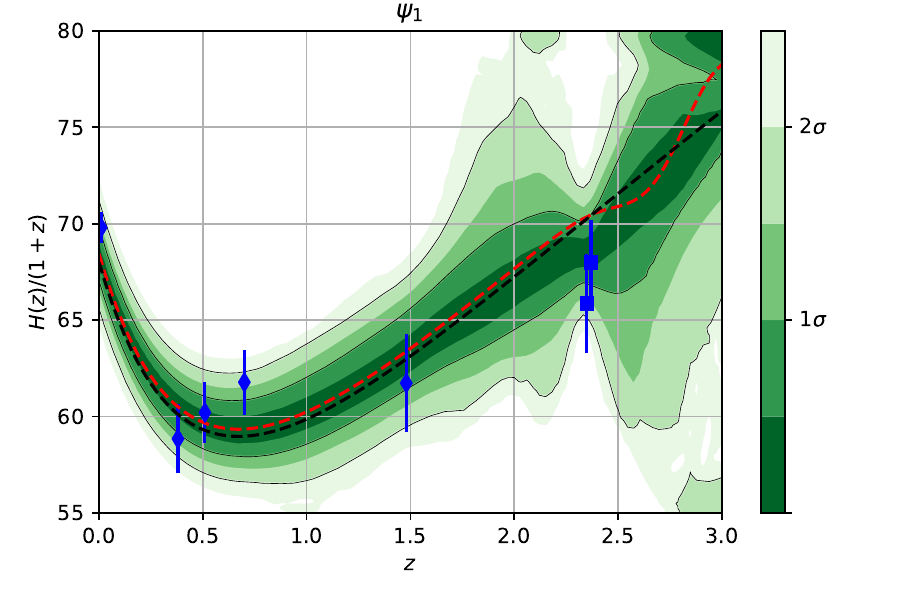}
    \includegraphics[trim = 5mm  0mm 25mm 0mm, clip, width=5.cm, height=4.5cm]{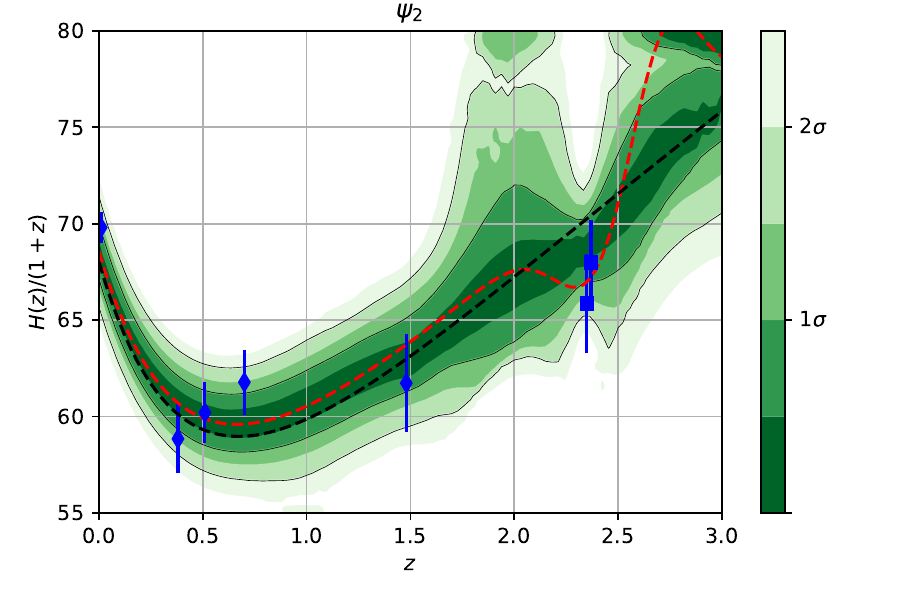}
    \includegraphics[trim = 5mm  0mm 25mm 0mm, clip, width=5.cm, height=4.5cm]{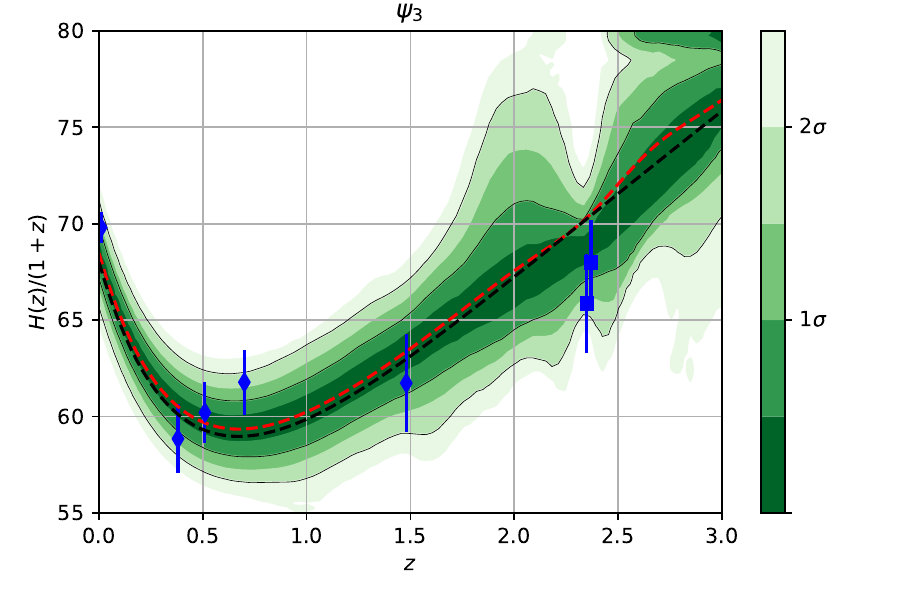}
    \includegraphics[trim = 5mm  0mm 5mm 0mm, clip, width=5.5cm, height=4.5cm]{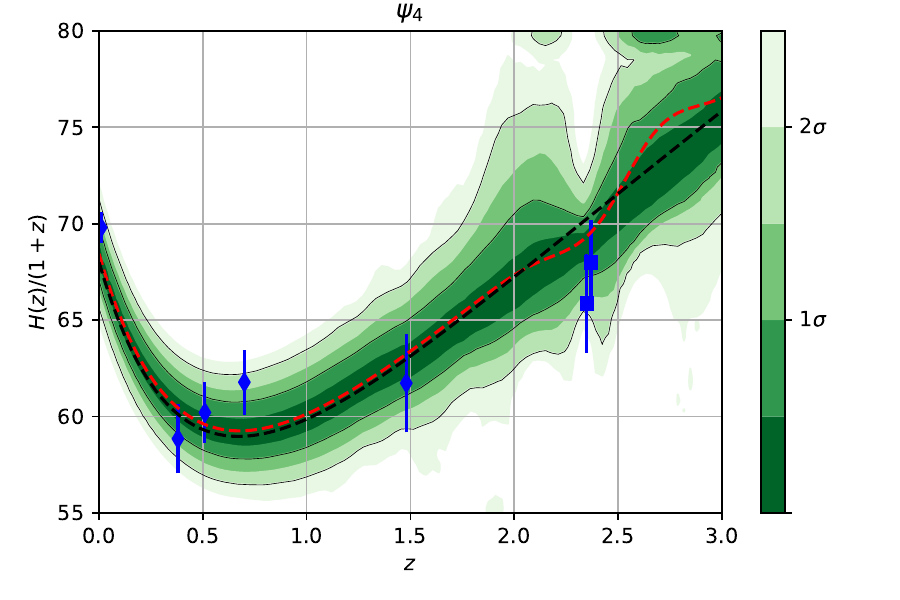}
    }

    \makebox[11cm][c]{
    \includegraphics[trim = 5mm  0mm 25mm 0mm, clip, width=5.cm, height=4.5cm]{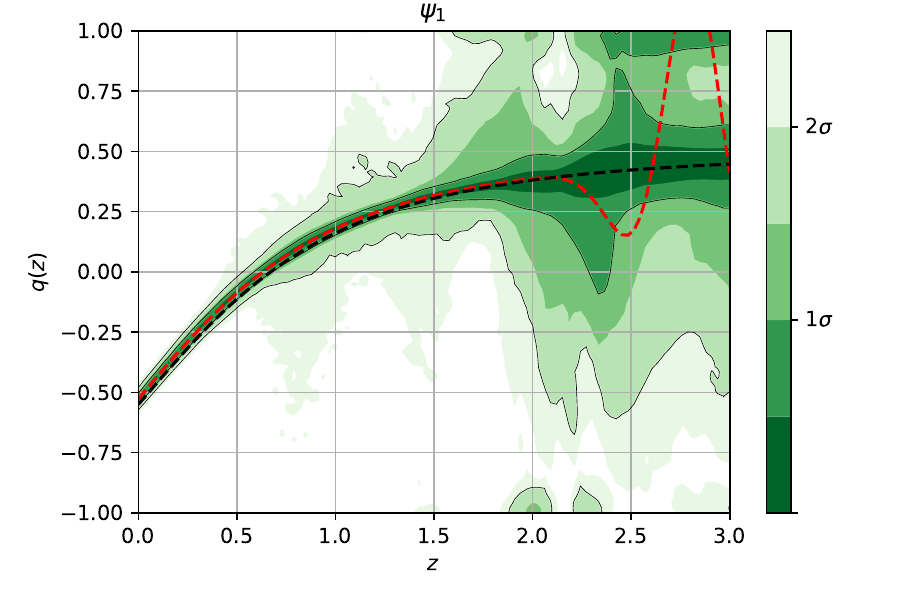}
    \includegraphics[trim = 5mm  0mm 25mm 0mm, clip, width=5.cm, height=4.5cm]{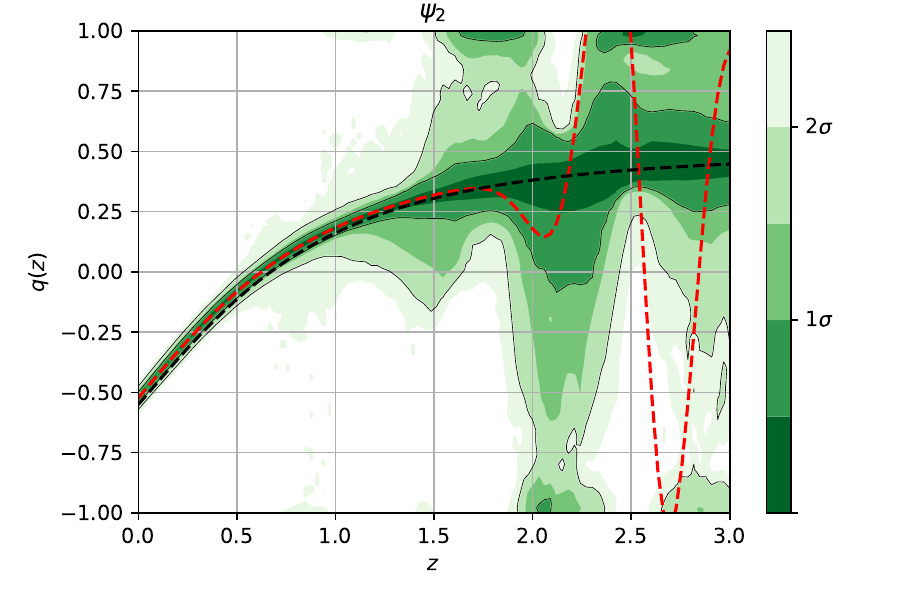}
    \includegraphics[trim = 5mm  0mm 25mm 0mm, clip, width=5.cm, height=4.5cm]{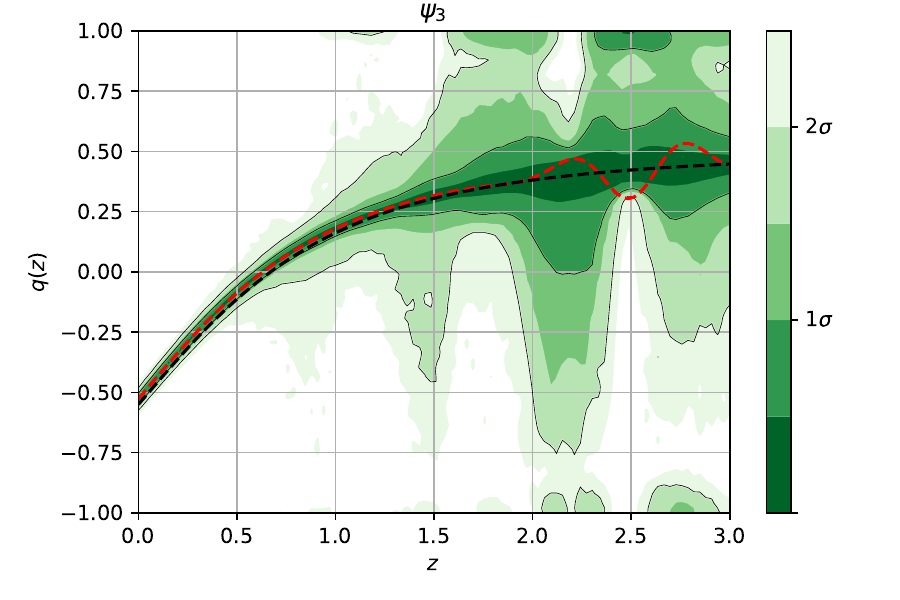}
    \includegraphics[trim = 5mm  0mm 5mm 0mm, clip, width=5.5cm, height=4.5cm]{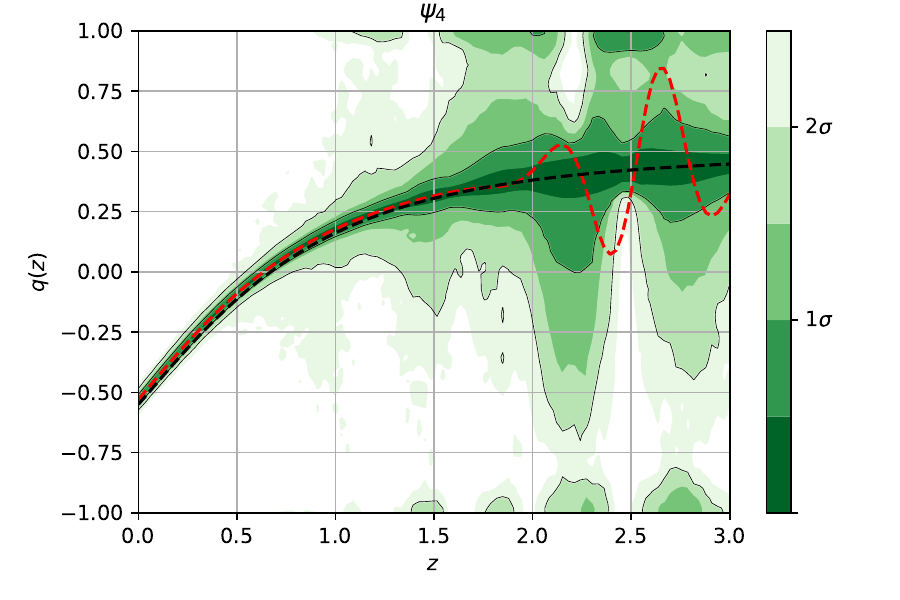}
    }

    \makebox[11cm][c]{
    \includegraphics[trim = 5mm  0mm 25mm 0mm, clip, width=5.cm, height=4.5cm]{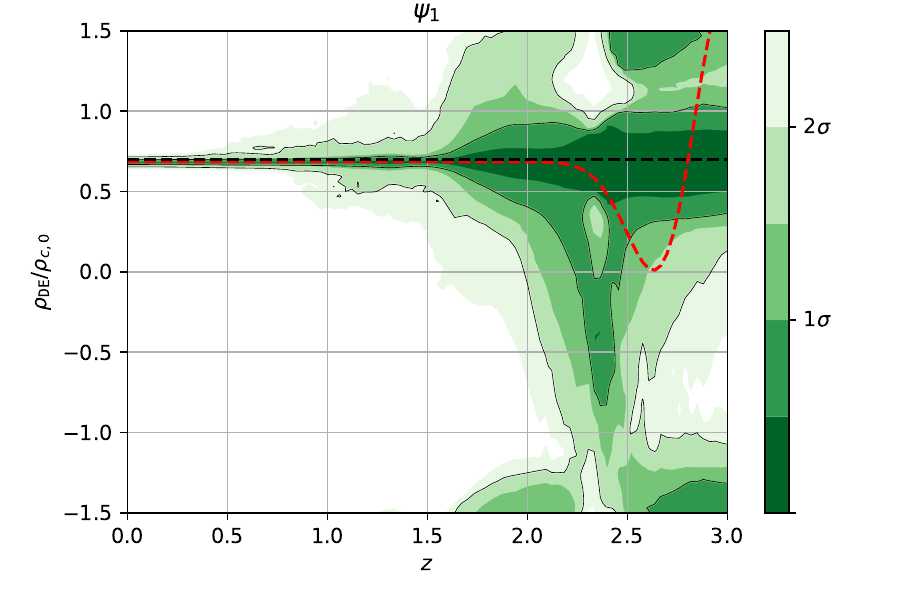}
    \includegraphics[trim = 5mm  0mm 25mm 0mm, clip, width=5.cm, height=4.5cm]{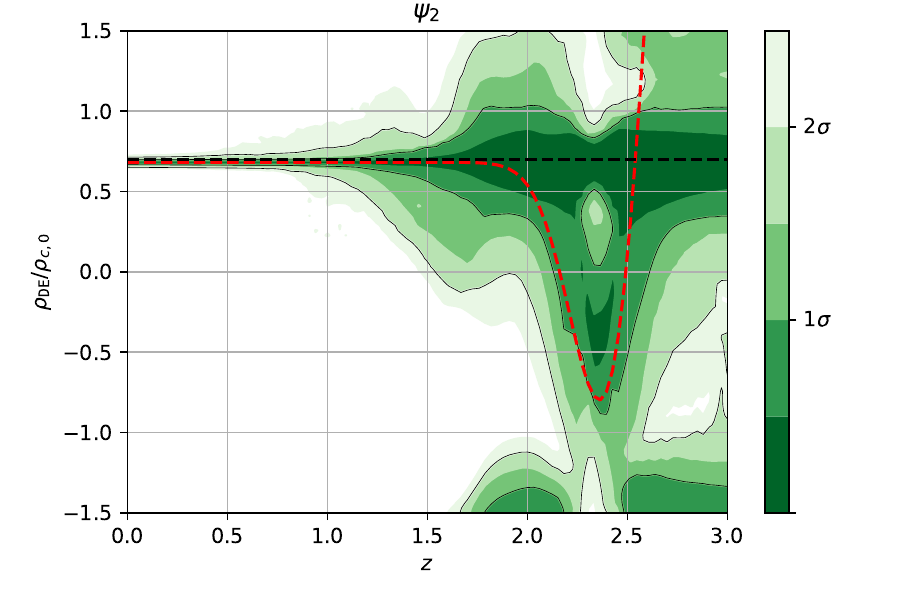}
    \includegraphics[trim = 5mm  0mm 25mm 0mm, clip, width=5.cm, height=4.5cm]{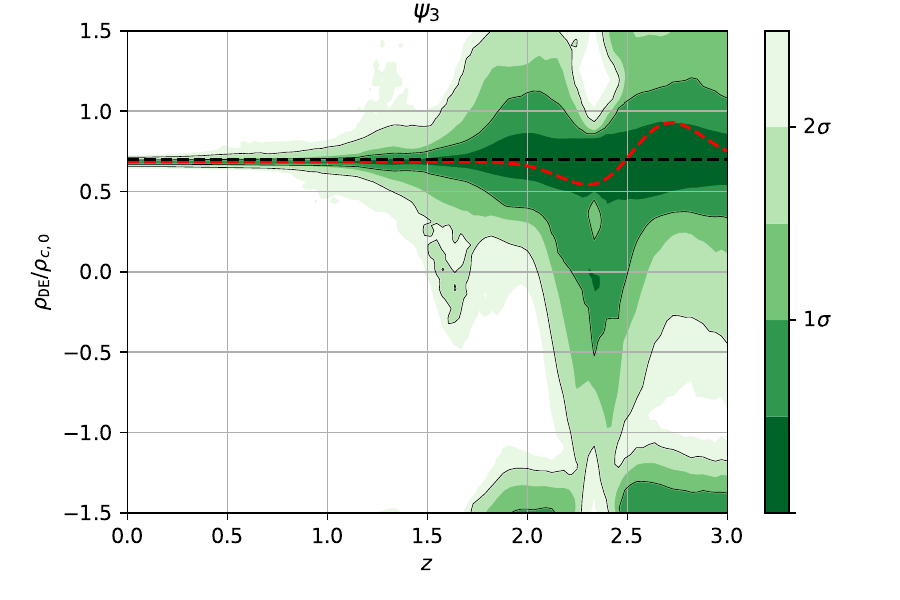}
    \includegraphics[trim = 5mm  0mm 5mm 0mm, clip, width=5.5cm, height=4.5cm]{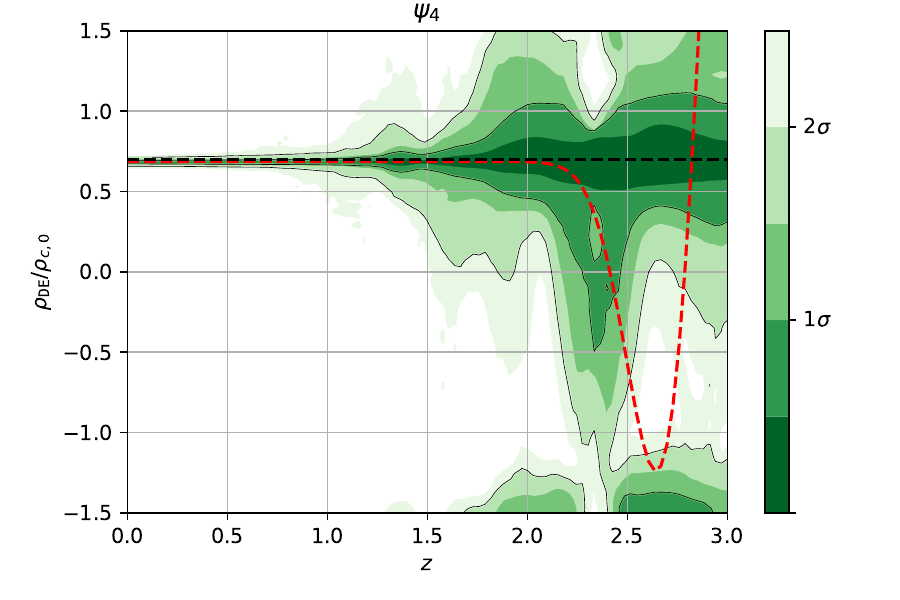}
    }

    \makebox[11cm][c]{
    \includegraphics[trim = 5mm  0mm 25mm 0mm, clip, width=5.cm, height=4.5cm]{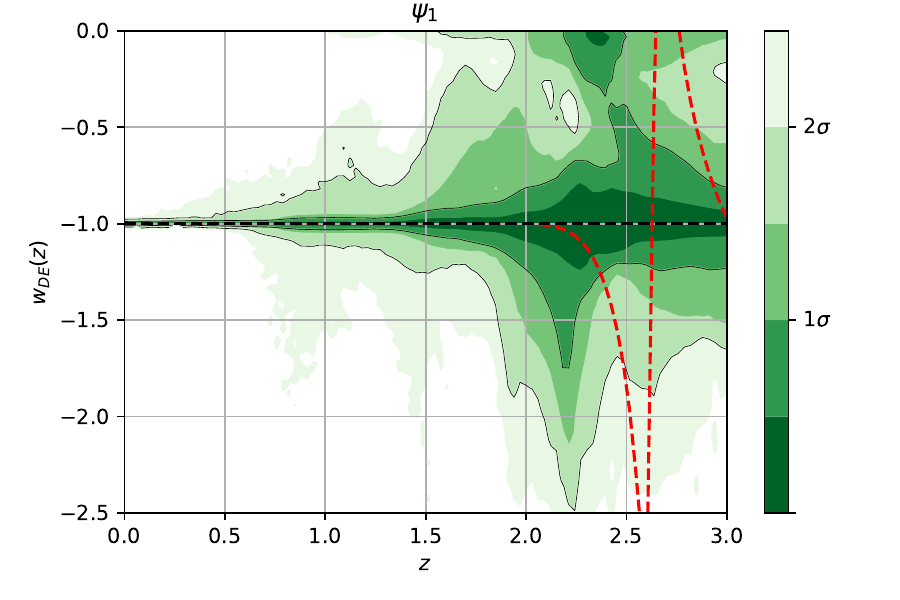}
    \includegraphics[trim = 5mm  0mm 25mm 0mm, clip, width=5.cm, height=4.5cm]{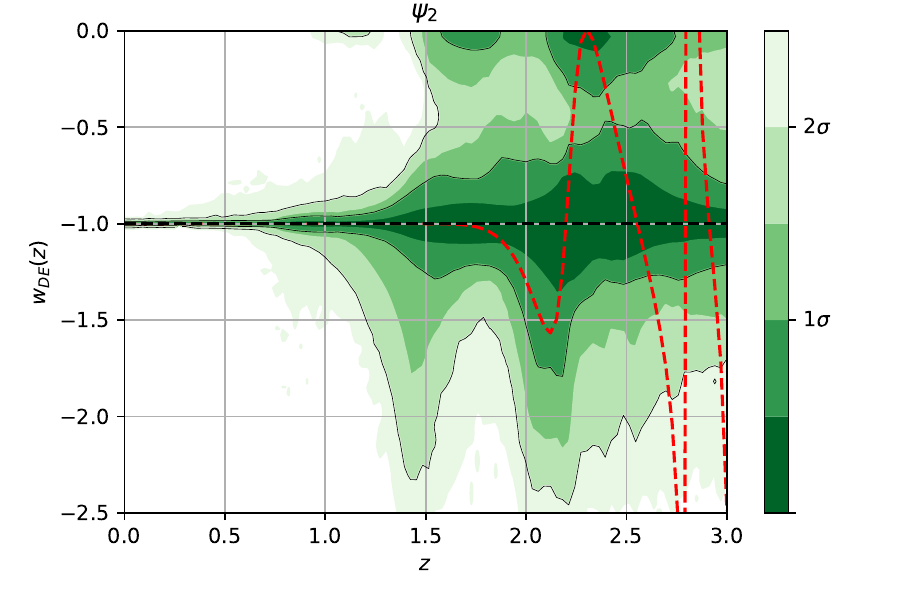}
    \includegraphics[trim = 5mm  0mm 25mm 0mm, clip, width=5.cm, height=4.5cm]{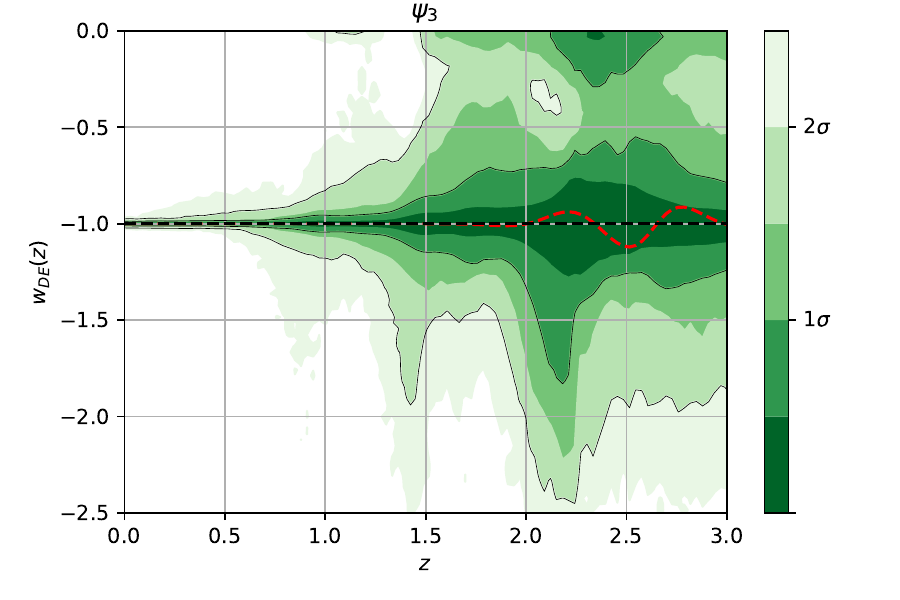}
    \includegraphics[trim = 5mm  0mm 5mm 0mm, clip, width=5.5cm, height=4.5cm]{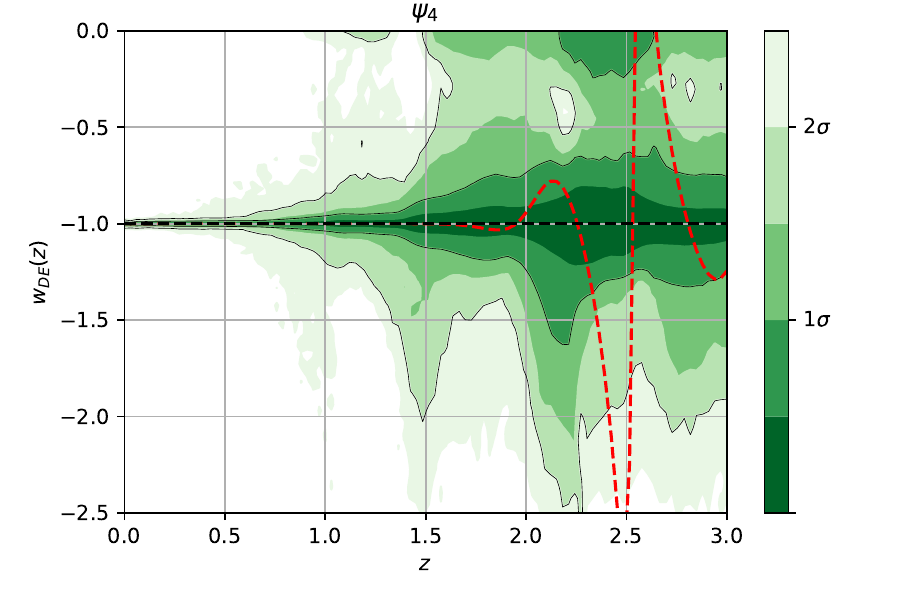}
    }

    \makebox[11cm][c]{
    \includegraphics[trim = 5mm  0mm 25mm 0mm, clip, width=5.cm, height=4.5cm]{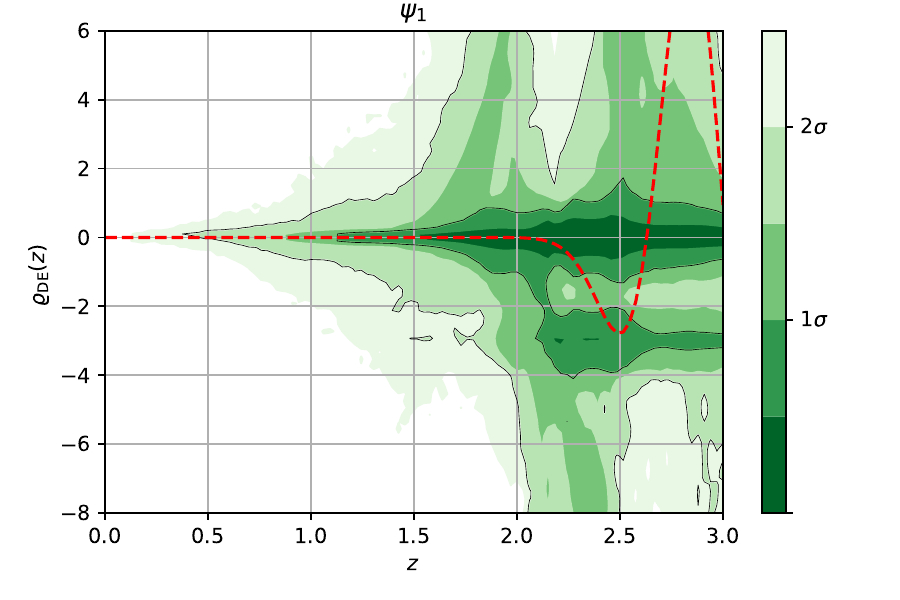}
    \includegraphics[trim = 5mm  0mm 25mm 0mm, clip, width=5.cm, height=4.5cm]{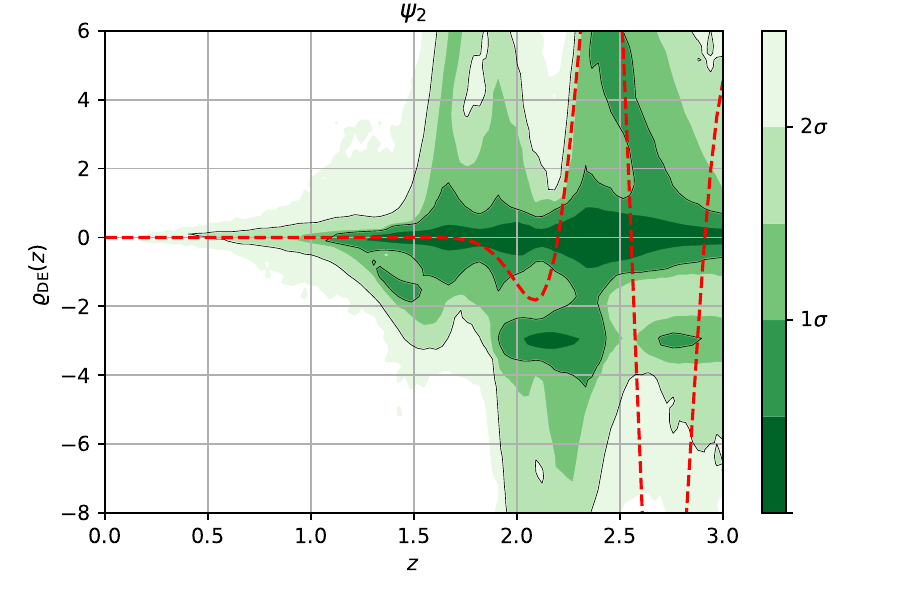}
    \includegraphics[trim = 5mm  0mm 25mm 0mm, clip, width=5.cm, height=4.5cm]{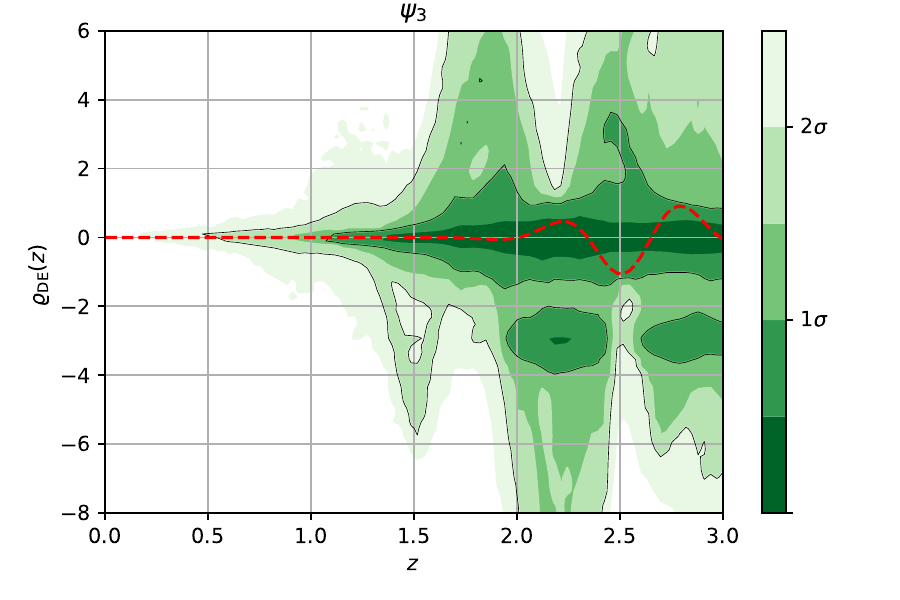}
    \includegraphics[trim = 5mm  0mm 5mm 0mm, clip, width=5.5cm, height=4.5cm]{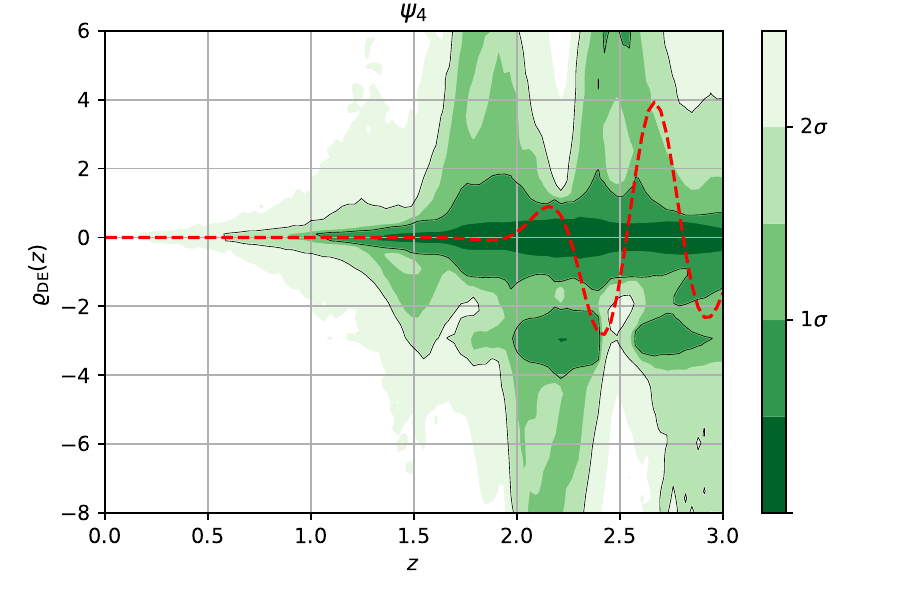}
    }

    \caption{Functional posterior probability of $H(z)/(1+z)$, $q(z)$, $\rho_{\rm DE}(z)/\rho_{\rm c,0}$, $w_{\rm DE}(z)$, and $\varrho_{\rm DE}(z)$ for the datasets \textbf{SB+SN}. 
    The probability, normalized in each slice of constant $z$, is shown with a color scale representing confidence interval values. The 68\% ($1\sigma$) and 95\% ($2\sigma$) confidence intervals are plotted as black lines. The dashed black line corresponds to the standard $\Lambda$CDM values, and the dashed red line represents the best-fit values of the wavelets. These plots were made using the Python library \texttt{fgivenx}~\cite{Handley_2018}.
}\label{fig:derived_wavelets}
\end{figure*}

\begin{figure}
\captionsetup{justification=Justified,font=footnotesize}
    \centering
    \includegraphics[trim = {0mm 0mm 0mm 0mm}, clip, width=8.5cm, height=5.cm]{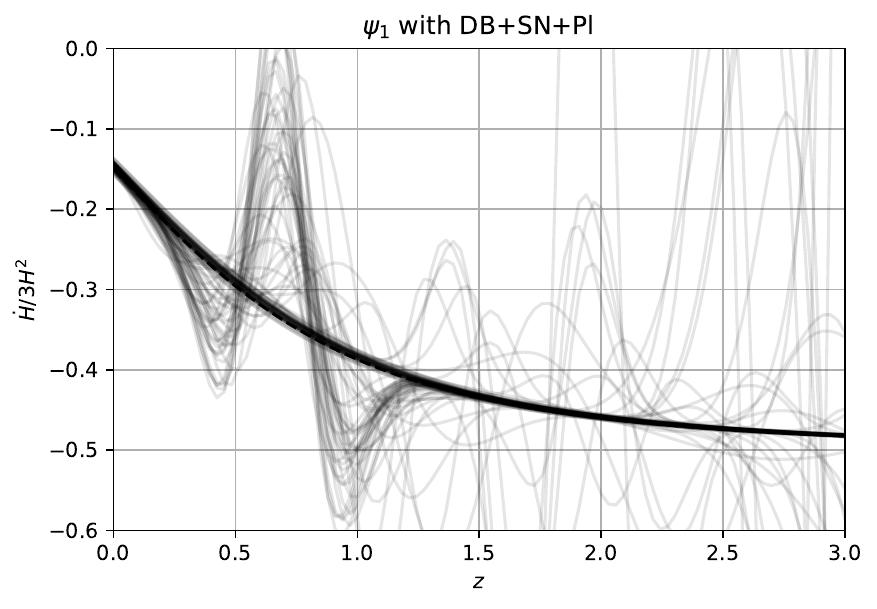}
    \caption{Functional posterior of the time derivative of the Hubble parameter, $\dot{H}$, normalized by the critical density, $3H^2$. The functional posterior was made with the reconstructed wavelet of first order, $\psi_1$, using the dataset combination \textbf{DB+SN+Pl}. The plot was made using the Python library \texttt{fgivenx}~\cite{Handley_2018}.
    }
    \label{fig:hdot_psi1}
\end{figure}

Let us investigate the results for a specific dataset combination to see the effect that the SDSS-BAO likelihood has, and to show the similar behavior of our wavelets of different orders. This case will be \textbf{SB+SN}, which presents a great fit of the data (the best out of all the \textbf{SB} cases) and relatively good Bayes' factors for every wavelet (all negative). 
The parameter $z^\dagger$ strongly favors values around $z = 2.3$. The specific values (with their $1\sigma$ uncertainties) are: $2.4 \pm 1.0$ for $\psi_1$, $2.44 \pm 0.92$ for $\psi_2$, $2.48 \pm 0.96$ for $\psi_3$, and $2.52 \pm 0.98$ for $\psi_4$.
When examining the functional posteriors depicted in~\cref{fig:derived_wavelets} concerning $H(z)/(1+z)$, the impact of the wavelets around $z \sim 2.4$ becomes evident. It can be deduced that the primary dataset influencing the positioning of our wavelets is the high-redshift Lyman-$\alpha$ BAO data given that it is localized at $z = 2.34$. This behavior directly translates to the DE density as a possible transition from positive at low redshifts to negative at high redshifts as seen in~\cref{fig:derived_wavelets}. 
This is particularly accentuated by the best-fits (red-dashed lines in~\cref{fig:derived_wavelets}) of wavelets $\psi_2$ and $\psi_4$, which have the lowest $-2\Delta\ln \mathcal{L_{\rm max}}$ overall (for the cases with \textbf{SB} in the datasets). 
This transition to a negative value is accentuated by the DE EoS parameter since it presents a pole ($\lim_{z\to z_{\rm p}^{\pm}} w_{\rm DE}(z)=\pm\infty$ with $z_{\rm p}$ being the point where DE density vanishes) when using the best-fit values (dashed red lines in~\cref{fig:derived_wavelets}) close to $z = 2.5$. This pole and asymptotic behavior have been observed in previous works considering effective DE densities changing sign~\cite{Sahni:2002dx,Sahni:2004fb,Tsujikawa:2008uc,Zhou:2009cy,Bauer:2010wj,Sahni:2014ooa,Gomez-Valent:2015pia,Wang:2018fng,Akarsu:2019hmw,Akarsu:2019ygx,DiValentino:2020naf,Akarsu:2019pvi,Acquaviva:2021jov,Akarsu:2021fol,Escamilla:2021uoj,Akarsu:2022lhx,Ozulker:2022slu,Wen:2023wes,Akarsu:2024eoo,Akarsu:2024qsi,Ong:2022wrs,DiGennaro:2022ykp,Tiwari:2023jle,Dwivedi:2024okk,Manoharan:2024thb} and its presence is required when the DE density changes sign for a conserved DE~\cite{Ozulker:2022slu}. This feature of the DE density being able to attain negative values while also being able to oscillate makes the DE source derived from the wavelets a member of the omnipotent DE family~\cite{Adil:2023exv}.

However, even though Hermitian wavelets seem to be a good extension for the standard model, we cannot ignore how their influence translates into the deceleration parameter $q(z)$, which exhibits strong oscillatory behavior, especially with $\psi_2$. This would imply significant changes in the dynamics of the Universe, directly affecting structure formation~\cite{SupernovaCosmologyProject:1998vns,SupernovaSearchTeam:1998fmf,Bolotin:2015dja,delCampo:2012ya,Naik:2023yhl}. These heavy oscillations are also evident in the DE density. The primary cause for this is the fact that the parameter $\beta_h$ (which has a direct impact on the width and amplitude of the oscillations) is not constrained due to the lack of data beyond $z=2.34$.
To fully resolve this issue, we would require data that covers a larger region of redshift so that $\beta_h$ can be fully characterized. Beyond $z=2.5$, we can only speculate about the possible behavior of the wavelets. Despite this issue, it is clear that \textbf{SB} prefers the Hermitian wavelet to exert its influence beyond $z \sim 1.5$.

Finally, we comment on the time derivative of the Hubble function as plotted in~\cref{fig:hdot_psi1}. When all the modifications to the Friedmann equations can be absorbed in the total energy density ($\rho$) and pressure ($p$) components of the energy-momentum tensor, this time derivative reads
\begin{equation}
    \dot{H}=-\frac{1}{2}(\rho+p),\label{eq:hdot}
\end{equation}
where the dot denotes differentiation with respect to cosmic time.
Notice that the total inertial mass density $\varrho=\rho+p$ we discuss here is different from the DE inertial mass density, $\varrho_{\rm DE}$, shown in the last row of~\cref{fig:derived_wavelets}. Plotting~\cref{eq:hdot} is of interest since if $\dot{H}$ is positive, the null energy condition is violated by the total energy-momentum content (which might include effective components originating from, say, a modified theory of gravity) of the Universe. The violation of the null energy condition and even the averaged null energy condition are abundant in physics; see the discussions and references in Refs.~\cite{Epstein:1965zza,Wald:1991xn,Visser:1999de,Curiel:2014zba,Ozulker:2022slu}. As seen in the example case presented in~\cref{fig:hdot_psi1} (corresponding to the case in~\cref{fig:dm_psi1} and the second panel of~\cref{fig:hz_different_datasets}), some of the plotted lines from the posterior become positive at certain redshifts, indicating a violation of the null energy condition. However, this behavior is not present in most of the samples and is not accentuated enough to make any claims regarding a preference for this behavior from the available data. Moreover, we would expect the presence of this violation within the wavelet framework to be highly dependent on the type of wavelet that is chosen; e.g., a more squarely shaped wavelet would induce more abrupt oscillations in the Hubble parameter, causing this violation to be much more common in its constrained parameter space.

\section{Conclusions}\label{section:conclusions}

In this paper, we used a parameter inference procedure to analyze the behaviors of the Hubble function as preferred by background cosmological data. In this regard, we parameterized the Hubble function by considering deviations from the Hubble radius of $\Lambda$CDM in the form of wavelets as discussed in Ref.~\cite{Akarsu:2022lhx}. Wavelets provide viable extensions of the $\Lambda$CDM model while preserving some features of it, such as the size of the comoving sound horizon and the fit to the measurement of its angular scale at last scattering.

All our reconstructions present a significant improvement in the fit to the data, reaching more than $3\sigma$ indication for wavelets when including DESI-BAO data. Most cases even present a negative Bayes' factor, indicating inconclusive or weak evidence in favor of wavelets (according to Jeffreys' Scale), despite having three extra parameters. In line with the discussions in Ref.~\cite{Akarsu:2022lhx}, a dataset that exerts great influence on the behavior of the wavelets is BAO, which can dictate where the wavelet will be localized. DESI-BAO prefers the center of the wavelets to be at $z\sim0.7$, while SDSS-BAO pushes it to $z\sim2.3$. Our results indicate that this behavior is driven by differences between the $D_H(z)/r_{\rm d}$ measurements of the two BAO datasets at $z<1$ and $z\sim2.3$, in particular, the discrepancy between the SDSS and DESI measurements at $z=0.51$ and in their Lyman-$\alpha$ BAO measurements. This discrepancy between BAO datasets could be due to statistical fluctuations, but it is beyond the scope of this work to explore its cause. We also note that the improvement in the fit compared to $\Lambda$CDM is significantly larger for the DESI-BAO compared to SDSS, and the addition of our $\textbf{SN}$ dataset slightly ameliorates these improvements in line with the findings of the DESI collaboration for a preference of dynamical DE with BAO and SN Ia data~\cite{DESI:2024mwx}.

We also showed how other derived quantities such as $q(z)$, $\rho_{\rm DE}$, and $w_{\rm DE}$ would behave and what effects the wavelets could have, particularly when SDSS-BAO is used. When the wavelet deviations are assumed to originate from a dynamical DE component, this DE presents some previously studied behaviors such as oscillations and a possible transition from a positive DE energy density to a negative one as redshift increases, making this DE a member of the omnipotent DE~\cite{Adil:2023exv} family characterized by the combination of these two behaviors. While oscillations are present in $\rho_{\rm DE}$ by construction due to the oscillatory behaviors of the wavelets, the DE density attaining negative values in the past is a consequence of the data.

Despite the promising results obtained in this study, there is a problem involving the lack of data beyond a certain redshift value. For some cases that involve the use of SDSS-BAO data, the center of the wavelet, determined by the parameter $z^\dagger$, is located near $z \approx 2.4$, and our data only reaches up to $z = 2.34$. Thus, there might be some features of the wavelets that simply cannot be constrained, such as the parameter $\beta_h$. To address this, our only recourse is to await the availability of more data beyond this range, at which point we can revisit and re-perform the reconstructions done in this paper.

We conclude by recognizing that Hermitian wavelets, on top of the Hubble radius of the standard $\Lambda$CDM model, in the late Universe show a lot of promise in capturing the dynamics of deviations from the standard model of cosmology at the background level without altering the early Universe and the constraints on the matter density and the Hubble constant.
These behaviors can then be attributed to underlying physics, such as a dynamical dark energy component, and different underlying physics with the same background phenomena can be tested at the perturbative level with full data.

\appendix
\counterwithin{figure}{section}
\section{Additional Constraints and Figures}\label{appendix_a}

\begin{figure*}[t!]
\captionsetup{justification=Justified,singlelinecheck=false,font=footnotesize}
    \centering
    \makebox[11cm][c]{
    \includegraphics[trim = 0mm  0mm 0mm 0mm, clip, width=6.0cm, height=6.0cm]{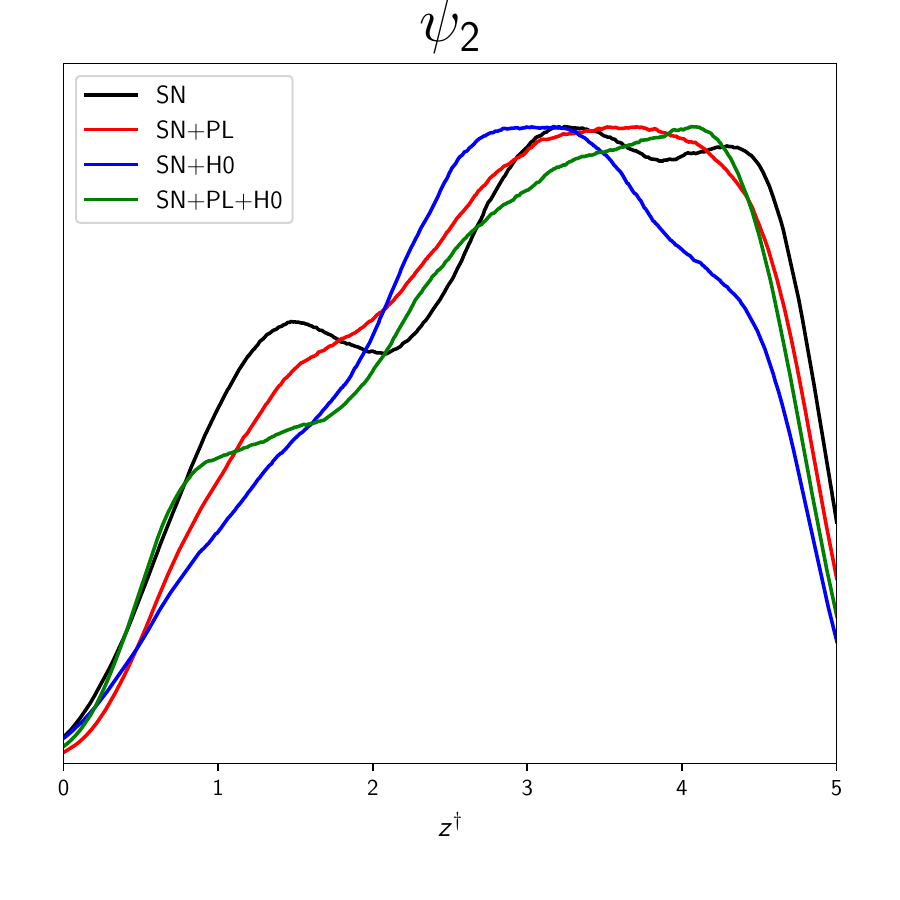}
    \includegraphics[trim = 0mm  0mm 0mm 0mm, clip, width=6.0cm, height=6.0cm]{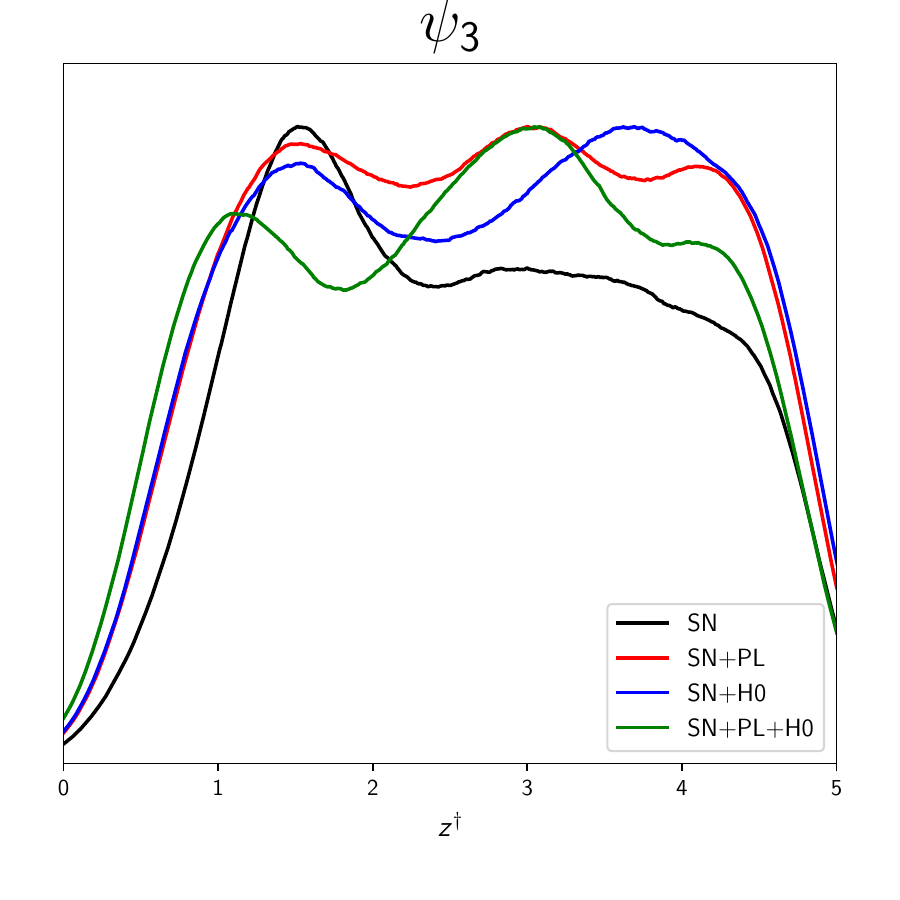}
    \includegraphics[trim = 0mm  0mm 0mm 0mm, clip, width=6.0cm, height=6.0cm]{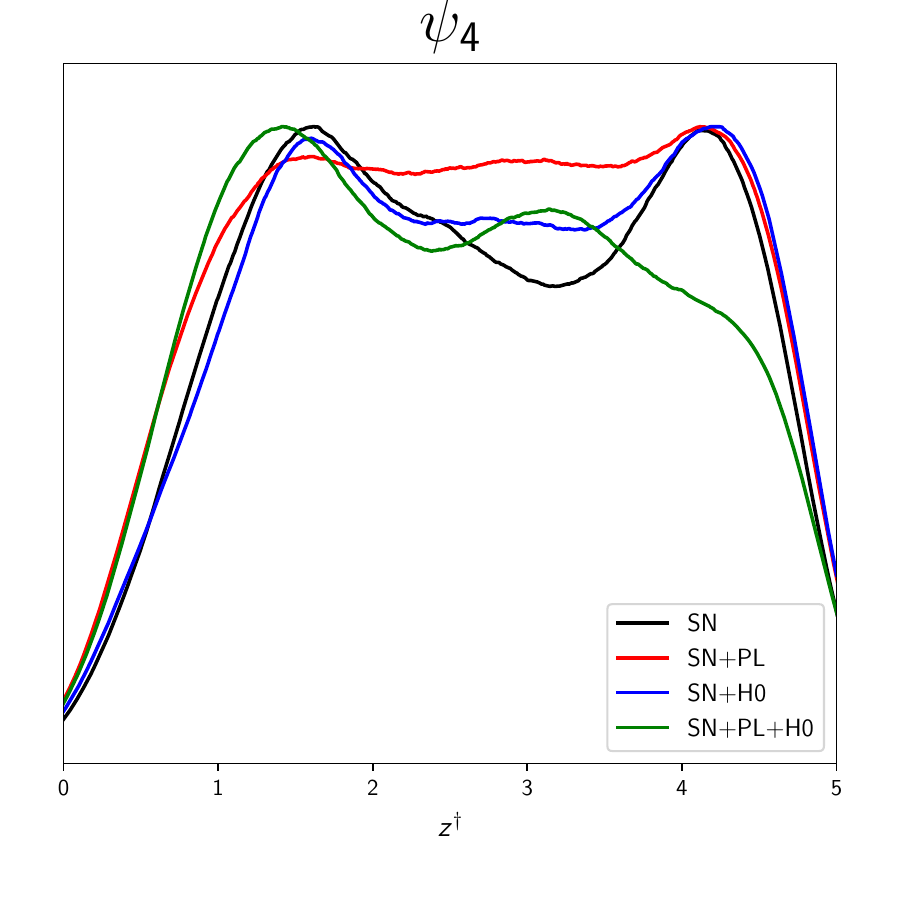}}

    \makebox[11cm][c]{
    \includegraphics[trim = 0mm  0mm 0mm 0mm, clip, width=6.0cm, height=6.0cm]{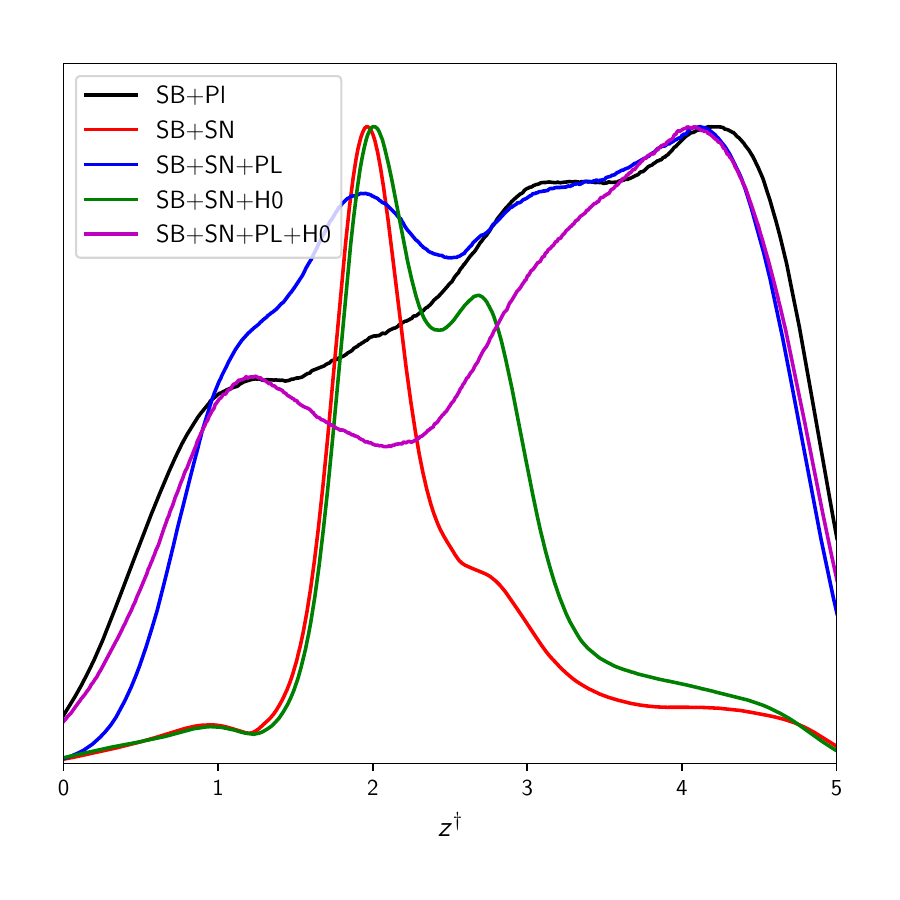}
    \includegraphics[trim = 0mm  0mm 0mm 0mm, clip, width=6.0cm, height=6.0cm]{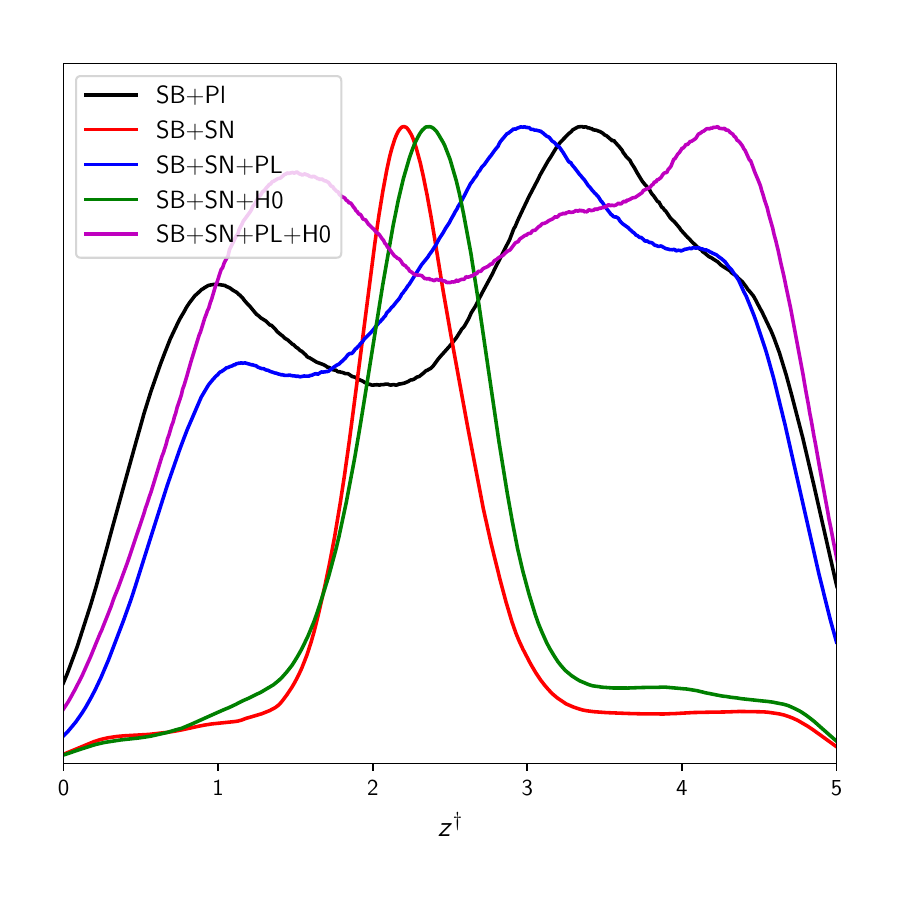}
    \includegraphics[trim = 0mm  0mm 0mm 0mm, clip, width=6.0cm, height=6.0cm]{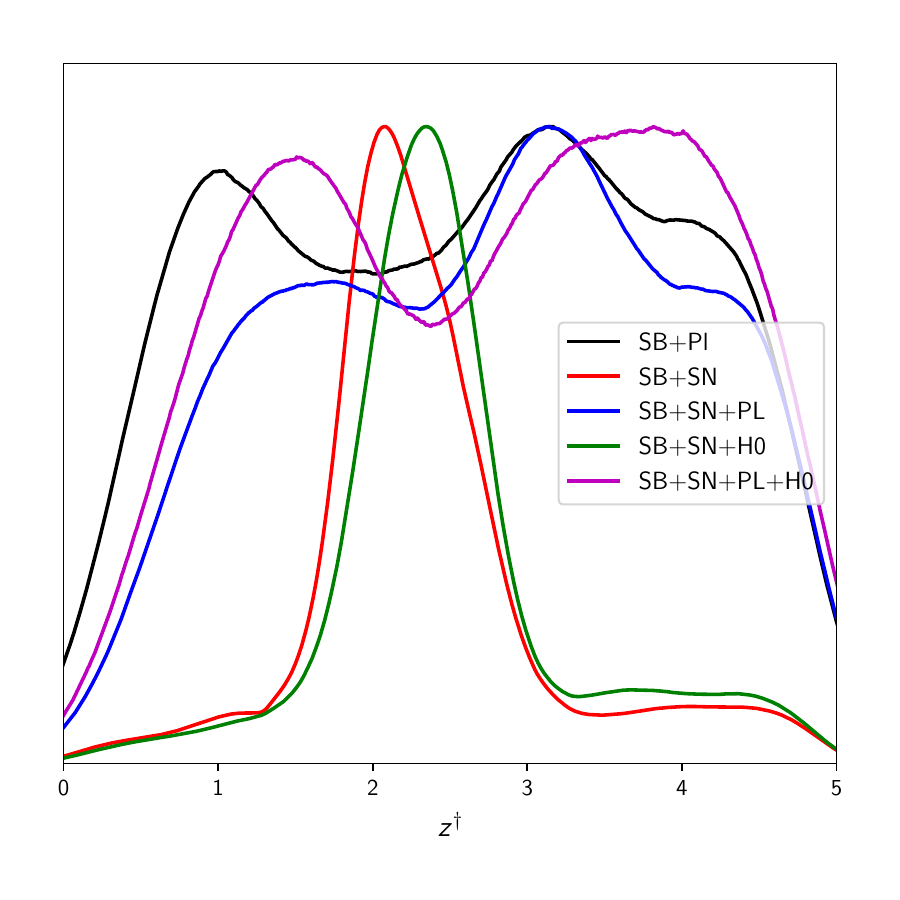}}
    
    \makebox[11cm][c]{
    \includegraphics[trim = 0mm  0mm 0mm 0mm, clip, width=6.0cm, height=6.0cm]{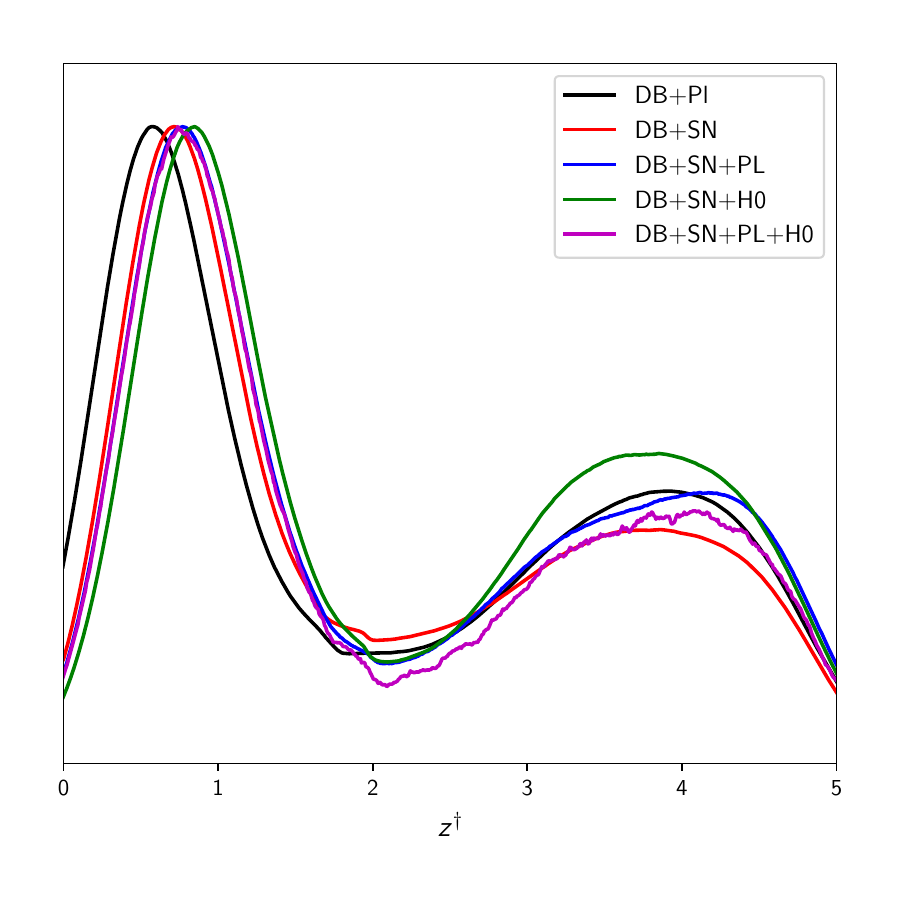}
    \includegraphics[trim = 0mm  0mm 0mm 0mm, clip, width=6.0cm, height=6.0cm]{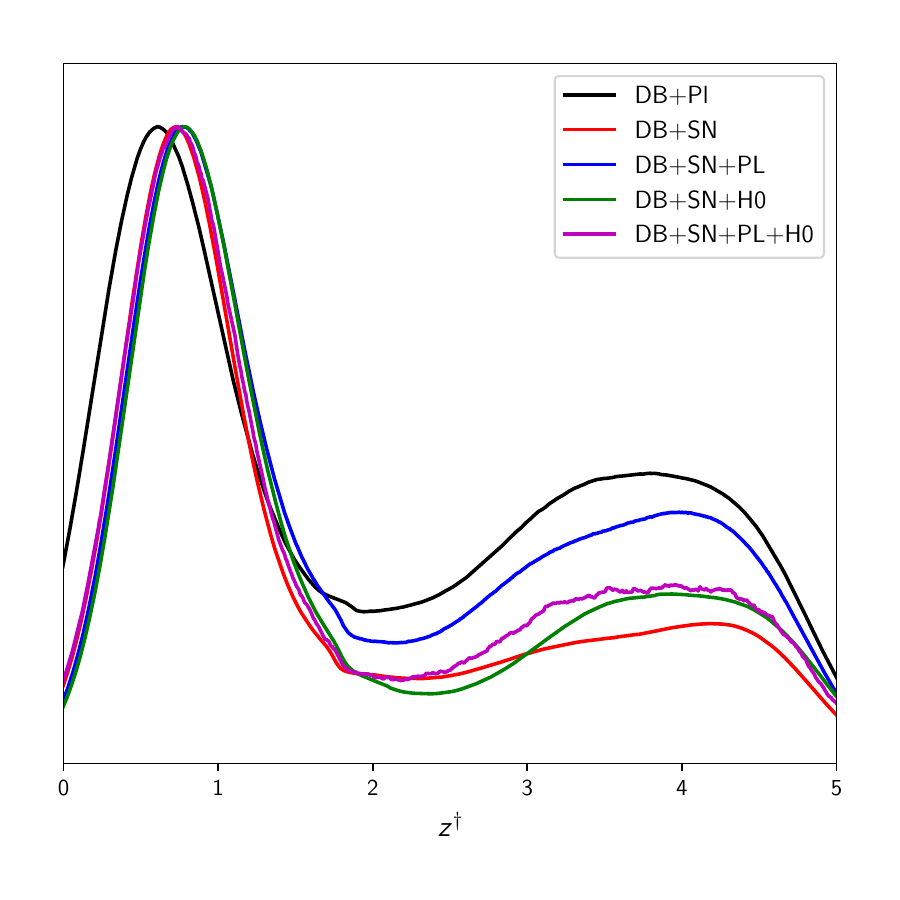}
    \includegraphics[trim = 0mm  0mm 0mm 0mm, clip, width=6.0cm, height=6.0cm]{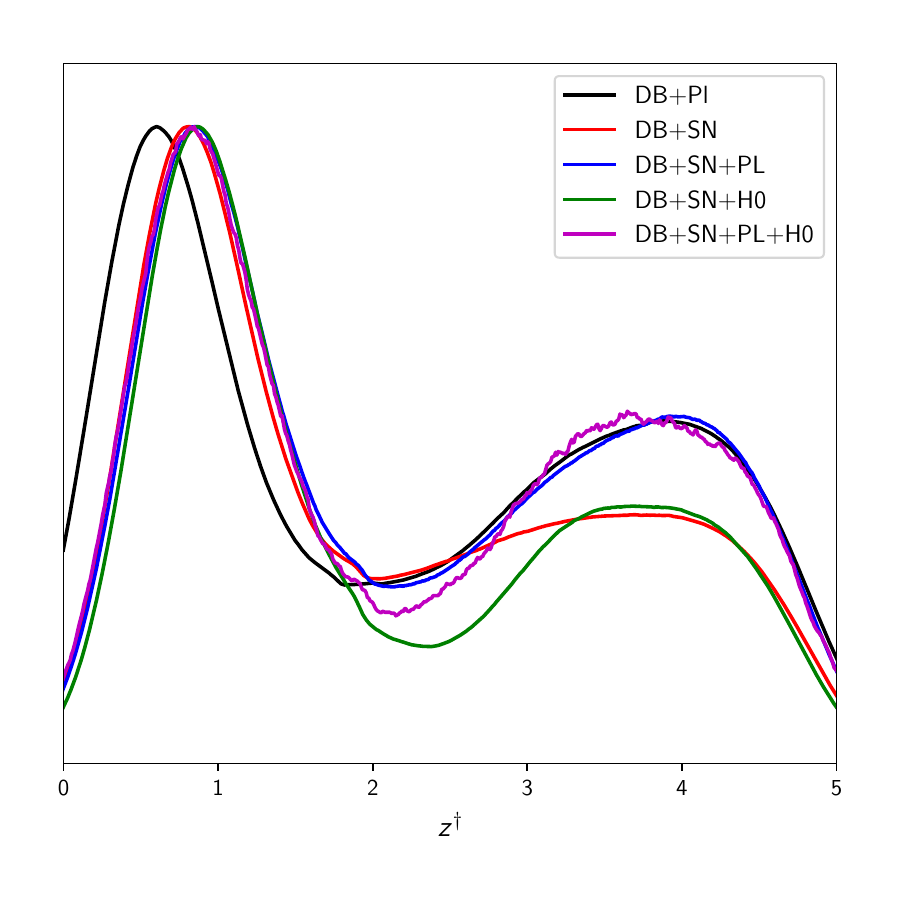}}
    \caption{1D marginalized posterior distributions for the reconstructed parameter $z^\dagger$ for (from left to right) $\psi_2(z)$, $\psi_3(z)$, and $\psi_4(z)$. We present the results in three different panels to illustrate the effect of the BAO datasets on the posteriors. These plots were created using the Python library \texttt{getdist}~\cite{Lewis:2019xzd}.
}\label{fig:1d_comparison_other_psi}
\end{figure*}

\begin{figure*}[t!]
\captionsetup{justification=Justified,singlelinecheck=false,font=footnotesize}
    \centering
    \makebox[11cm][c]{
    \includegraphics[trim = 0mm  0mm 0mm 0mm, clip, width=5.cm, height=4.cm]{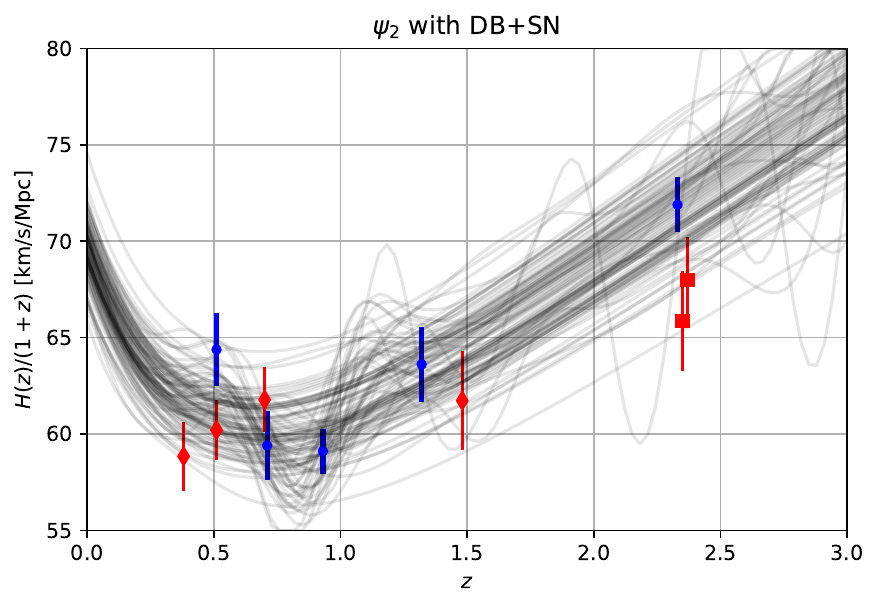}
    \includegraphics[trim = 12mm  0mm 0mm 0mm, clip, width=4.88cm, height=4.cm]{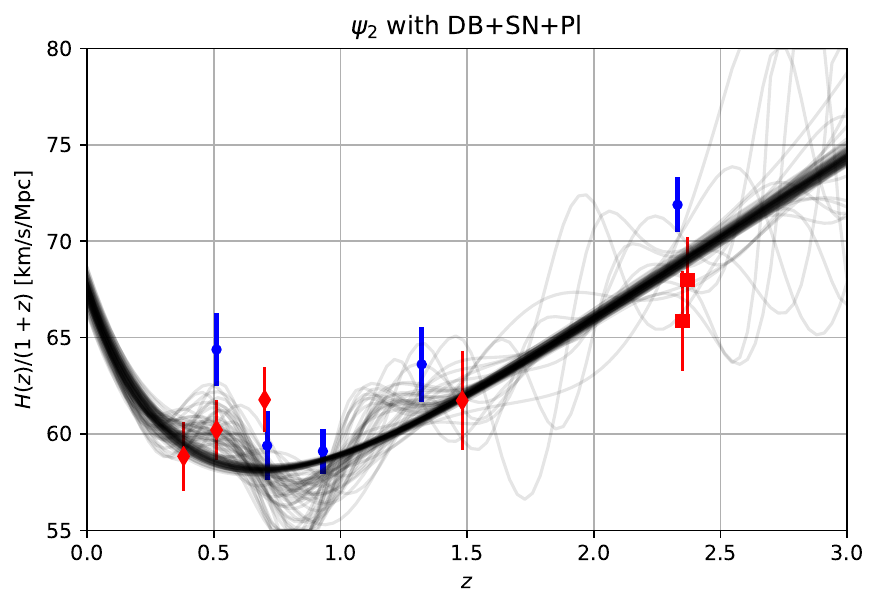}
    \includegraphics[trim = 12mm  0mm 0mm 0mm, clip, width=4.88cm, height=4.cm]{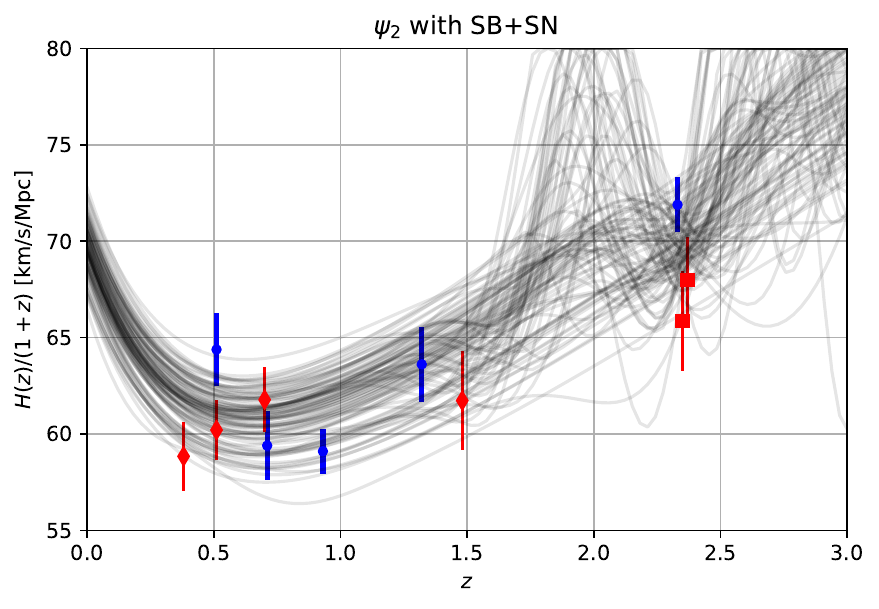}
    \includegraphics[trim = 12mm  0mm 0mm 0mm, clip, width=4.88cm, height=4.cm]{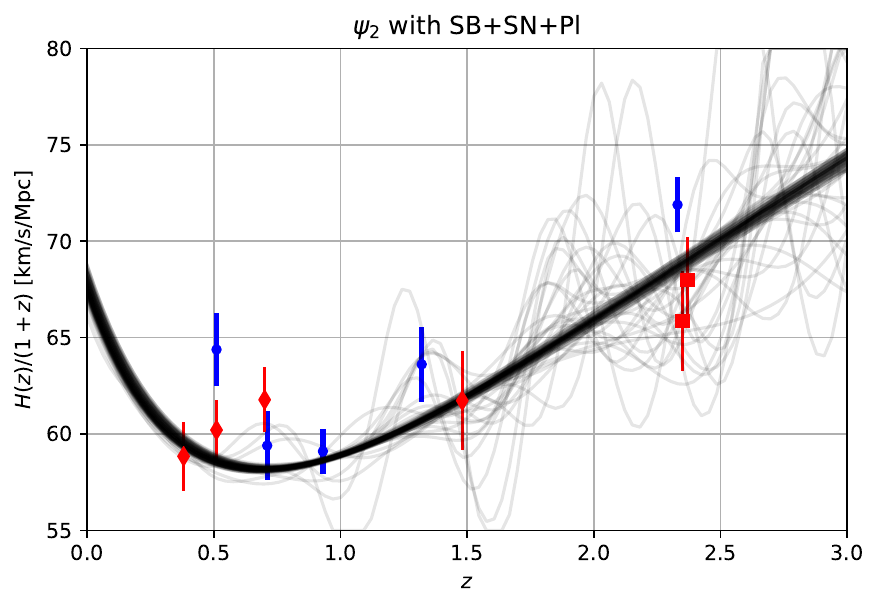}
    }

    \makebox[11cm][c]{
    \includegraphics[trim = 0mm  0mm 0mm 0mm, clip, width=5.cm, height=4.cm]{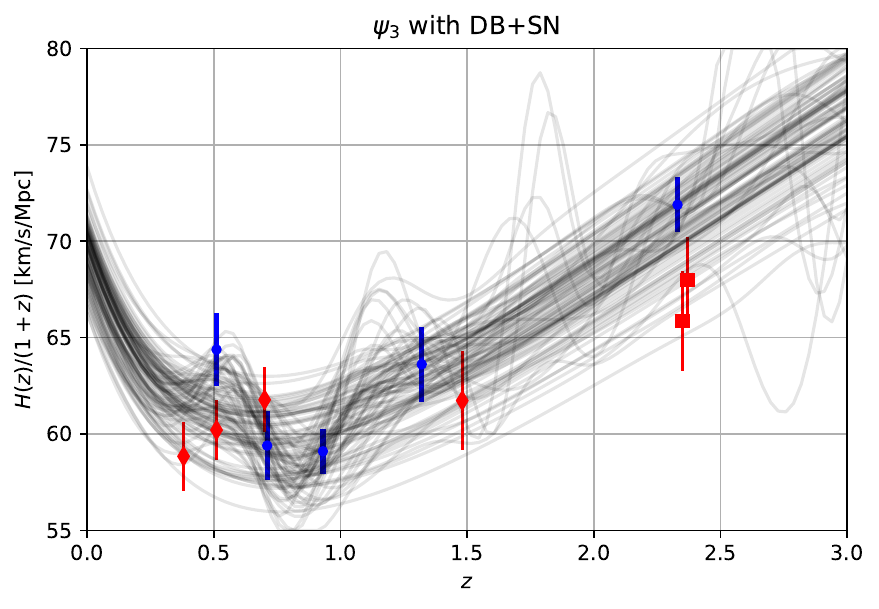}
    \includegraphics[trim = 12mm  0mm 0mm 0mm, clip, width=4.88cm, height=4.cm]{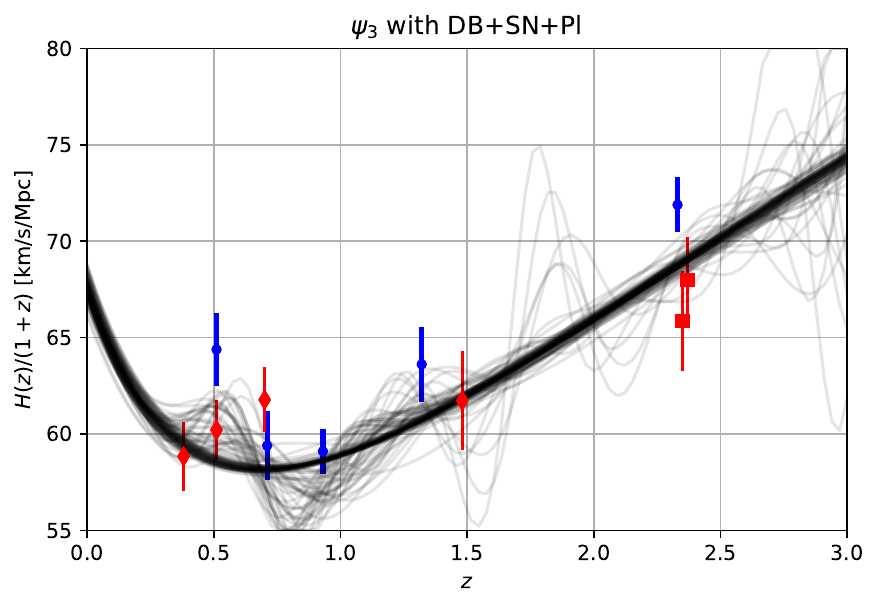}
    \includegraphics[trim = 12mm  0mm 0mm 0mm, clip, width=4.88cm, height=4.cm]{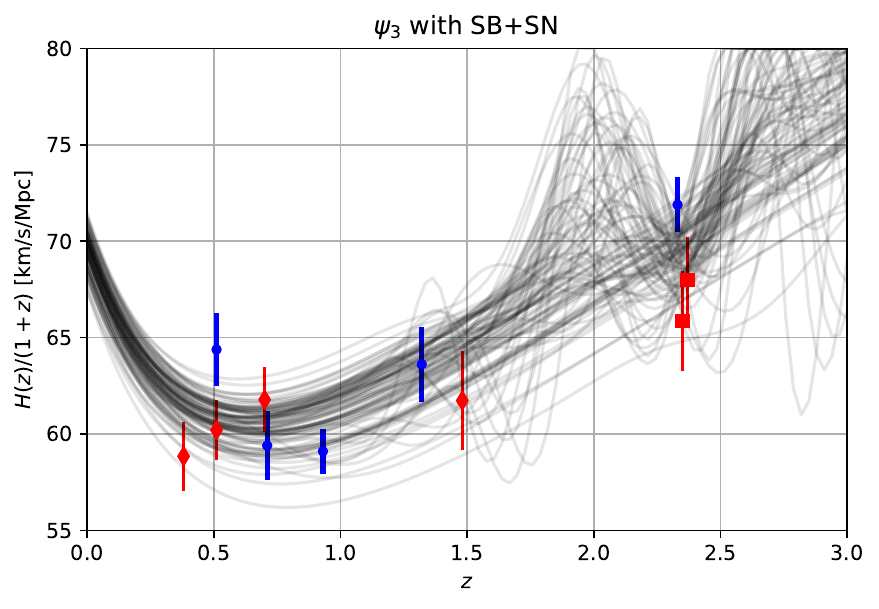}
    \includegraphics[trim = 12mm  0mm 0mm 0mm, clip, width=4.88cm, height=4.cm]{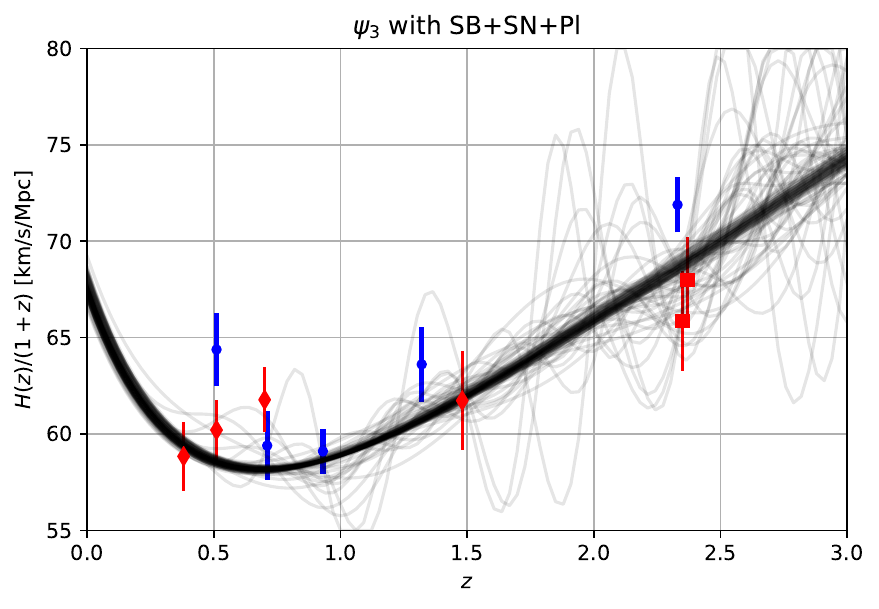}
    }

    \makebox[11cm][c]{
    \includegraphics[trim = 0mm  0mm 0mm 0mm, clip, width=5.cm, height=4.cm]{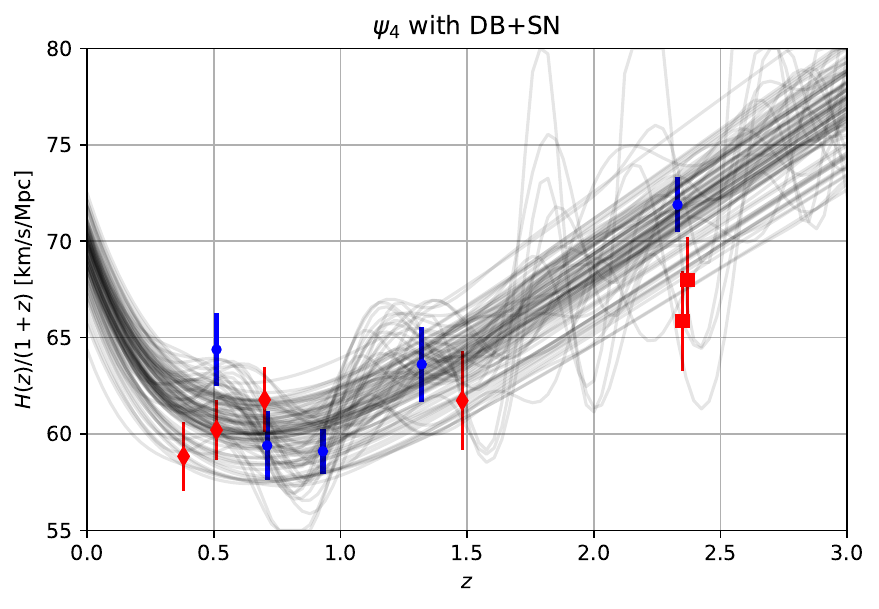}
    \includegraphics[trim = 12mm  0mm 0mm 0mm, clip, width=4.88cm, height=4.cm]{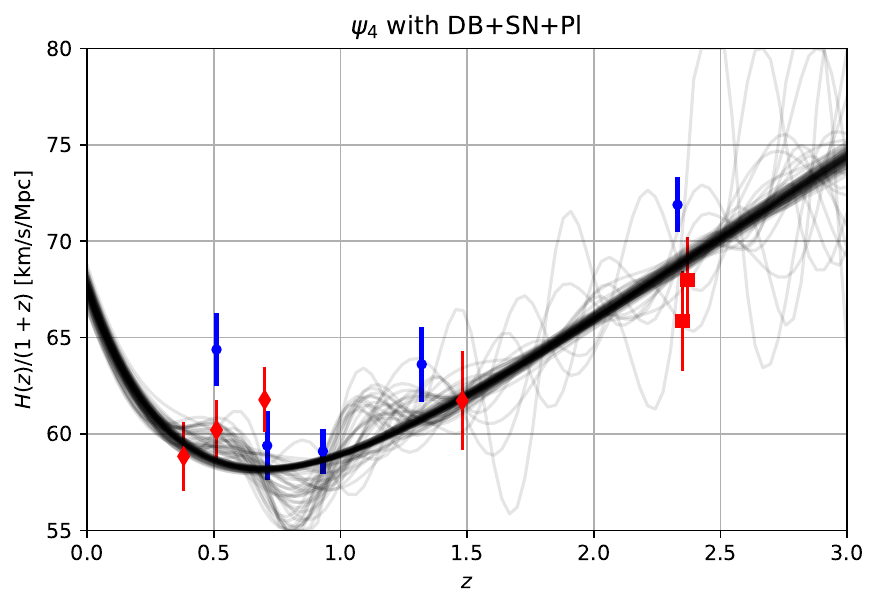}
    \includegraphics[trim = 12mm  0mm 0mm 0mm, clip, width=4.88cm, height=4.cm]{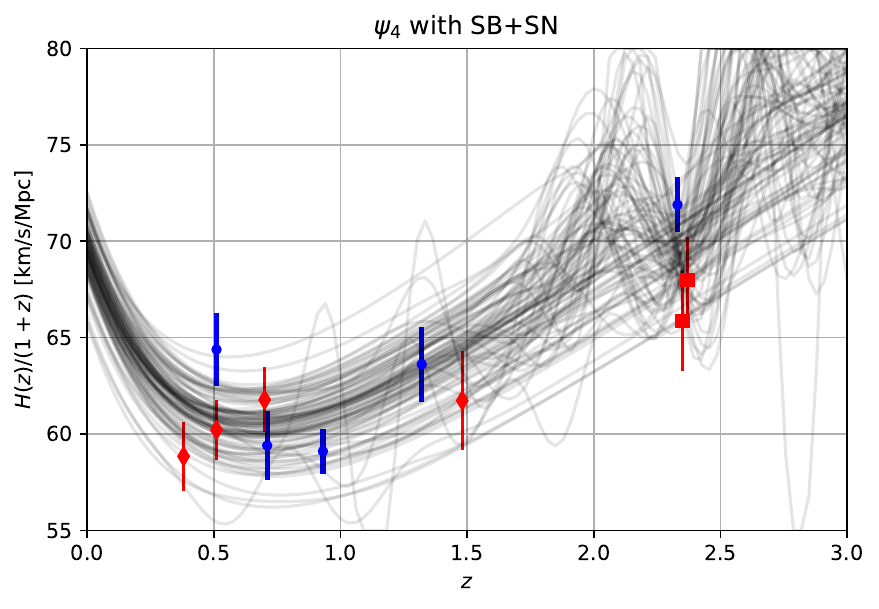}
    \includegraphics[trim = 12mm  0mm 0mm 0mm, clip, width=4.88cm, height=4.cm]{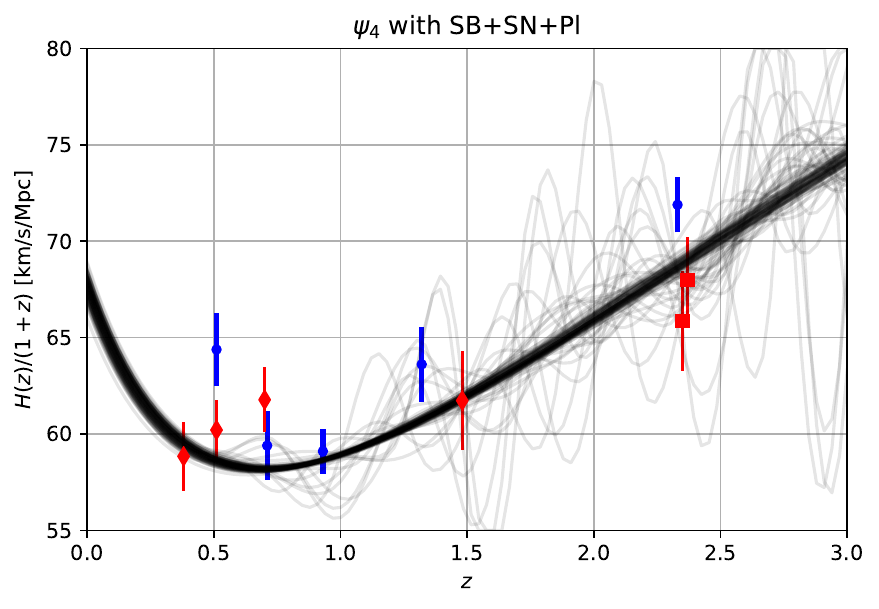}
    }
    
    \caption{Functional posterior for $H(z)/(1+z)$ for $\psi_1(z)$ using four distinct data combinations: \textbf{DB+SN}, \textbf{DB+SN+Pl}, \textbf{SB+SN}, and \textbf{SB+SN+Pl}. The red (blue) error bars correspond to the SDSS (DESI) BAO distance $D_H(z)/r_{\rm d}$ measurements. The size of the sound horizon was fixed at the robust value of $147$ Mpc to highlight the BAO data in the figure. The plots were created using the Python library \texttt{fgivenx}~\cite{Handley_2018}.
}\label{fig:hz_different_datasets_appendix}
\end{figure*}

For completeness, we provide some results for the other reconstructed wavelets, although this might seem redundant given how similar they are. It is no surprise that most of the outcomes and behaviors reconstructed for $\psi_1$ are shared by the rest of the Hermitian wavelets, given their similarities. This is easily verifiable by comparing~\cref{fig:1d_comparison} with~\cref{fig:1d_comparison_other_psi} and~\cref{fig:hz_different_datasets} with~\cref{fig:hz_different_datasets_appendix}.
Despite using different Hermitian wavelets, the constraints on the parameter $z^\dagger$ remain essentially the same, and the Hubble functions show only minor variations, primarily driven by the choice of datasets. Therefore, we have opted not to extend the discussion to the other wavelets.

\begin{figure*}[t!]
\captionsetup{justification=Justified,singlelinecheck=false,font=footnotesize}
    \centering
    \makebox[11cm][c]{
    \includegraphics[trim = 0mm  0mm 0mm 0mm, clip, width=5.0cm, height=5.0cm]{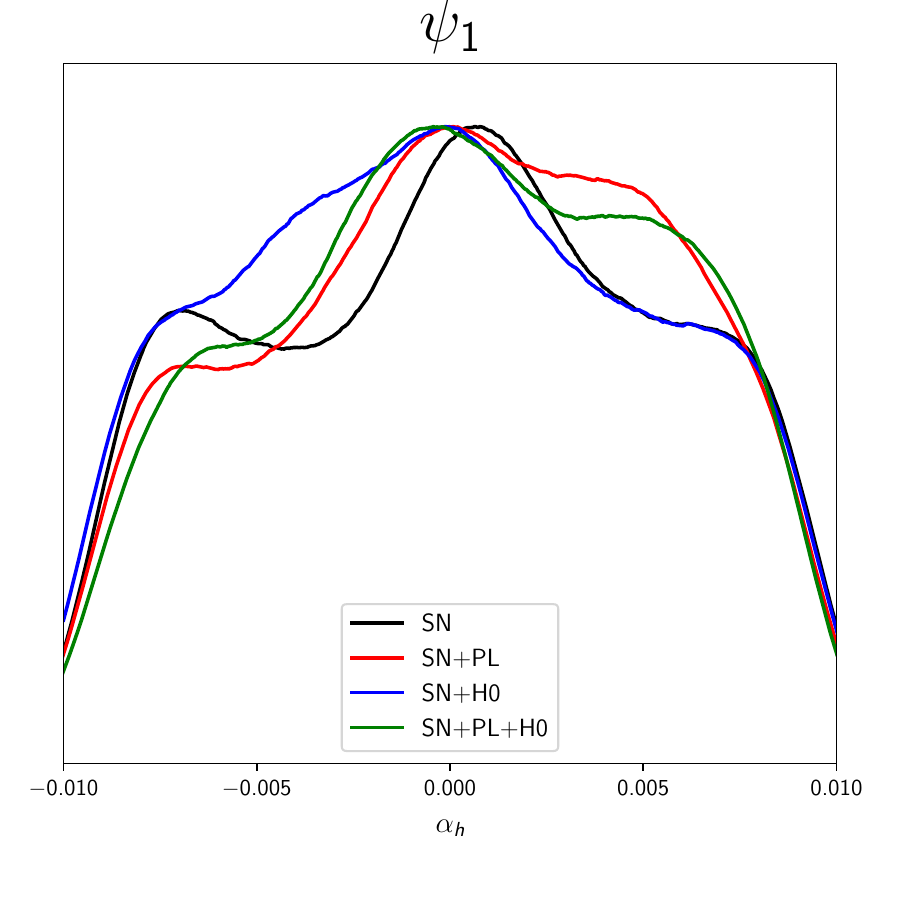}
    \includegraphics[trim = 0mm  0mm 0mm 0mm, clip, width=5.0cm, height=5.0cm]{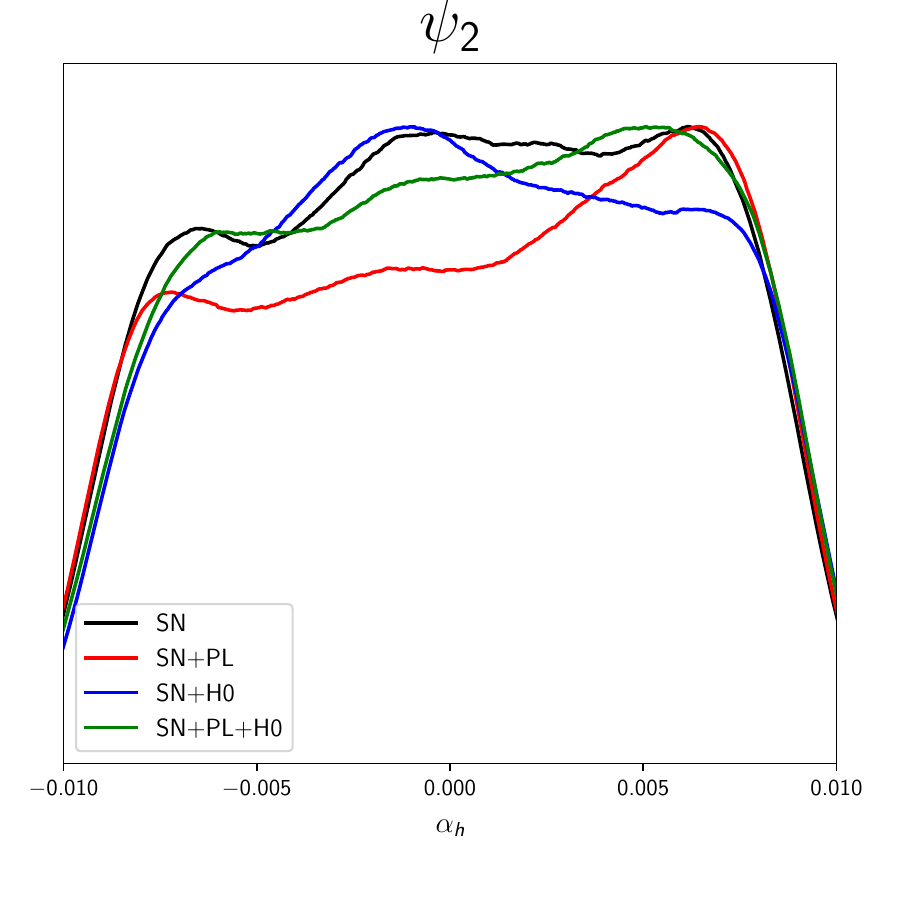}
    \includegraphics[trim = 0mm  0mm 0mm 0mm, clip, width=5.0cm, height=5.0cm]{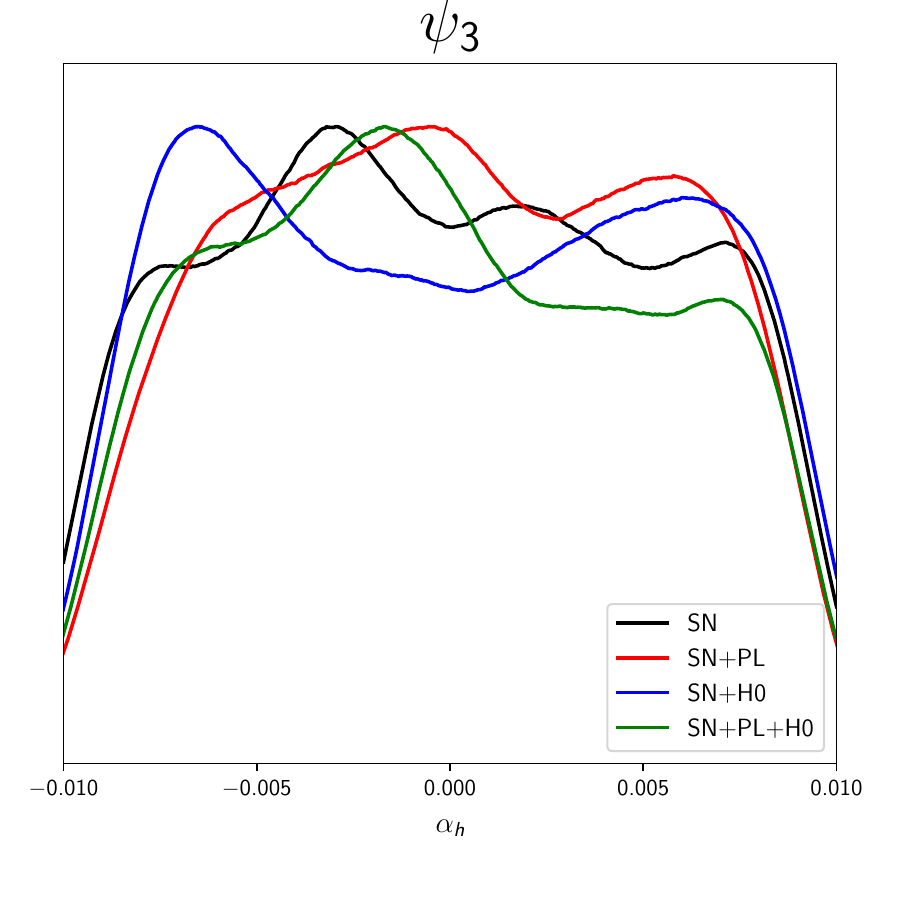}
    \includegraphics[trim = 0mm  0mm 0mm 0mm, clip, width=5.0cm, height=5.0cm]{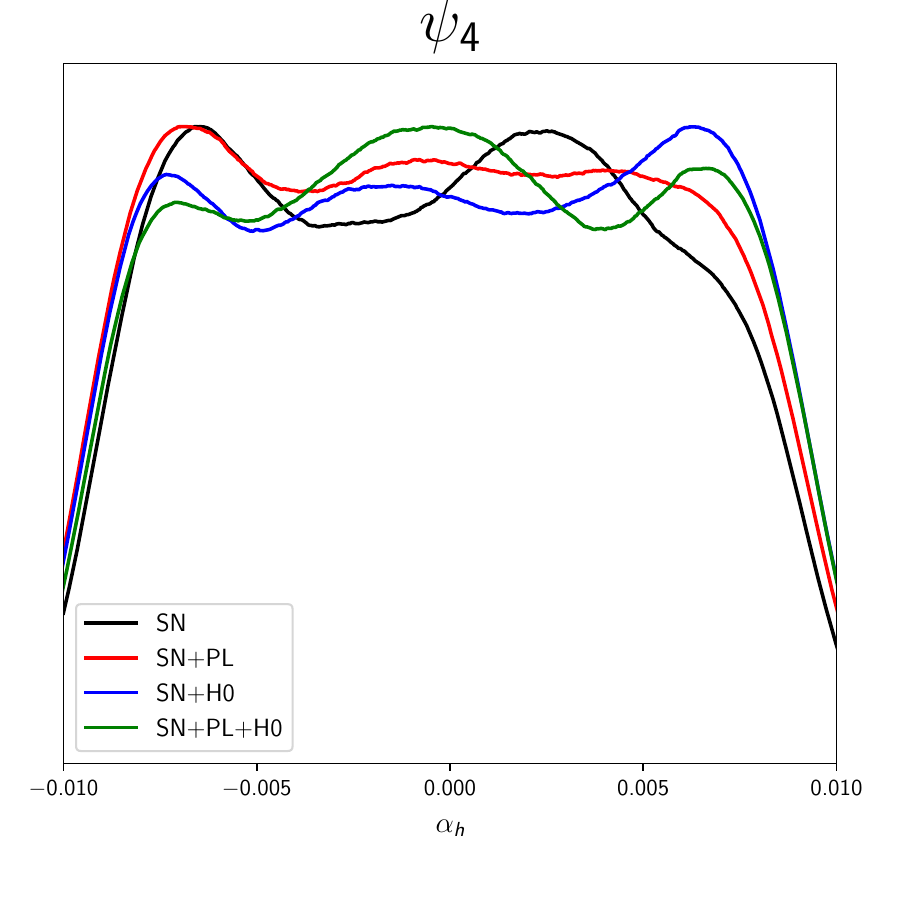}}

    \makebox[11cm][c]{
    \includegraphics[trim = 0mm  0mm 0mm 0mm, clip, width=5.0cm, height=5.0cm]{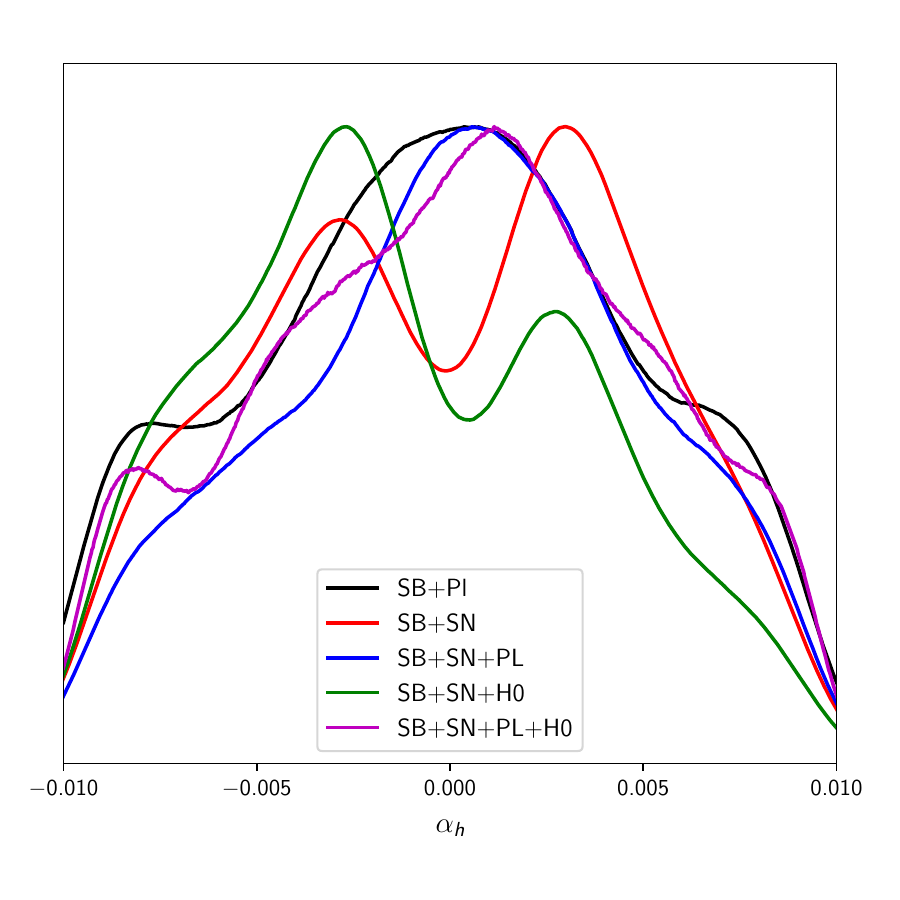}
    \includegraphics[trim = 0mm  0mm 0mm 0mm, clip, width=5.0cm, height=5.0cm]{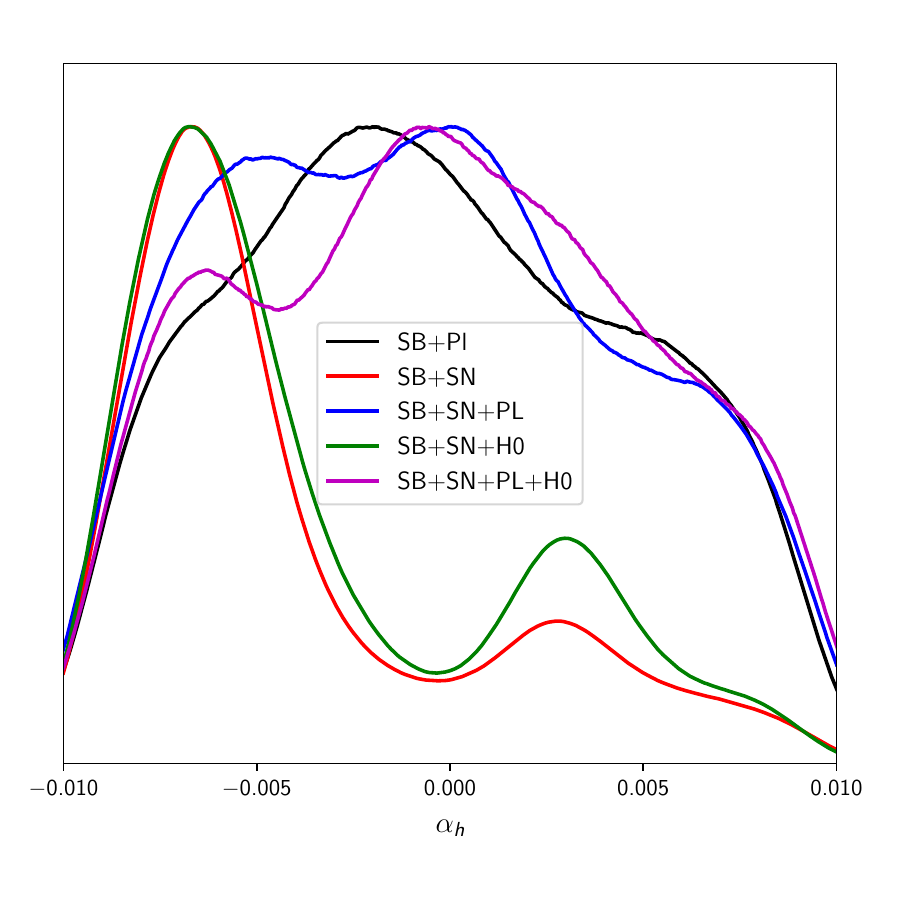}
    \includegraphics[trim = 0mm  0mm 0mm 0mm, clip, width=5.0cm, height=5.0cm]{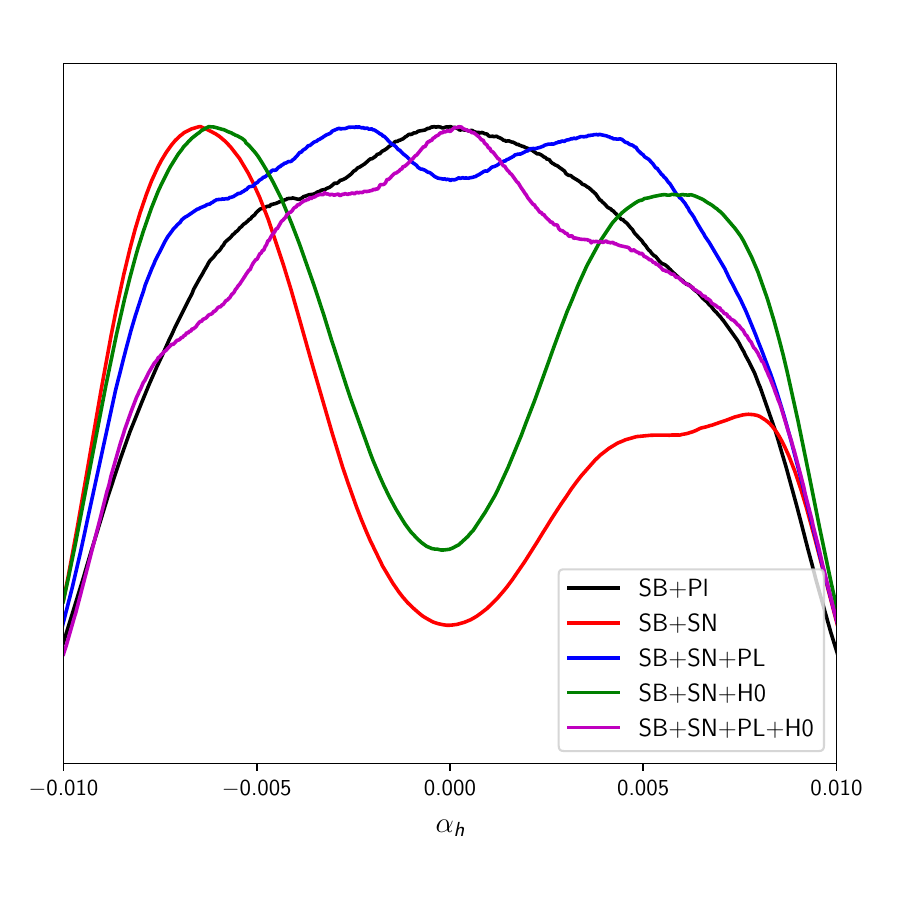}
    \includegraphics[trim = 0mm  0mm 0mm 0mm, clip, width=5.0cm, height=5.0cm]{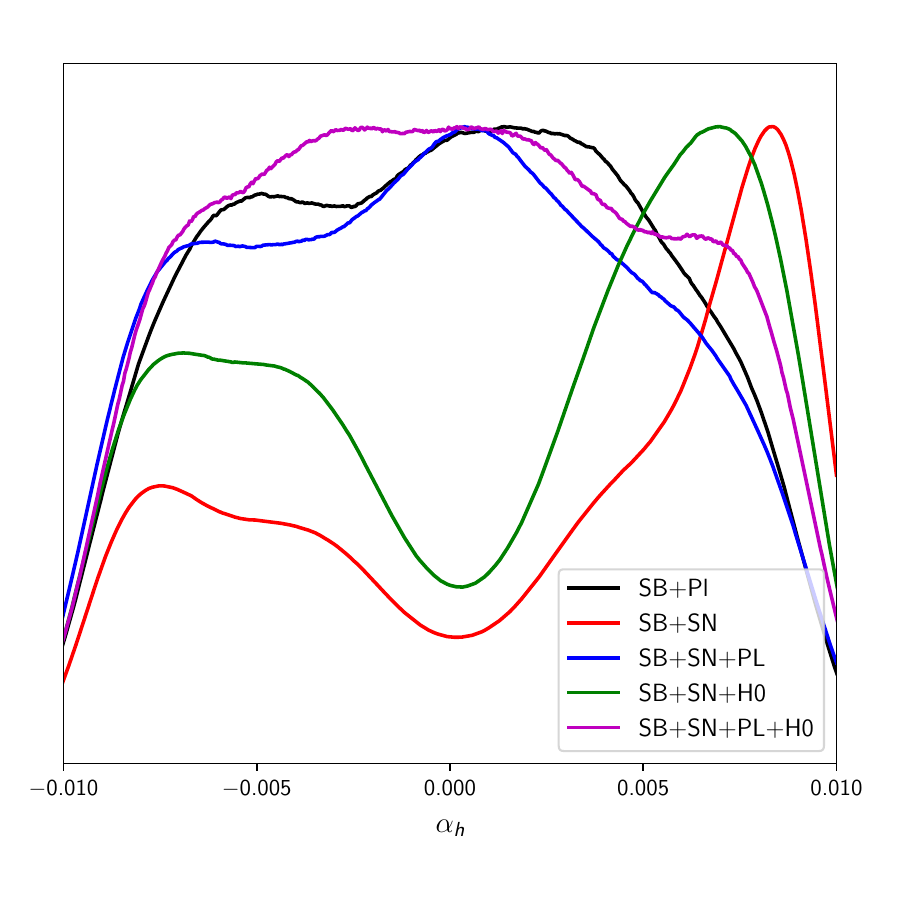}}
    
    \makebox[11cm][c]{
    \includegraphics[trim = 0mm  0mm 0mm 0mm, clip, width=5.0cm, height=5.0cm]{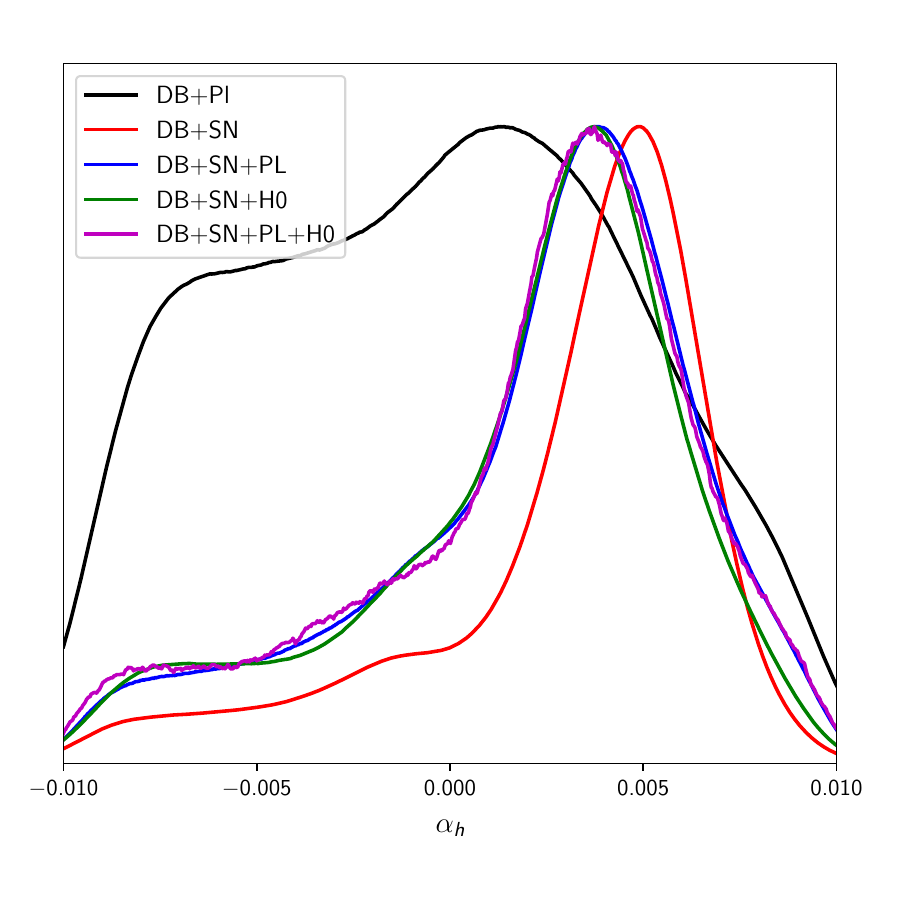}
    \includegraphics[trim = 0mm  0mm 0mm 0mm, clip, width=5.0cm, height=5.0cm]{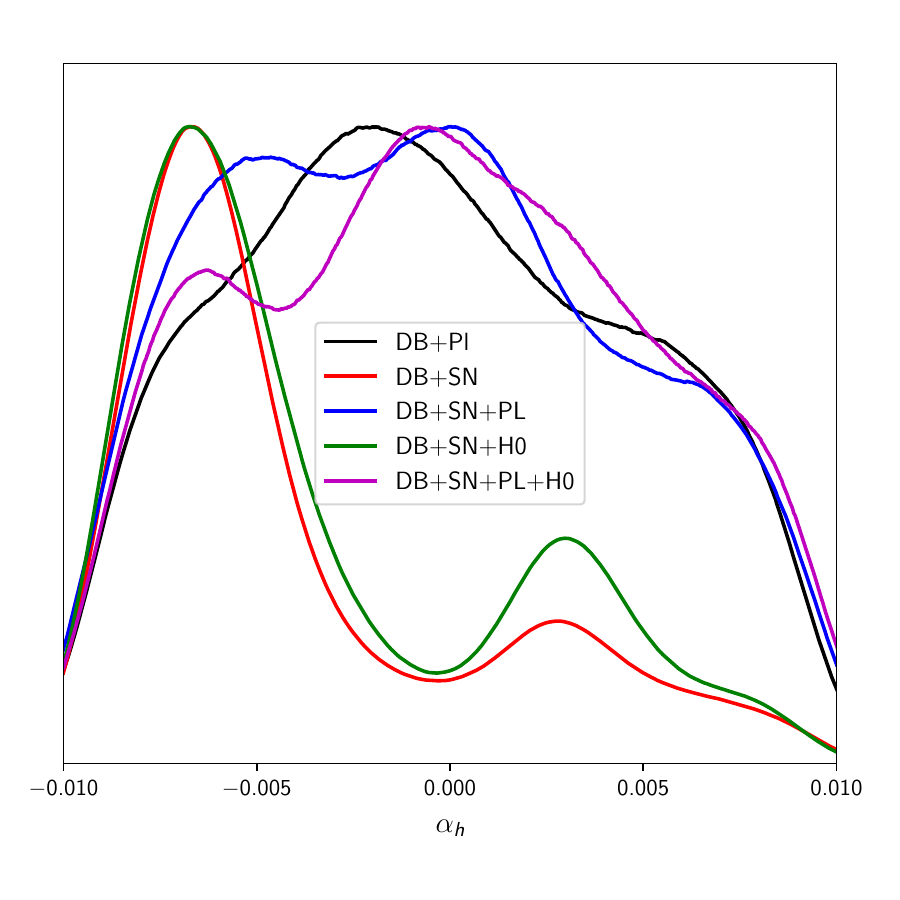}
    \includegraphics[trim = 0mm  0mm 0mm 0mm, clip, width=5.0cm, height=5.0cm]{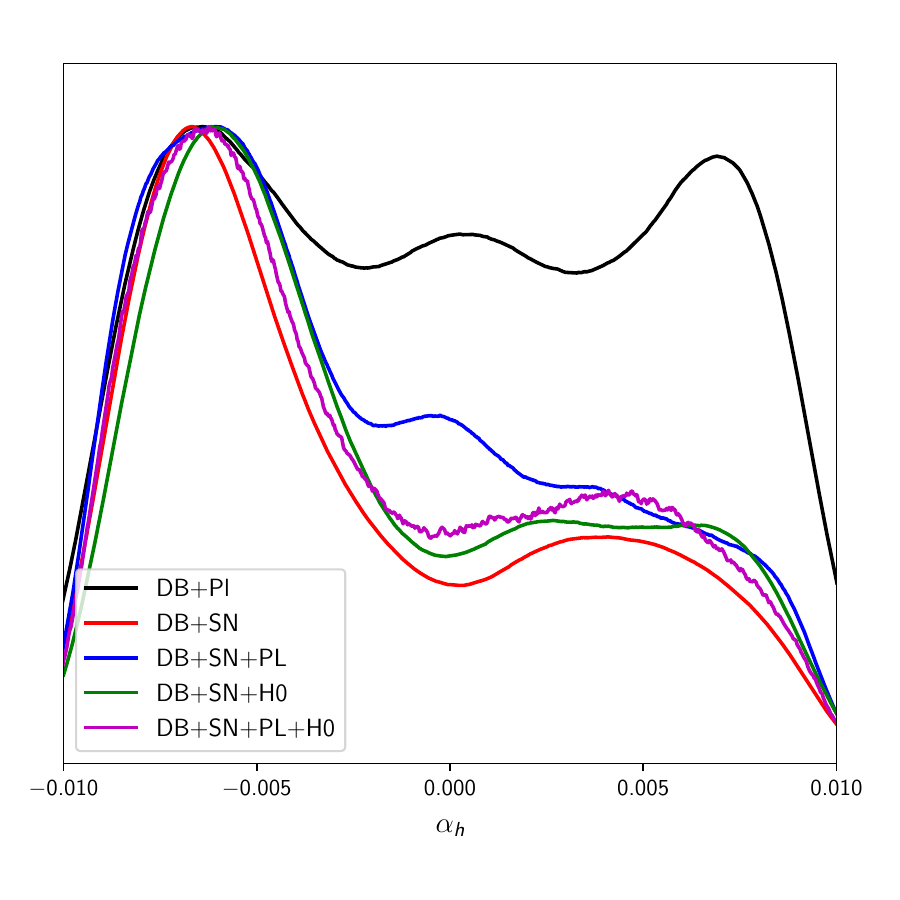}
    \includegraphics[trim = 0mm  0mm 0mm 0mm, clip, width=5.0cm, height=5.0cm]{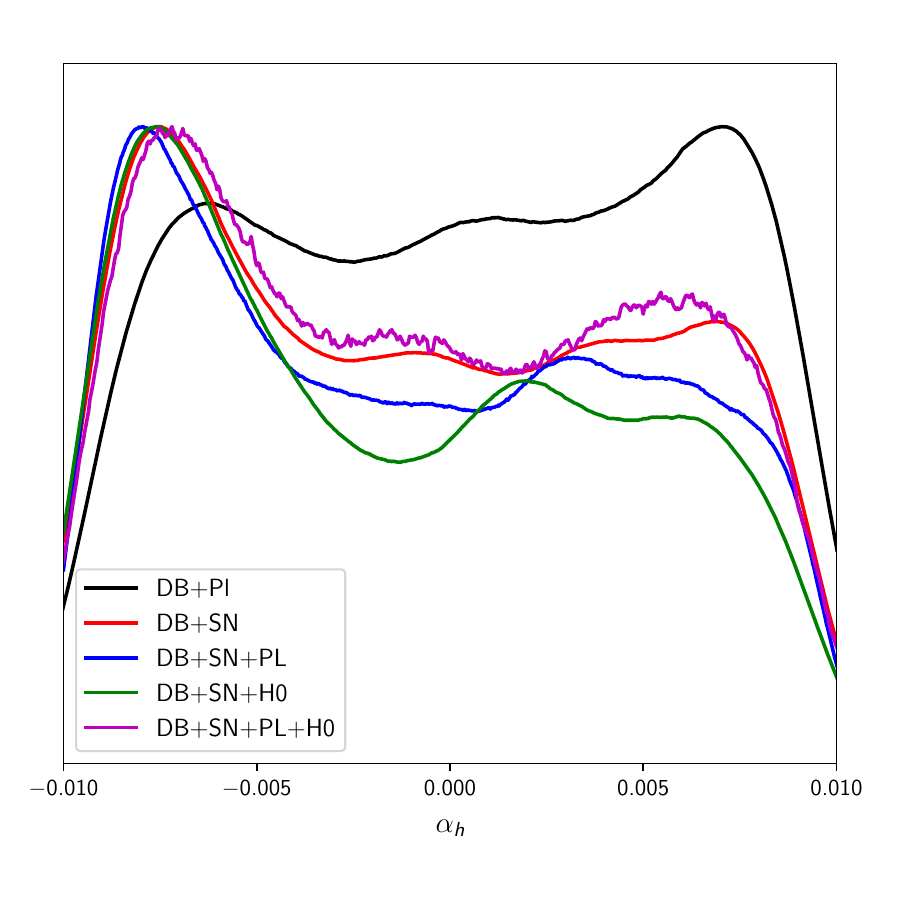}}
    \caption{1D marginalized posterior distributions for the reconstructed parameter $\alpha_h$ for, from left to right, $\psi_2(z)$, $\psi_3(z)$, and $\psi_4(z)$. We present the results in three different panels to illustrate the effect of the BAO datasets on the posteriors. These plots were created using the Python library \texttt{getdist}~\cite{Lewis:2019xzd}.
}\label{fig:1d_comparison_alpha}
\end{figure*}

\begin{figure*}[t!]
\captionsetup{justification=Justified,singlelinecheck=false,font=footnotesize}
    \centering
    \makebox[11cm][c]{
    \includegraphics[trim = 0mm  0mm 0mm 0mm, clip, width=5.0cm, height=5.0cm]{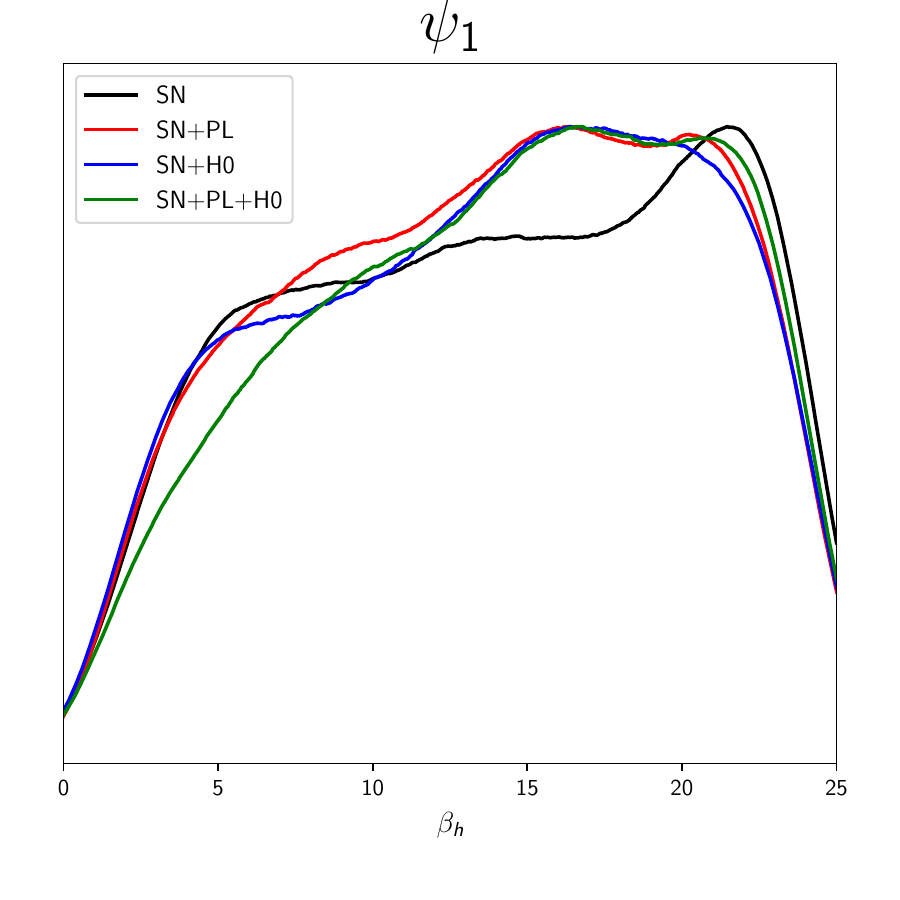}
    \includegraphics[trim = 0mm  0mm 0mm 0mm, clip, width=5.0cm, height=5.0cm]{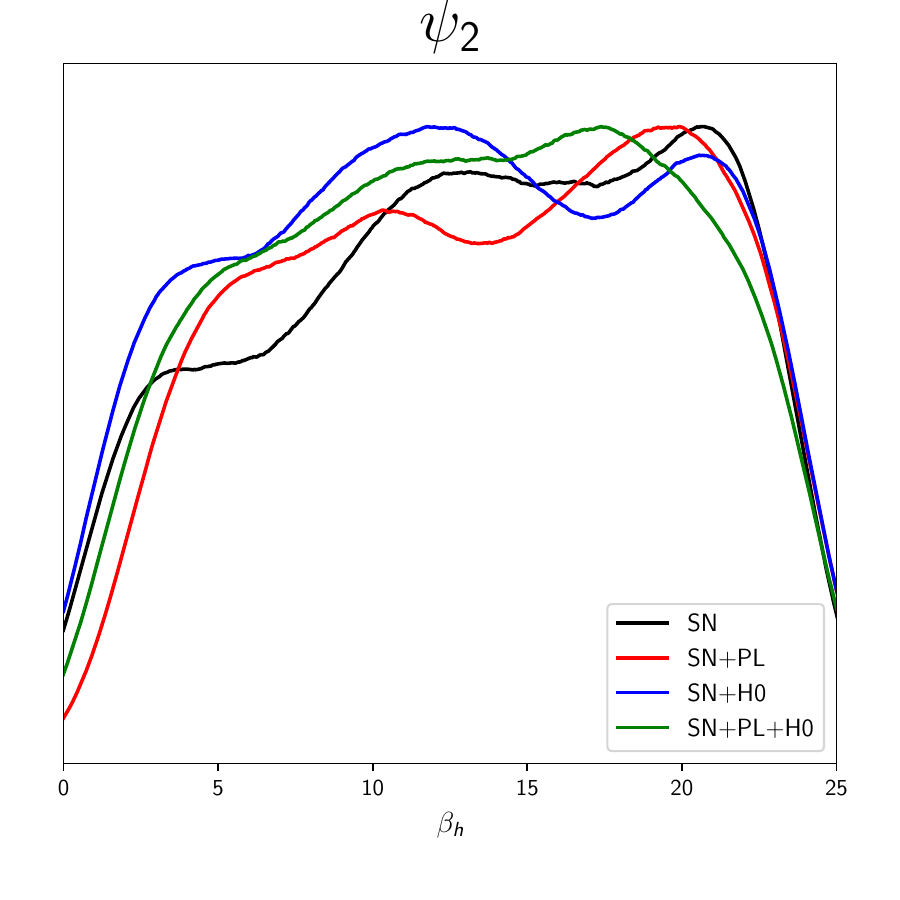}
    \includegraphics[trim = 0mm  0mm 0mm 0mm, clip, width=5.0cm, height=5.0cm]{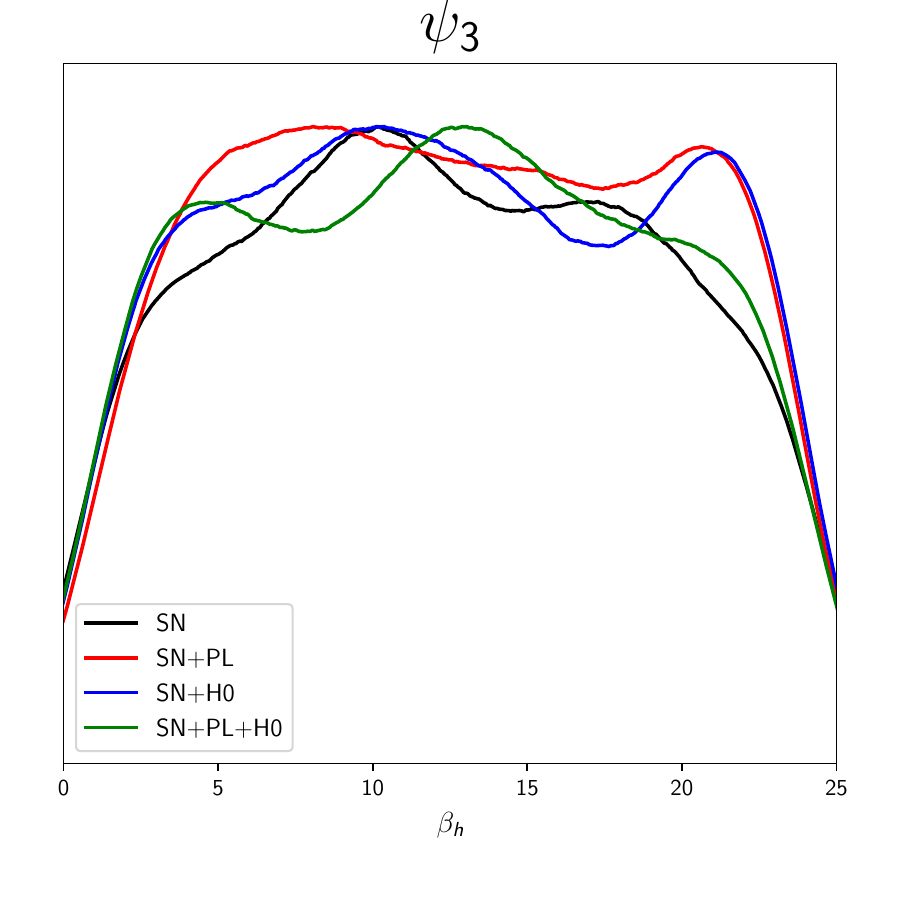}
    \includegraphics[trim = 0mm  0mm 0mm 0mm, clip, width=5.0cm, height=5.0cm]{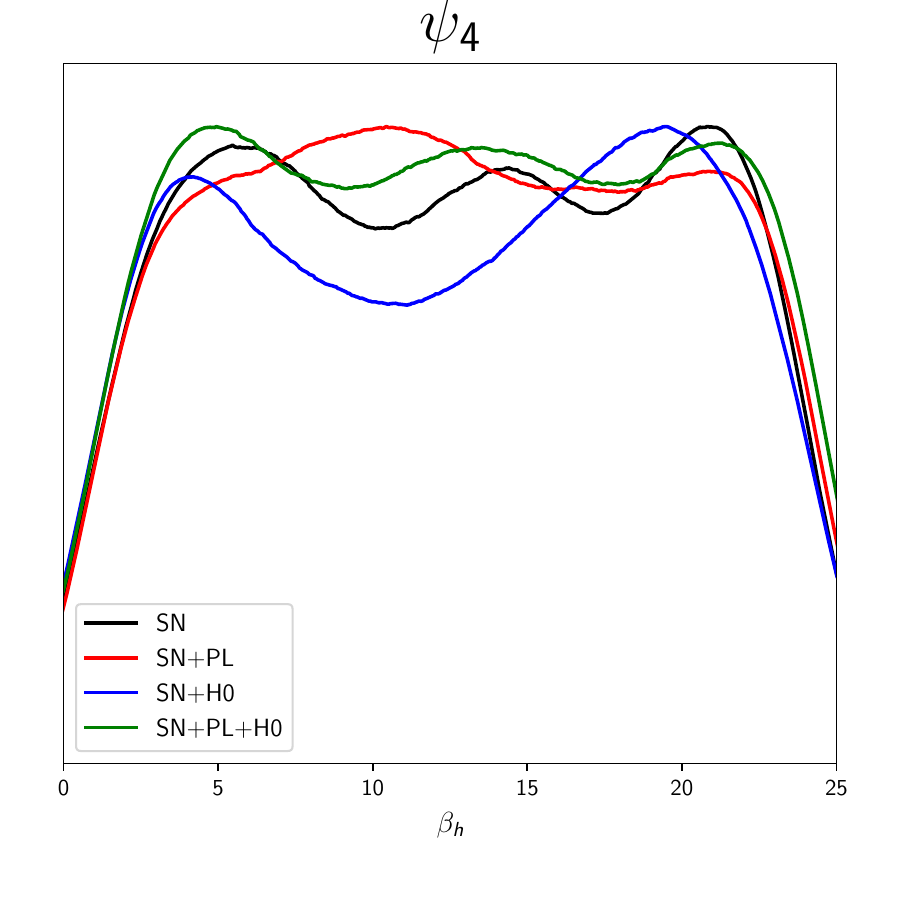}}

    \makebox[11cm][c]{
    \includegraphics[trim = 0mm  0mm 0mm 0mm, clip, width=5.0cm, height=5.0cm]{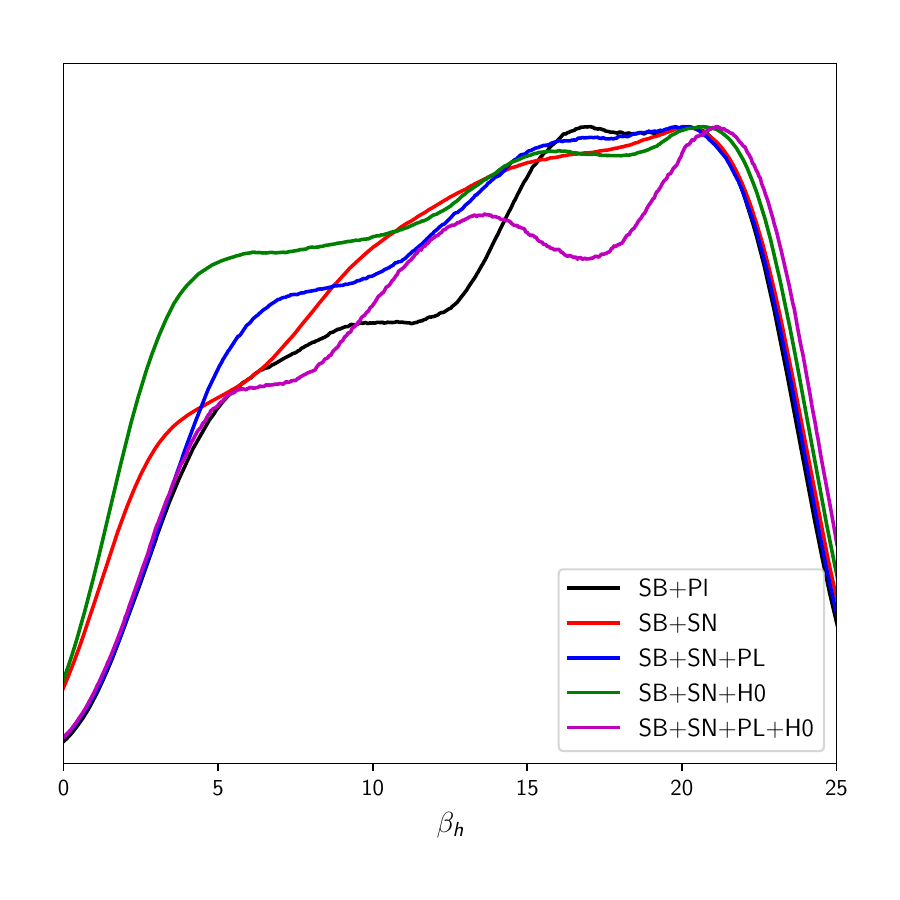}
    \includegraphics[trim = 0mm  0mm 0mm 0mm, clip, width=5.0cm, height=5.0cm]{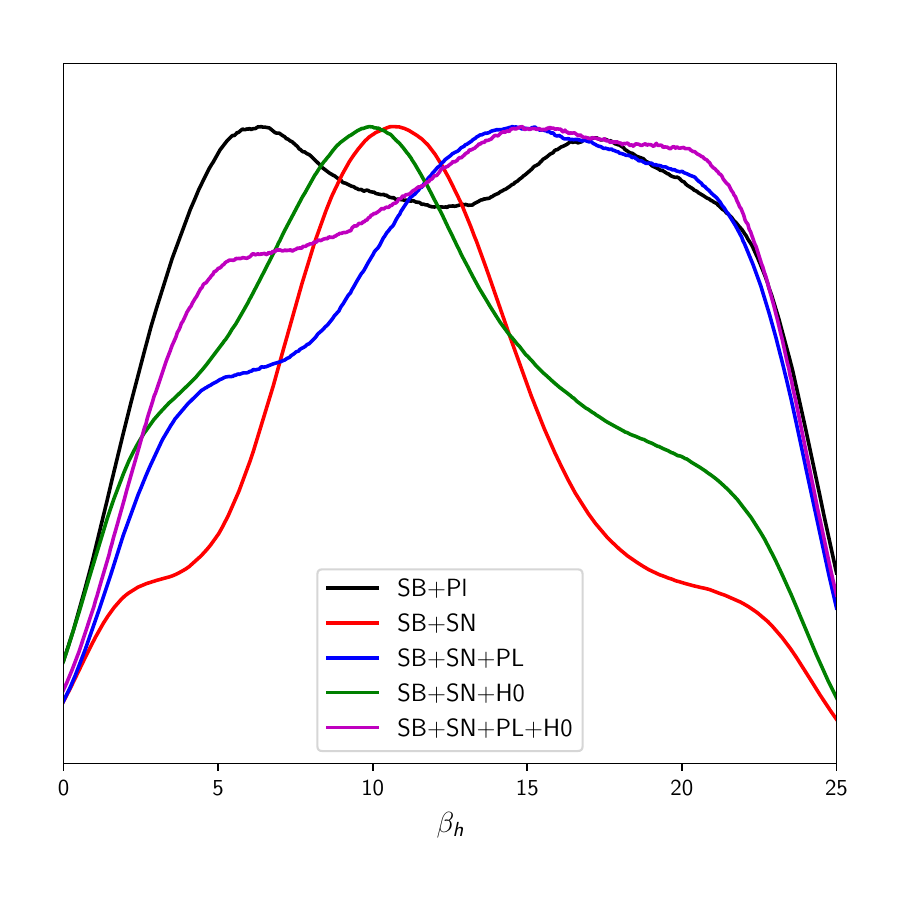}
    \includegraphics[trim = 0mm  0mm 0mm 0mm, clip, width=5.0cm, height=5.0cm]{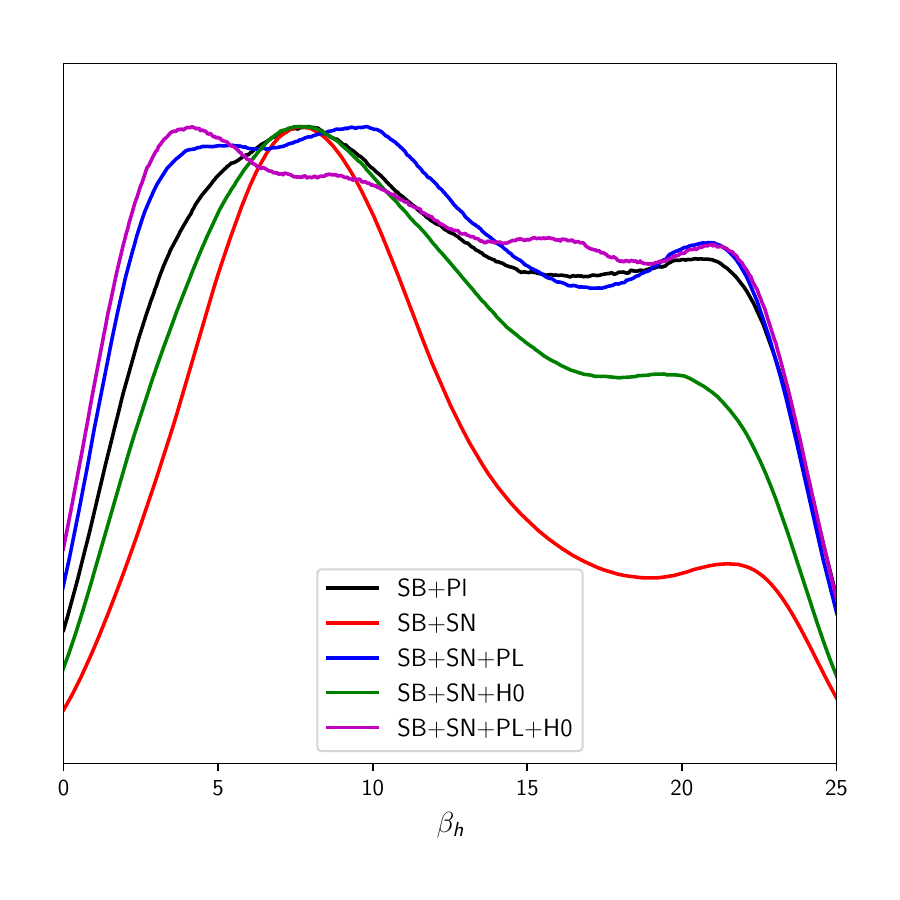}
    \includegraphics[trim = 0mm  0mm 0mm 0mm, clip, width=5.0cm, height=5.0cm]{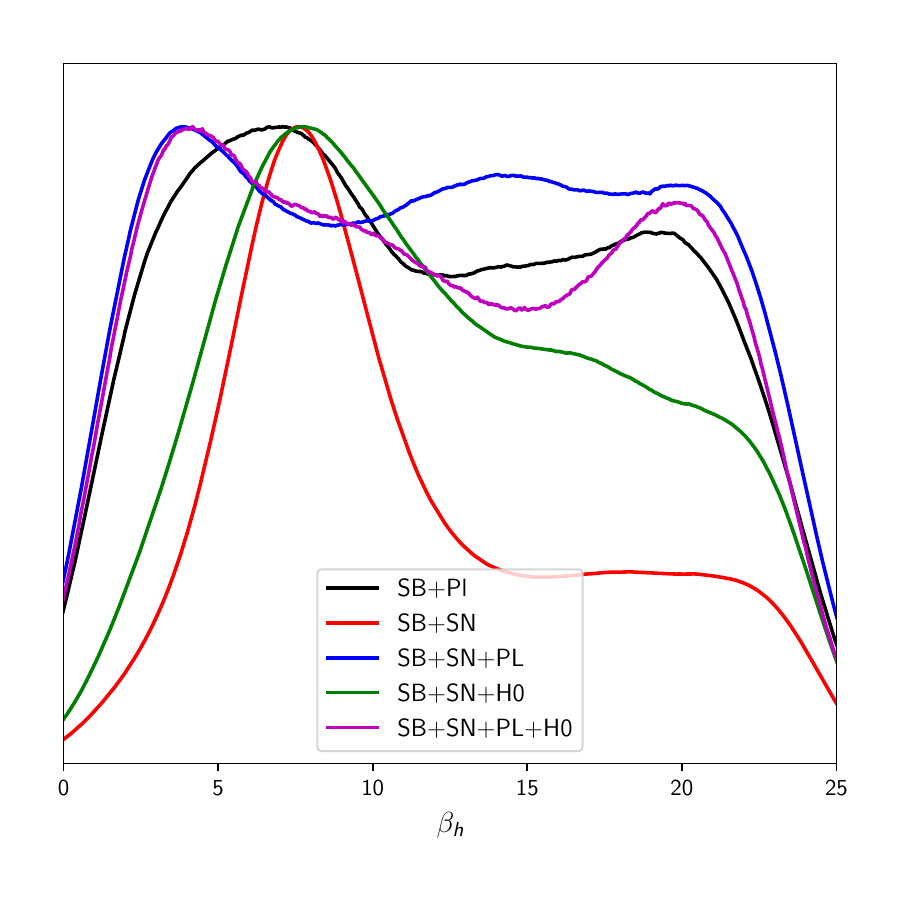}}
    
    \makebox[11cm][c]{
    \includegraphics[trim = 0mm  0mm 0mm 0mm, clip, width=5.0cm, height=5.0cm]{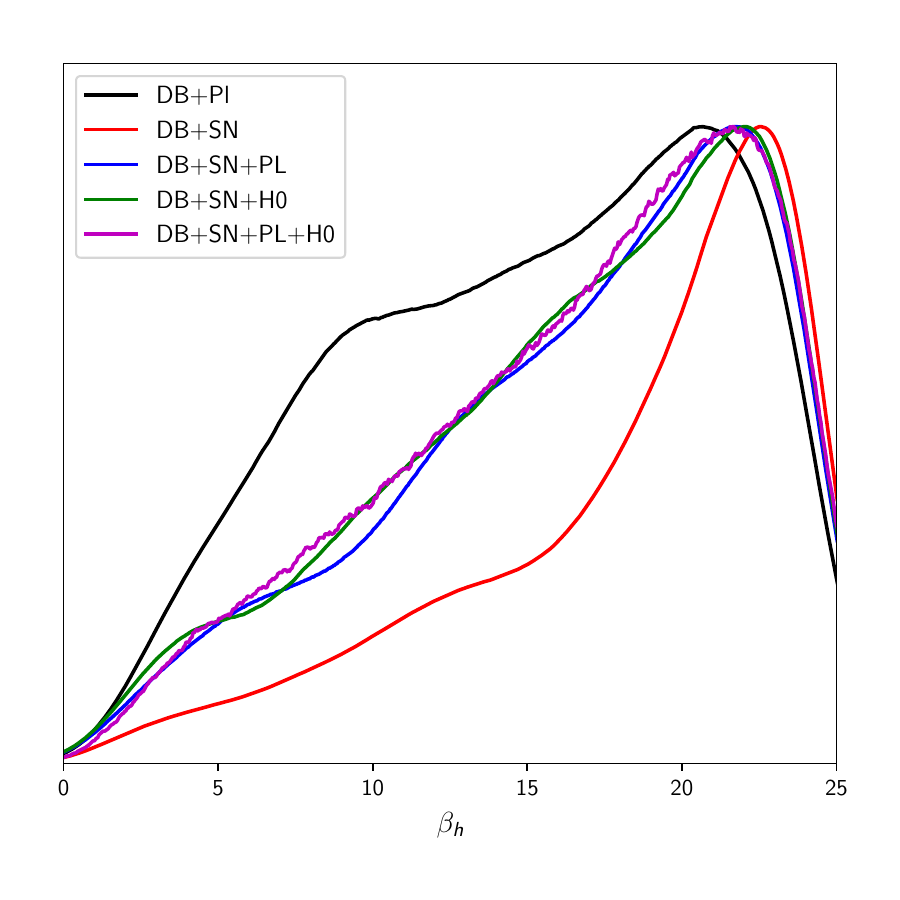}
    \includegraphics[trim = 0mm  0mm 0mm 0mm, clip, width=5.0cm, height=5.0cm]{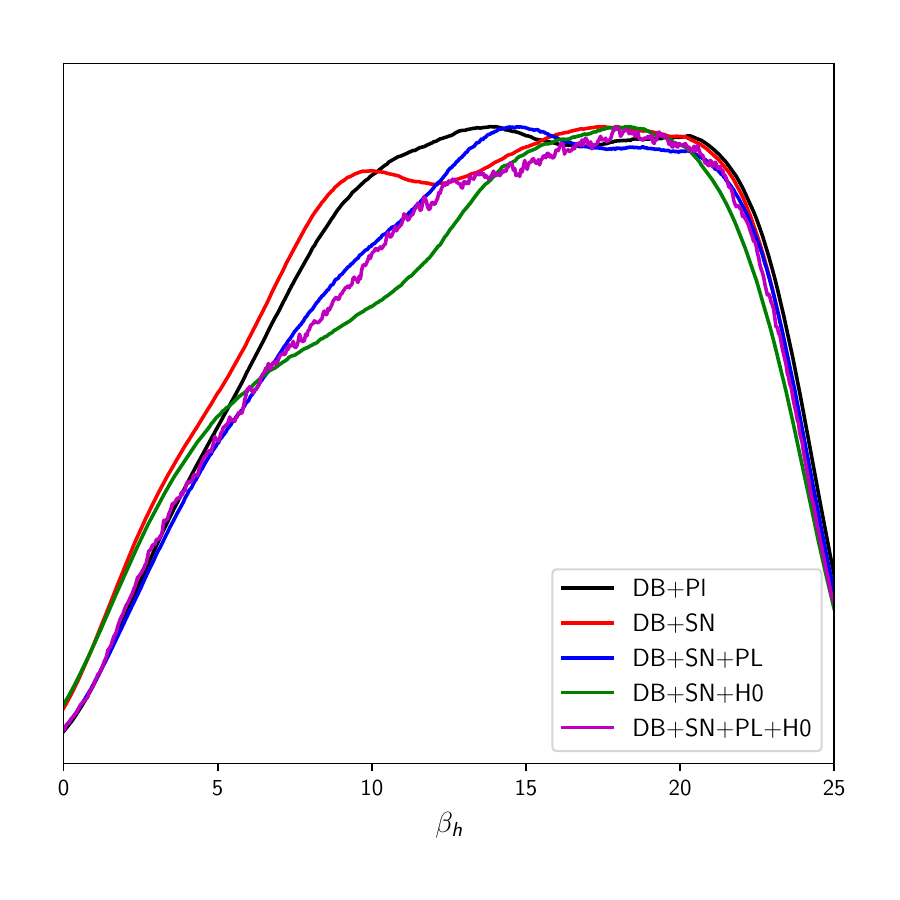}
    \includegraphics[trim = 0mm  0mm 0mm 0mm, clip, width=5.0cm, height=5.0cm]{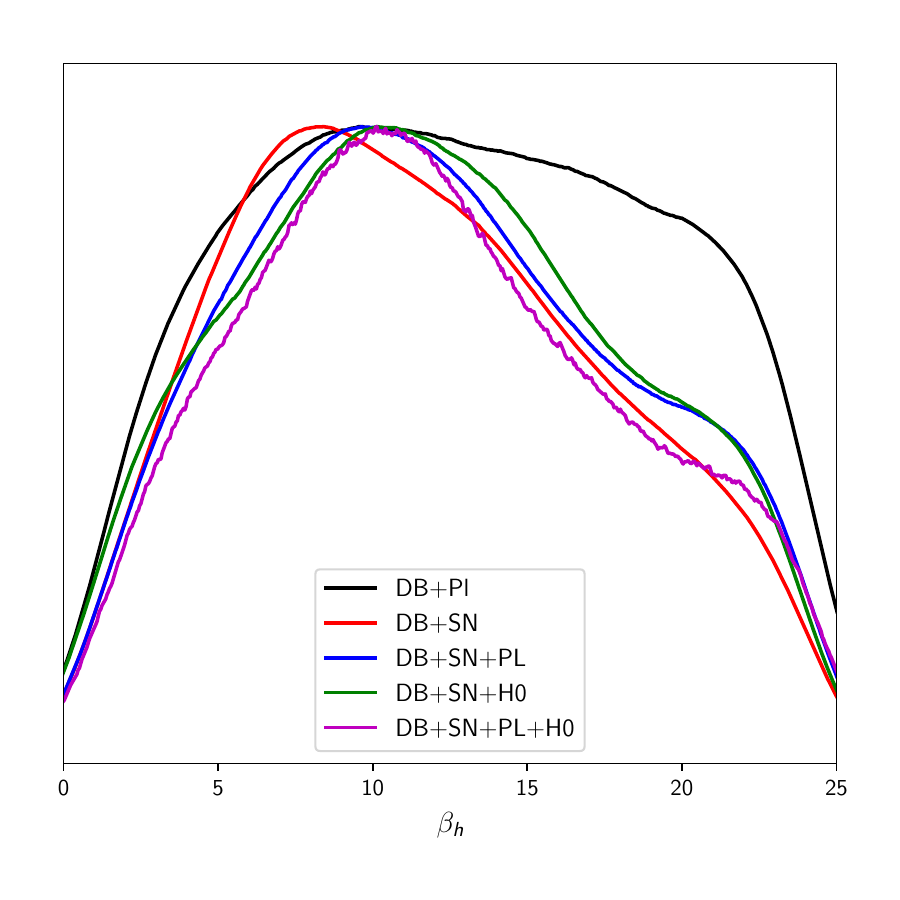}
    \includegraphics[trim = 0mm  0mm 0mm 0mm, clip, width=5.0cm, height=5.0cm]{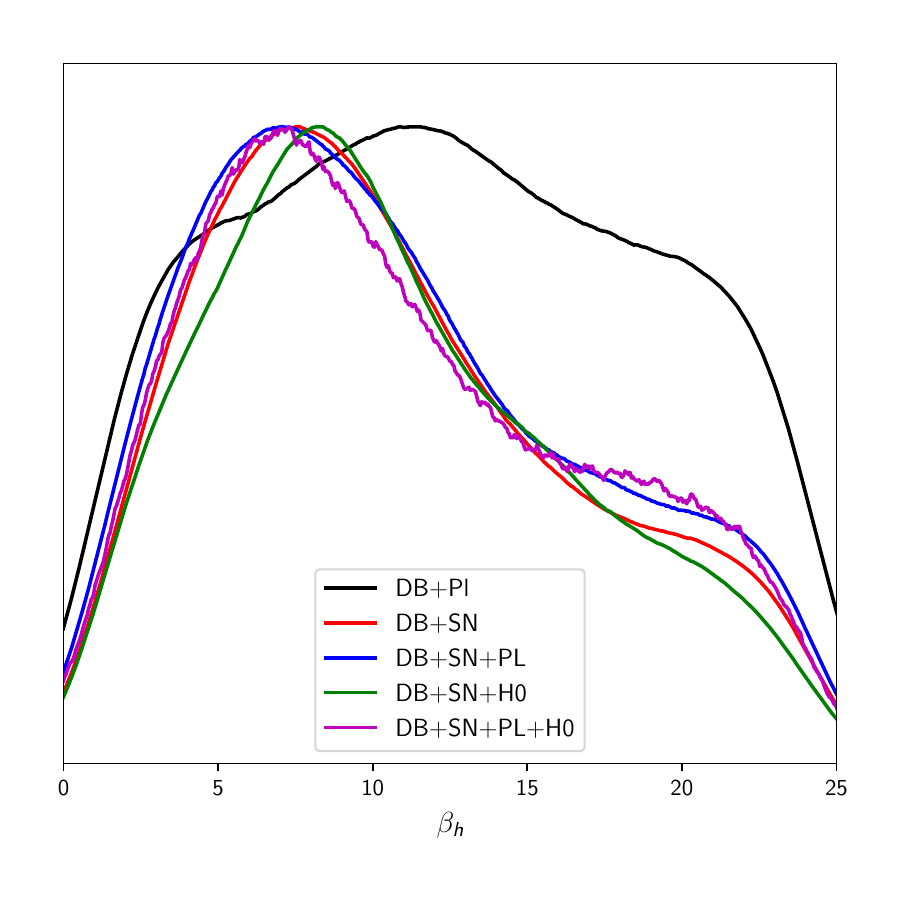}}
    \caption{1D marginalized posterior distributions for the reconstructed parameter $\beta_h$ for, from left to right, $\psi_1(z)$, $\psi_2(z)$, $\psi_3(z)$, and $\psi_4(z)$. We present the results in three different panels to illustrate the effect of the BAO datasets on the posteriors. These plots were created using the Python library \texttt{getdist}~\cite{Lewis:2019xzd}.
}\label{fig:1d_comparison_beta}
\end{figure*}

A topic that requires some discussion is the result of the parameter inference procedure on the other two wavelet parameters, $\alpha_h$ and $\beta_h$, found in~\cref{fig:1d_comparison_alpha} and~\cref{fig:1d_comparison_beta}, respectively. For $\alpha_h$, we observe a similar behavior to the one displayed by $z^\dagger$, in the sense that the BAO dataset is the primary driver of the parameter inference, not \textbf{SN} or \textbf{Pl}. We also observe bimodality in some cases and a lack of convergence in others. The parameter $\beta_h$, on the other hand, has very little impact on the datasets used. In~\cref{fig:1d_comparison_beta}, it is evident that $\beta_h$ remains largely unconstrained across most cases within the prior range considered. A few exceptions exist, where a faint maximum appears, such as in $\psi_2$ or $\psi_4$ with \textbf{SB+SN}. This lack of a well-defined posterior is expected for this particular parameter, given its minimal impact on the overall wavelet behavior (as shown in~\cref{fig:wavelet_varying_params}) and the limited data available beyond $z\sim2.3$, as discussed in~\cref{section:results}. 
While these parameters do play a role in shaping the wavelet, their overall impact is less significant than that of $z^\dagger$. Therefore, their discussion has been reserved for this appendix.

Finally, and to aid in the understanding and clarity of the results presented in this study with regards to the reconstructed $H(z)/(1+z)$, we include additional figures that provide enhanced visualizations of the wavelet behavior in the late Universe. These figures are artistically adapted versions of the previously shown plots in~\cref{fig:hz_different_datasets}, designed to improve readability and facilitate a clearer interpretation of the data.

\begin{figure*}[htb!]
\captionsetup{justification=Justified,singlelinecheck=false,font=footnotesize}
    \centering
    \makebox[11cm][c]{
    \includegraphics[trim = 0mm  0mm 0mm 0mm, clip, width=5.cm, height=4.cm]{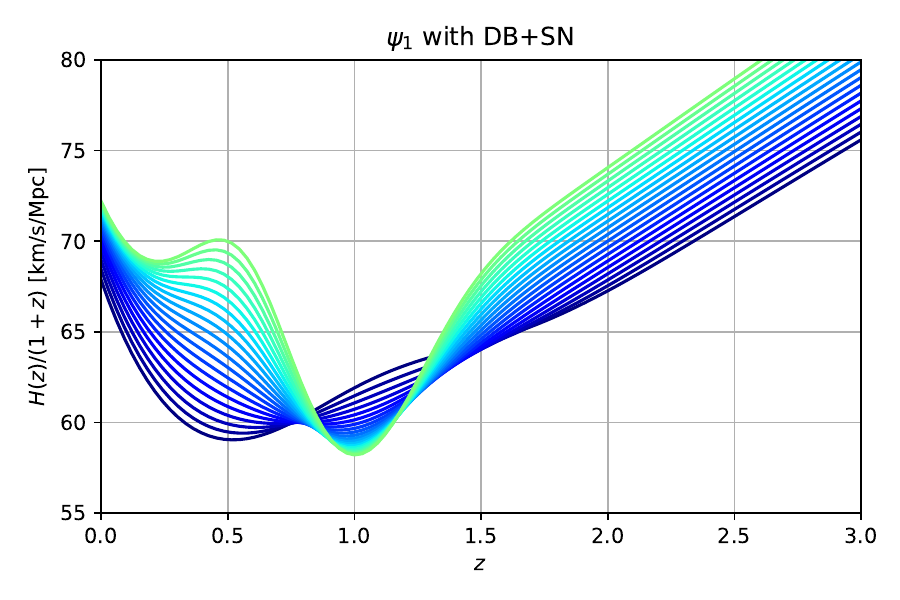}
    \includegraphics[trim = 10mm  0mm 0mm 0mm, clip, width=4.88cm, height=4.cm]{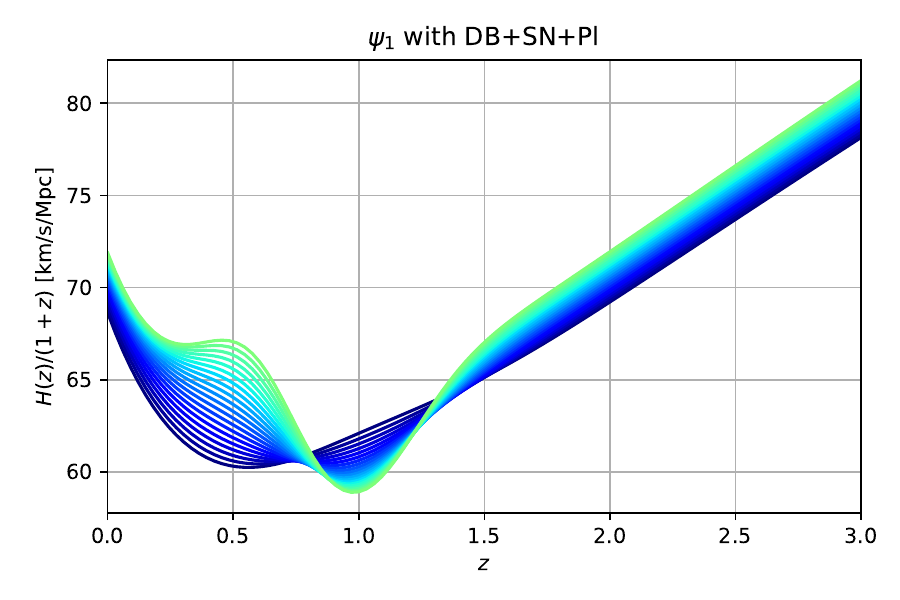}
    \includegraphics[trim = 10mm  0mm 0mm 0mm, clip, width=4.88cm, height=4.cm]{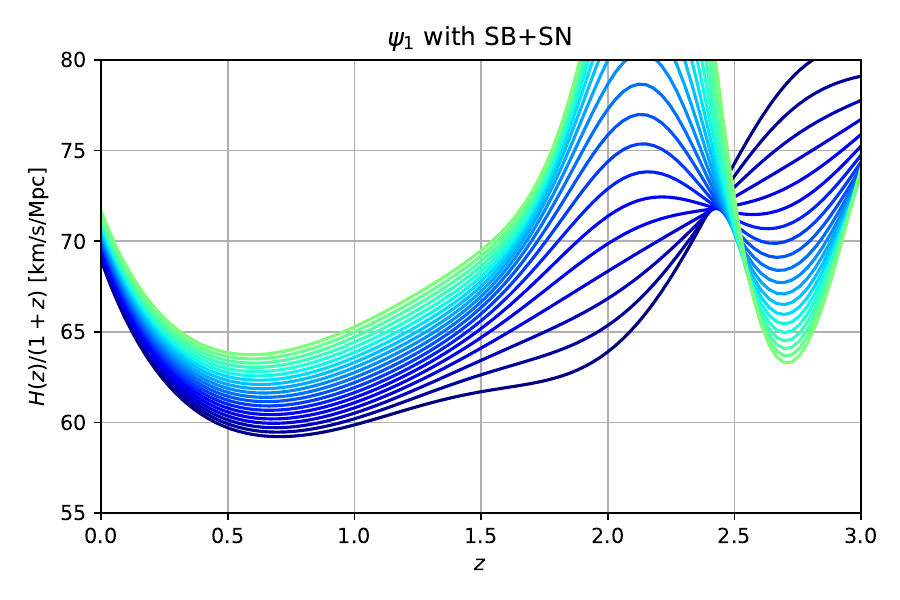}
    \includegraphics[trim = 10mm  0mm 0mm 0mm, clip, width=4.88cm, height=4.cm]{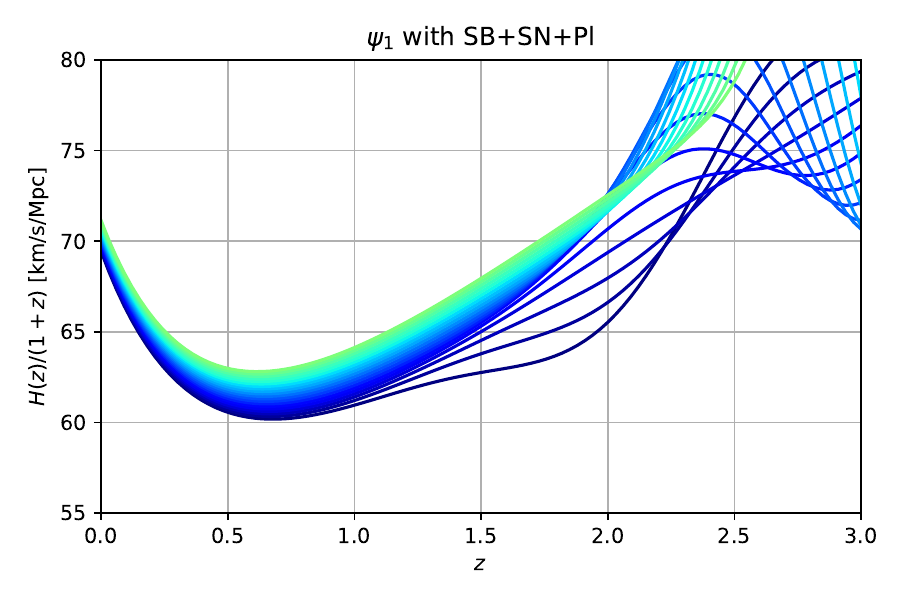}
    }
    \caption{
    Artistically adapted version of~\cref{fig:hz_different_datasets} to enhance the readability of wavelet behavior in the late Universe.
}\label{fig:hz_different_datasets_v2}
\end{figure*}

\section*{Acknowledgements}
L.A.E. was supported by CONAHCyT M\'exico. E.\"{O}.\ acknowledges support from the T\"{u}rkiye Bilimsel ve Teknolojik Ara\c{s}t{\i}rma Kurumu (T\"{U}B\.{I}TAK, Scientific and Technological Research Council of Turkey) through the 2214/A National Graduate Scholarship Program.
\"{O}.A. acknowledges the support by the Turkish Academy of Sciences in scheme of the Outstanding Young Scientist Award  (T\"{U}BA-GEB\.{I}P).  E.D.V. is supported by a Royal Society Dorothy Hodgkin Research Fellowship.
J.A.V. acknowledges the support provided by FOSEC SEP-CONACYT Investigaci\'on B\'asica A1-S-21925, FORDECYT-PRONACES-CONACYT/304001/2020 and UNAM-DGAPA-PAPIIT IN117723. We also want to thank the Unidad de C\'omputo of ICF-UNAM for their assistance in the maintenance and use of the computing equipment. This article is based upon work from COST Action CA21136 Addressing observational tensions in cosmology with systematics and fundamental physics (CosmoVerse) supported by COST (European Cooperation in Science and Technology).

\bibliographystyle{apsrev4-2_mod}
\bibliography{bibliography}


\end{document}